\documentclass{article}

\PassOptionsToPackage{numbers, compress}{natbib}

\usepackage[preprint]{neurips_2026}

\usepackage[utf8]{inputenc} 
\usepackage[T1]{fontenc}    
\usepackage[colorlinks=true, linkcolor=red, urlcolor=magenta, citecolor=blue]{hyperref}
\usepackage{url}            
\usepackage{booktabs}       
\usepackage{amsfonts}       
\usepackage{amsmath}        
\usepackage{nicefrac}       
\usepackage{microtype}      
\usepackage{xcolor}         
\usepackage{graphicx}       
\usepackage{colortbl}       

\definecolor{best}{HTML}{FFDAB9}    
\definecolor{second}{HTML}{D6EAF8}  

\usepackage{xspace}
\makeatletter
\DeclareRobustCommand\onedot{\futurelet\@let@token\@onedot}
\def\@onedot{\ifx\@let@token.\else.\null\fi\xspace}
\def\eg{\emph{e.g}\onedot}

\def\etal{\emph{et al}\onedot}
\makeatother

\newcommand{\xg}[1]{#1}

\title{ClothTransformer: Unified Latent-Space Transformers for Scalable Cloth Simulation}

\author{%
  Yu Zhang$^{1}$ \quad Yidi Shao$^{2}$ \quad Wenqi Ouyang$^{1}$ \quad Yushi Lan$^{3}$ \quad Zhexin Liang$^{1}$ \\
  \bf Chengrui Wu$^{4}$ \quad Xudong Xu$^{5}$ \quad Xingang Pan$^{1}$
  \\[0.5ex]
  $^{1}$S-Lab, Nanyang Technological University, Singapore \quad
  $^{2}$Feeling AI \\
  $^{3}$University of Oxford \quad
  $^{4}$Nanyang Technological University \quad
  $^{5}$Shanghai AI Laboratory
}

\begin{document}

\maketitle

\begin{figure}[h]
  \centering
  \includegraphics[width=\linewidth]{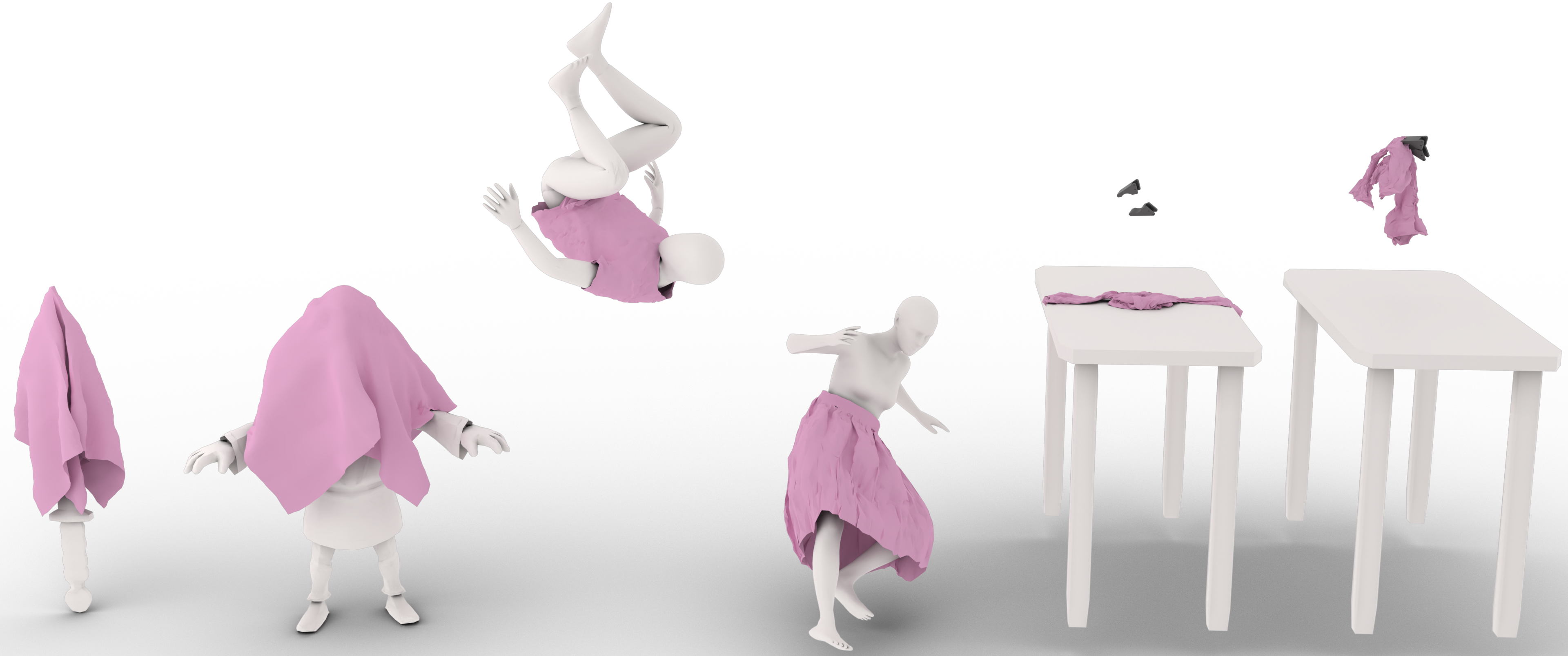}
  \vspace{-5mm}
  \caption{\textbf{ClothTransformer} generalizes to unseen test cases across three diverse scenarios. \textit{Left two:} Diverse Object Collision---cloth falling onto unseen rigid objects (sword, character). \textit{Middle two:} Human Garment---unseen body, garment, and animation combinations (front-flip, dancing). \textit{Right two:} Robotic Manipulation---unseen cloth meshes grasped and lifted by a robotic gripper.}
  \label{fig:teaser}
\end{figure}

\begin{abstract}
Unified and scalable Transformers have recently achieved remarkable success in modeling diverse phenomena traditionally associated with computer graphics, such as 3D visual effects, rendering processes, and motion in videos.
In this work, we take a step further by investigating whether modern Transformer techniques can tackle the challenging task of cloth simulation.
To this end, we present ClothTransformer, a framework that reformulates cloth simulation as autoregressive sequence modeling in a learned latent space. Existing neural cloth simulators are largely specialized to single scenarios, intrinsically coupled to the mesh discretization, and lack robust collision handling. Our approach addresses these limitations through three contributions: (1) a unified Transformer architecture that handles diverse scenarios---body-driven garments, robotic manipulation, and free-fall collisions---under a single model and achieves approximately $4$--$9{\times}$ lower error than prior state-of-the-art methods across all scenarios; (2) a scalable latent-space formulation that compresses arbitrary-resolution meshes into a fixed-size set of latent tokens, making temporal dynamics computation independent of mesh resolution; and (3) a diverse-scenario high-fidelity penetration-free dataset of ${\sim}$493.4k frames  spanning all three settings, which enables a differentiable Continuous Collision Detection (CCD) module to suppress penetration artifacts. Project Page: \href{https://yucrazing.github.io/clothtransformer/}{https://yucrazing.github.io/clothtransformer/}

\end{abstract}

\section{Introduction}
\label{sec:intro}


Realistic cloth simulation is essential for a wide range of applications.
In film and visual effects, convincing fabric motion brings digital characters to life; in gaming and virtual reality, interactive garments are key to immersion; and in embodied AI, \xg{the recent rapid development further intensifies the demand for efficient and physically plausible simulation.}  Despite decades of progress, however, simultaneously achieving high fidelity and real-time performance remains challenging. Physically Based Simulation (PBS) methods~\cite{DBLP:conf/siggraph/BaraffW98}, including advanced variational contact solvers such as IPC~\cite{DBLP:journals/tog/LiFSLZPJK20}, can produce highly accurate results; yet even with modern GPU acceleration~\cite{DBLP:journals/tog/HuangCLK24}, high-resolution cloth can still take tens of seconds per frame---far beyond real-time budgets.

Learning-based neural simulators offer a promising alternative. Most recent progress is driven by Graph Neural Networks (GNNs)~\cite{DBLP:conf/iclr/PfaffFSB21,DBLP:conf/cvpr/GrigorevBH23}, which predict vertex dynamics via message passing on mesh edges. Still, existing approaches face three fundamental limitations:

\textbf{Lack of Generalization.} Existing learning-based cloth simulators are largely specialized to a single setting---typically human-garment dressing on an animated body---or require training a separate model for each scenario. In both cases, they lack a single unified architecture and model capable of handling diverse scenarios such as robotic manipulation or free-fall collisions, which hinders their applicability to broader simulation tasks.

\textbf{The Resolution Bottleneck.} GNN simulators are tightly coupled to mesh discretization: inference cost grows with vertex/edge count. This creates a direct conflict between visual fidelity (dense meshes) and efficiency (fast, memory-light inference), undermining the core motivation of neural simulation.

\textbf{The Penetration Problem.} Almost all existing learning-based methods rely on Discrete Collision Detection (DCD) for collision handling during training, which only checks for intersections at discrete time steps and leads to the tunneling problem under fast motions (see Figure~\ref{fig:tunneling} for an illustration). Continuous Collision Detection (CCD) resolves this by sweeping the inter-frame trajectory, but demands high-quality penetration-free supervision that public datasets do not provide.



\xg{Motivated by the recent success of Transformers~\cite{vaswani2017attention} with minimal inductive bias in vision and graphics tasks, we present \textbf{ClothTransformer}, a unified Transformer-based framework that reformulates cloth simulation as autoregressive sequence modeling in a learned latent space. Transformers have proven effective at capturing complex physical rules in computer graphics, such as 3D-to-2D projection~\cite{liu2023zero} and rendering~\cite{DBLP:conf/siggraph/00010PW025}; here, we study their potential in the more challenging cloth simulation task.} 

Our framework jointly addresses all three aforementioned limitations with several key designs. The Transformer's minimal inductive bias enables a single unified architecture that handles diverse scenarios---body-driven garments, robotic manipulation, and free-fall collisions---without per-scenario tuning (Figure~\ref{fig:teaser}). To overcome the resolution bottleneck, we compress the cloth state into a compact, fixed-size set of latent vectors via cross-attention and evolve dynamics entirely in latent space, making temporal computation effectively independent of mesh resolution. 
To suppress penetration artifacts, we construct a diverse-scenario penetration-free dataset spanning all three settings, which enables a differentiable CCD loss during training and CCD post-processing at inference.
Our results highlight the strong potential of Transformer-based autoregressive models for learning-based physical simulation.

In summary, our contributions are:
\begin{itemize}
    \item We propose \textbf{ClothTransformer}, a unified Transformer architecture that handles diverse cloth simulation scenarios---body-driven garments, robotic manipulation, and free-fall collisions---under a single model, achieving approximately $4$--$9{\times}$ lower error than prior state-of-the-art methods across all scenarios.
    \item Our latent-space formulation compresses arbitrary-resolution meshes into a fixed-size set of latent tokens, making temporal dynamics computation efficient and independent of mesh resolution.
    \item We construct a diverse-scenario high-quality penetration-free dataset of ${\sim}$493.4k frames spanning body-driven garments, robotic manipulation, and free-fall collisions, which enables a differentiable Continuous Collision Detection (CCD) loss with CCD post-processing to suppress penetration artifacts.
\end{itemize}


\section{Related Work}

\subsection{Physics-Based Cloth Simulation}

Cloth simulation has a long history in computer graphics. Early mass-spring models were intuitive but numerically stiff, requiring very small time steps. Baraff and Witkin~\cite{DBLP:conf/siggraph/BaraffW98} addressed this with implicit integration, enabling large stable steps. Finite Element Methods (FEM) further improved accuracy by treating cloth as a continuum with well-defined constitutive models for stretching, bending, and shearing.

A central challenge in traditional cloth simulation is collision handling. Bridson \etal~\cite{DBLP:journals/tog/BridsonFA02} combined geometric intersection tests with repulsion forces, while IPC~\cite{DBLP:journals/tog/LiFSLZPJK20} unified contact into a variational framework guaranteeing intersection-free results. The current state-of-the-art solver GIPC~\cite{DBLP:journals/tog/HuangCLK24} further accelerated IPC with GPU-based Gauss-Newton optimization, yet a single high-resolution frame can still take tens of seconds. This trade-off between fidelity and cost motivates data-driven alternatives.

\subsection{Learning-Based Cloth Simulation}

Learning-based approaches have become increasingly popular for cloth simulation~\cite{DBLP:conf/cvpr/PatelLP20, DBLP:conf/iccv/BerticheMTE21, DBLP:conf/cvpr/SantestebanOC22, DBLP:journals/vc/HeCGLHLYLH25, DBLP:conf/cvpr/MaYRPPTB20, DBLP:conf/cvpr/SantestebanTOC21, bertiche2022neural}. 
Among them, GNN-based and Transformer-based methods are the two dominant paradigms.

\textbf{Graph Neural Networks.}
Graph Neural Network (GNN)–based methods \cite{DBLP:conf/iclr/PfaffFSB21, DBLP:conf/cvpr/GrigorevBH23, DBLP:conf/siggraph/0002BBHT24, DBLP:conf/icml/Sanchez-Gonzalez20, DBLP:conf/mig/LibaoLKL23, DBLP:journals/cgf/VidaurreSGC20, DBLP:conf/icml/CaoCLJ23} are widely adopted for learning-based cloth simulation, representing vertices as graph nodes and propagating information along mesh edges. The current state-of-the-art GNN backbone for cloth simulation, adopted by both HOOD~\cite{DBLP:conf/cvpr/GrigorevBH23} and ContourCraft~\cite{DBLP:conf/siggraph/0002BBHT24}, introduces hierarchical message passing on a multi-resolution mesh graph, enabling faster long-range information flow across the garment.

\textbf{Transformers.}
Transformer-based models have also been applied to physics simulation~\cite{DBLP:conf/eccv/ShaoLD22, DBLP:conf/iccv/ShaoL023, DBLP:journals/cgf/LiWKCS24, DBLP:conf/icml/Holzschuh0KT25, DBLP:journals/tvcg/LiSZK24, DBLP:journals/cvm/LiQLSZ25, DBLP:journals/pami/LiSZZK25, DBLP:conf/iclr/HanGPWL22, DBLP:conf/iclr/ChenXCCPMCCG23, DBLP:conf/siggrapha/ChangCWCCG23}.
LayersNet~\cite{DBLP:conf/iccv/ShaoL023} groups mesh vertices into UV-derived patch tokens to reduce token count, but the fixed UV boundaries can cause spatial discontinuities and mesh collapse. Manifold-aware Transformer~\cite{DBLP:journals/cgf/LiWKCS24} tokenizes the garment at the mesh-face level and modulates self-attention with local mesh connectivity to predict per-frame deformation gradients, yet its per-face tokenization keeps the attention cost coupled to the mesh resolution.

\textbf{Limitations of Existing Paradigms.}
Existing GNN- and Transformer-based methods share two fundamental limitations: (i) they are developed and evaluated solely within the human-garment dressing setting, with no demonstrated capability on broader scenarios such as robotic manipulation and free-fall collisions; and (ii) their inference cost remains coupled to the input mesh resolution, whether through edge-message-passing in GNNs or UV-patch / per-face tokenization in Transformers. In contrast, our method learns a \emph{data-driven} compression through cross-attention with learnable queries, reducing the dynamics complexity to $O(N_{\text{latents}}^2)$ independent of the input mesh size and avoiding scenario-specific structural priors, thereby supporting diverse cloth simulation scenarios under a single unified architecture.


\textbf{Collision Handling in Neural Cloth Simulation.}
Most neural cloth simulators rely on DCD~\cite{DBLP:conf/iclr/PfaffFSB21, DBLP:conf/cvpr/GrigorevBH23, DBLP:conf/iccv/ShaoL023, DBLP:journals/tog/RomeroCPO21, DBLP:journals/tog/RomeroCCO22} or post-hoc penalty forces, both prone to tunneling under fast motions. Alternative strategies include repulsion units~\cite{DBLP:conf/eccv/TanZWCSM22} and auxiliary self-collision graphs~\cite{DBLP:conf/eccv/LiaoWK24}. ContourCraft~\cite{DBLP:conf/siggraph/0002BBHT24} uses DCD to detect intersecting triangle pairs, groups them into intersection contours, and learns to resolve multi-garment interpenetrations as a post-processing step. However, none integrates CCD into the training loop, leaving them vulnerable to tunneling. Moreover, CCD-based training requires high-quality penetration-free ground truth; if the training data itself contains residual intersections, CCD gradients become contradictory, yet most existing datasets are generated by solvers that tolerate such artifacts. In contrast, we construct a high-fidelity penetration-free dataset with strict intersection-free guarantees, enabling a differentiable CCD loss that penalizes trajectory-level intersections during training.

\section{Methodology}
\label{sec:method}

To handle the high dimensionality and varying topology of cloth meshes, we propose a \textbf{ClothTransformer} architecture. This framework compresses the geometric and dynamic state into a compact latent representation, modeling the temporal dynamics in this latent space, and subsequently reconstructing the mesh. 

\subsection{Problem Formulation}
\label{subsec:formulation}

We formulate cloth simulation as an autoregressive sequence modeling task. Let a cloth mesh be represented as $\mathcal{M} = (\mathcal{V}, \mathcal{E})$, where $\mathcal{V}$ denotes the set of $N_v$ vertices and $\mathcal{E}$ the set of edges. To fully capture the physical state of the system, we define the state at time step $t$ using both vertex positions $\mathbf{X}_t \in \mathbb{R}^{N_v \times 3}$ and their instantaneous velocities $\mathbf{V}_t \in \mathbb{R}^{N_v \times 3}$. Formally, the input is the current cloth state $\mathcal{S}_t = \{\mathbf{X}_t, \mathbf{V}_t\}$. The system is conditioned on the collision environment, represented by the collision object mesh $\mathbf{C}_{t+1}$ at the target frame. Our goal is to learn a mapping function $F_\theta$ parameterized by a neural network that predicts the future position state:
\begin{equation}
    \hat{\mathbf{X}}_{t+1} = F_\theta(\mathbf{X}_{t}, \mathbf{V}_{t}, \mathbf{C}_{t+1} \mid \mathbf{X}_{rest})
\end{equation}
where $\mathbf{X}_{rest}$ represents the rest shape of the cloth.

\subsection{Architecture Overview}
\begin{figure}[t]
  \centering
  \includegraphics[width=\linewidth]{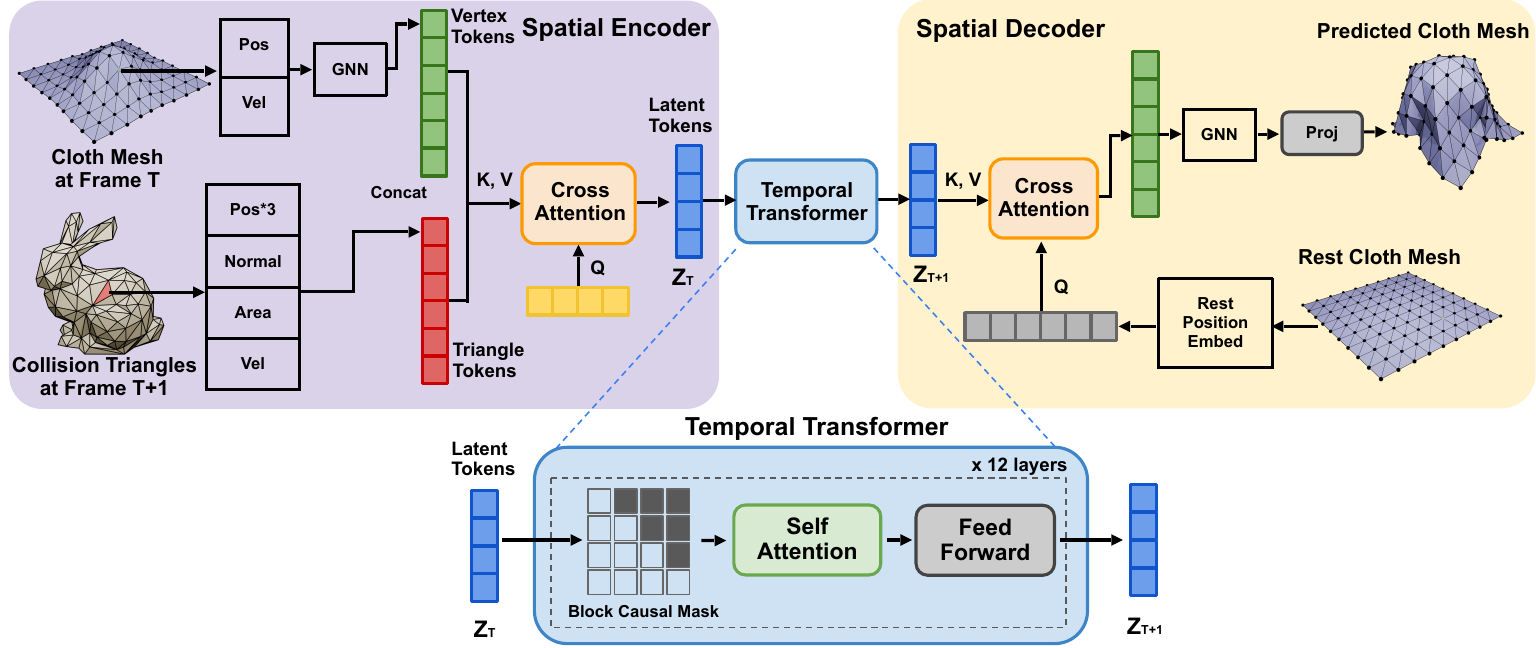}
  \caption{\textbf{Overview of the proposed auto-regressive cloth simulation architecture.}
  The framework consists of three main components:
  (1) A \textbf{Spatial Encoder} (left) that compresses the physical state of the history cloth mesh at frame \(T\) and the lookahead collision geometry at frame \(T+1\) into a compact set of latent tokens.
  (2) A \textbf{Temporal Transformer} (middle) that models the dynamics in the latent space, predicting the future latent state \(Z_{T+1}\) (representing the next logical step in the sequence) from the current state \(Z_T\).
  (3) A \textbf{Spatial Decoder} (right) that reconstructs the predicted cloth mesh from the latent representation. It queries the latent tokens using position-embedded vertices from the rest-shape mesh, followed by GNN refinement to ensure topological consistency in the final output.}
  \label{fig:architecture}
\vspace{-10pt}
\end{figure}

As illustrated in Figure~\ref{fig:architecture}, the framework consists of three primary components: (1) a Spatial Encoder that compresses the geometry and dynamics of the cloth and collision objects into latent tokens; (2) a Temporal Transformer that propagates dynamics in the latent space; and (3) a Spatial Decoder that reconstructs the vertex positions from the predicted latents. More details of our architecture can be found in the supplementary materials.

\textbf{Spatial Encoder.}
To efficiently process high-resolution meshes, we encode the physical state at frame $T$ into a fixed-size set of latent vectors $\mathbf{Z}_T \in \mathbb{R}^{K \times D}$. The encoder processes two distinct inputs: 1) The cloth mesh at frame $T$. Each input cloth vertex is processed to obtain both a position embedding and a velocity embedding. Then, these embeddings are fused and processed by a 2-layer GNN, yielding a set of topology-aware cloth vertex tokens.
2) The collision object is represented as a set of triangles from the lookahead frame $T+1$. Similar to the cloth mesh, we encode these using a combination of vertex embeddings (position and velocity) and explicit geometric descriptors, including the surface normal and triangle area. This produces collision triangle tokens. 

To decouple the latent representation from the mesh resolution, we employ a cross-attention mechanism. We initialize a set of $K$ learnable query tokens $\mathbf{Q}_{learn}$. These queries attend to the concatenated sequence of cloth vertex tokens and collision triangle tokens (acting as keys $\mathbf{K}$ and values $\mathbf{V}$). This operation compresses the variable-sized geometric and dynamic input into a fixed set of latent tokens $\mathbf{Z}_T$. The number of latent tokens $K$ is a hyperparameter that controls the trade-off between compression and accuracy.

\textbf{Temporal Transformer.}
The core dynamics are modeled by a Transformer operating on the latent tokens. The input to the transformer is the latent state $\mathbf{Z}_T$. The Transformer processes the latent representation to evolve the state forward in time. During training, we stack multi-frame past latents and use a block-causal masking (inter-frame) and self-attention layers (intra-frame) to model the complex dependencies between latent vectors. The output is the predicted latent state for the next frame $\mathbf{Z}_{T+1}$.

\textbf{Spatial Decoder.}
The decoder reconstructs the predicted cloth mesh $\hat{\mathbf{X}}_{next}$ from the evolved latent tokens output by the Temporal Transformer. The decoding process is conditioned on the cloth's rest shape to ensure the output maintains the material's intrinsic structure. We generate rest vertex tokens by applying the sinusoidal position embedding to the rest-shape vertices. These tokens serve as queries $\mathbf{Q}$ in a cross-attention layer, where the keys $\mathbf{K}$ and values $\mathbf{V}$ are the predicted latent tokens. This mechanism allows the model to retrieve the dynamic state corresponding to each specific vertex based on its canonical position. The output of the cross-attention layer represents the coarse predicted state per vertex. To ensure local surface smoothness and resolve high-frequency noise, these features are passed through a final GNN block. A projection layer then maps the refined features to 3D coordinates, yielding the final predicted cloth mesh.

\textbf{Unified Design for Diverse Scenarios.}
Notably, our architecture is \emph{scenario-agnostic}: cross-attention compression handles arbitrary numbers of cloth vertices and collision triangles, the Transformer imposes no scenario-specific priors (e.g., humanoid context or fixed UV layouts), and collision objects are encoded as generic triangle tokens that generalize across articulated bodies, robotic grippers, and rigid objects. A single instance of our model is therefore trained jointly across all three scenarios without per-scenario adaptation.

\subsection{Continuous Collision Detection Module}
\label{subsec:ccd}

\begin{figure}[t]
  \centering
  \includegraphics[width=0.85\linewidth]{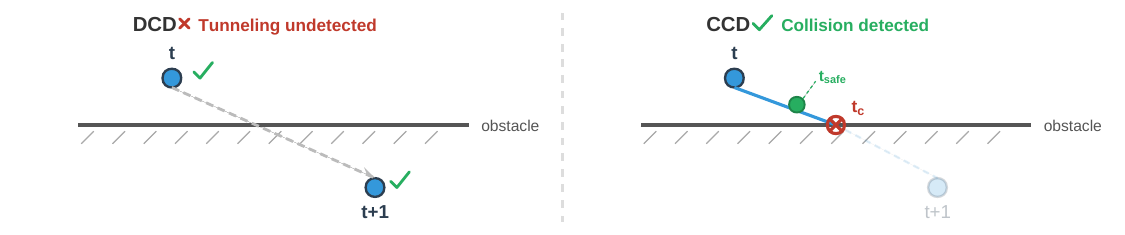}
  \caption{\textbf{DCD vs.\ CCD.} DCD checks only at discrete time steps and can miss mid-step penetrations (left). CCD sweeps the entire trajectory between two consecutive frames to locate the exact collision time \(t_c\), then corrects the vertex to a safe position \(t_{\text{safe}}\) before the collision occurs, eliminating tunneling by construction (right).}
  \label{fig:tunneling}
\vspace{-8pt}
\end{figure}

A learned simulator that supervises only discrete frame states inevitably produces inter-frame ``tunneling'': a vertex may lie on the correct side of a collider at both frame $T$ and frame $T{+}1$, yet pass straight through it during the intervening motion. Discrete Collision Detection (DCD), which inspects only the sampled frame states, is blind to such events. As illustrated in Figure~\ref{fig:tunneling}, Continuous Collision Detection (CCD) instead sweeps the entire linear trajectory between two consecutive frames, locates the exact collision time, and corrects the vertex to a safe pre-collision position, eliminating tunneling by construction. We therefore equip our framework with a CCD module that operates in two complementary stages: a \emph{differentiable CCD loss} $\mathcal{L}_{\text{CCD}}$ that shapes the trajectories during training, and a \emph{non-differentiable CCD post-processor} that removes residual penetrations at inference. Both stages are made possible by our penetration-free training data (Sec.~\ref{sec:dataset}), without which a strict collision objective would penalize the ground truth itself.

\textbf{Collision types.}
The module handles all five primitive-level contact types: Vertex-Face (a cloth vertex against a collider triangle), Edge-Edge (a cloth edge against a collider edge), Face-Vertex (a collider vertex against a cloth triangle), and the two self-collision types Self-VF and Self-EE (a cloth vertex/edge against another triangle/edge of the same mesh). During training, the differentiable CCD loss focuses on the two self-collision types, since cloth--object collisions are already supervised by the contact loss $\mathcal{L}_{\text{contact}}$ (Sec.~\ref{subsec:loss}); the inference-time post-processor covers all five types.

\textbf{Differentiable CCD loss.}
The boolean outcome of a geometric intersection test is non-differentiable, so we adopt a ``detect-then-regress'' strategy. A non-differentiable CCD pass first identifies the set of colliding primitive pairs $\mathcal{P}_{col}$, each with its collision time $t_c \in [0, 1]$ along the linear inter-frame trajectory. For every such pair we define a safe time $t_{safe} = \max(0,\, t_c - \epsilon)$ just before contact, and obtain the corresponding collision-free position of each involved vertex by linear interpolation,
\begin{equation}
    \mathbf{x}_{safe} = \mathbf{x}_{start} + t_{safe}\, (\mathbf{x}_{end} - \mathbf{x}_{start}).
    \label{eq:ccd_safe}
\end{equation}
The loss then regresses the predicted end position toward this safe position:
\begin{equation}
    \mathcal{L}_{\text{CCD}} = \frac{1}{|\mathcal{P}_{col}|} \sum_{i \in \mathcal{P}_{col}} \frac{1}{4}\sum_{j=1}^{4} \big\| \mathbf{x}_{end}^{(i,j)} - \mathbf{x}_{safe}^{(i,j)} \big\|^2,
    \label{eq:ccd_loss}
\end{equation}
where the inner sum runs over the four vertices involved in each colliding pair (one free vertex and three triangle vertices for Self-VF, or the two endpoints of each edge for Self-EE). Since $\| \mathbf{x}_{end} - \mathbf{x}_{safe} \|^2 = (1 - t_{safe})^2 \| \mathbf{x}_{end} - \mathbf{x}_{start} \|^2$, early collisions ($t_c \approx 0$) receive stronger gradients than late ones ($t_c \approx 1$), so the objective naturally prioritizes the most severe penetrations. Detecting collision times reduces to finding the roots of a cubic polynomial on $[0,1]$; we solve this efficiently and robustly following~\cite{yuksel2022fast}, and provide the full root-finding and inference-time post-processing details in the supplementary.

\subsection{Training Objective}
\label{subsec:loss}

We supervise the model with a combination of a reconstruction term, a contact term, and the differentiable CCD term introduced above, and train in two stages.

\textbf{Reconstruction loss.}
The primary supervision is the mean squared positional error between the predicted vertices $\hat{\mathbf{x}}_i$ and the ground truth $\mathbf{x}_i$ over all $N_v$ vertices:
\begin{equation}
\mathcal{L}_{mse} = \frac{1}{N_v} \sum_{i=1}^{N_v} \| \hat{\mathbf{x}}_i - \mathbf{x}_i \|_2^2 .
\label{eq:mse}
\end{equation}

\textbf{Contact loss.}
To instill basic cloth--object collision awareness already during pretraining, we add a contact term. For each cloth vertex we locate its nearest collision triangle via kNN, compute the signed distance along the face normal, and apply a cubic penalty on the penetration depth $d_i = \min(0,\, s_i)$, where $s_i$ is the signed distance (negative inside the collider):
\begin{equation}
\mathcal{L}_{contact} = \frac{1}{N_v} \sum_{i=1}^{N_v} |d_i|^3 .
\label{eq:contact}
\end{equation}
The cubic form leaves shallow contacts nearly unpenalized while sharply punishing deep penetrations.

\textbf{Two-stage training.}
We optimize the network in two stages. During \emph{pretraining}, we use
\begin{equation}
\mathcal{L}_{pretrain} = \lambda_{mse}\, \mathcal{L}_{mse} + \lambda_{contact}\, \mathcal{L}_{contact},
\label{eq:pretrain}
\end{equation}
which yields physically plausible trajectories with only mild residual penetrations. We then \emph{finetune} with the differentiable CCD loss added,
\begin{equation}
\mathcal{L}_{finetune} = \mathcal{L}_{pretrain} + \lambda_{ccd}\, \mathcal{L}_{ccd}.
\label{eq:finetune}
\end{equation}
Introducing $\mathcal{L}_{ccd}$ only at the finetuning stage is deliberate: the ``detect-then-regress'' gradients then act on already-accurate predictions rather than on the noisy outputs of an untrained model, where the estimated collision times would be unreliable and the gradients dominated by noise. The detailed training schedule is reported in Sec.~\ref{subsec:implementation}.

\section{Penetration-Free Dataset}
\label{sec:dataset}
\vspace{-3mm}

We present a high-fidelity cloth dataset free of interpenetrations. Existing datasets often rely on approximate solvers that allow minor penetrations, which makes the training of a strict CCD loss impossible, as the ground truth itself would be penalized.

\textbf{Simulation Settings.}
We generate ground-truth data using GIPC~\cite{DBLP:journals/tog/HuangCLK24}, a state-of-the-art penetration-free GPU cloth solver based on incremental potential contact methods. Each sequence spans 240 frames (4\,s) at $\Delta t = 1/60$\,s. Detailed physical parameters are provided in the supplementary.

\begin{table}[t]
    \centering
    \caption{Mesh complexity statistics of our penetration-free dataset.}
    \label{tab:dataset_stats}
    \resizebox{\linewidth}{!}{%
    \begin{tabular}{l|ccc|cc|cc}
    \toprule
    & \multicolumn{3}{c|}{\textbf{Cloth Mesh}} & \multicolumn{2}{c|}{\textbf{Collision Mesh}} & \multicolumn{2}{c}{\textbf{Scale}} \\
    Subset & \#Vertices & \#Faces & \#Meshes & \#Faces & \#Meshes & Sequences & Frames \\
    \midrule
    Human Garment & 1k--3.6k & 2k--7.1k & 14 & 1k--5.1k & 5 & 56 & 13.4k \\
    Robotic Manip. & 1k--4k & 1.9k--7.9k & 1000 & 0.6k & 1 & 1000 & 240k \\
    Diverse Object Collision & 3.6k & 7k & 1 & 1k--4k & 1000 & 1000 & 240k \\
    \midrule
    \textbf{Total} & 1k--4k & 1.9k--7.9k & 1015 & 0.6k--5.1k & 1006 & 2056 & 493.4k \\
    \bottomrule
    \end{tabular}%
    }
    \vspace{-2mm}
\end{table}

\vspace{-1mm}

\textbf{Simulation Scenarios.}
To ensure the model generalizes across diverse topologies and interaction types, we construct three distinct subsets. 1) \textbf{Human Garment}, consists of T-shirts and skirts dressed on animated SMPL~\cite{loper2023smpl} avatars. The avatars perform a variety of complex motions~\cite{guo2025make}---including walking, running, dancing, and jumping---that induce rich cloth dynamics such as large-amplitude swinging, body-cloth contact, and self-folding in regions like the armpits and waist. 2) \textbf{Robotic Manipulation}, features over 1000 diverse cloth meshes sourced from~\cite{korosteleva2021generating}, each grasped and lifted by a robotic gripper. This scenario introduces localized external forces and asymmetric deformation patterns that differ fundamentally from body-driven motion, testing the model's ability to handle point-contact manipulation and gravitational draping simultaneously. 3) \textbf{Diverse Object Collision}, simulates cloths falling freely onto rigid objects. We randomly sample over 1000 collision meshes from the Objaverse~\cite{deitke2023objaverse} dataset, covering a wide range of geometric features including sharp edges, concavities, and thin structures. This subset stresses the model's capacity to generalize across highly varied collision geometries unseen during training.

\textbf{Dataset Statistics.}
Table~\ref{tab:dataset_stats} summarizes the dataset. Note that the ranges reflect the use of multiple mesh templates: in Human Garment, most cloth meshes have ${\sim}$3.6k vertices (${\sim}$7k faces) with collision bodies at ${\sim}$5k faces, while the lower end corresponds to a few simpler garments; in Robotic Manipulation, most cloths have ${\sim}$4k vertices (${\sim}$7.9k faces).

\vspace{-2mm}
\section{Experiments}
\label{sec:experiments}

We evaluate ClothTransformer with a \emph{single unified model} trained jointly on all three scenarios (no per-scenario fine-tuning), demonstrating that our architecture learns shared latent dynamics across  different interaction modes. We present quantitative and qualitative comparisons against state-of-the-art baselines, ablations of each design choice, and a scalability analysis of our latent-space formulation. Additional architecture, training, and evaluation details are in the supplementary.

\subsection{Implementation Details}
\label{subsec:implementation}

\textbf{Network Architecture.}
Our Spatial Encoder utilizes a 2-layer GNN with 1024 hidden units to extract local features, followed by a cross-attention layer compressing the mesh into $N_{latents} = 1024$ latent tokens. The Temporal Transformer consists of 12 layers with 12 attention heads, an embedding dimension of 768, and a feed-forward dimension of 3072 with SwiGLU activation. The Spatial Decoder mirrors the encoder with a cross-attention layer followed by a 2-layer GNN for topological refinement.

\textbf{Training Settings.}
We train our model end-to-end using the AdamW optimizer with a learning rate of $1 \times 10^{-4}$ and a cosine annealing schedule decaying to $1 \times 10^{-7}$. The batch size is 32. Following the two-stage objective defined in Sec.~\ref{subsec:loss}, we first pretrain with $\mathcal{L}_{pretrain}$ (Eq.~\ref{eq:pretrain}) for 160k steps, then finetune with $\mathcal{L}_{finetune}$ (Eq.~\ref{eq:finetune}) for 40k steps, taking approximately 300 NVIDIA H200 GPU hours in total;
 Gradient norms are clipped to 1.0. We employ a rollout curriculum strategy, starting with single-step predictions and linearly increasing the horizon to 5 steps over the first 180,000 training steps to mitigate error accumulation. The dataset is randomly split into training, validation, and test sets in an 8:1:1 ratio (per subset).

\subsection{Comparative Results}
\label{subsec:comparison}

\begin{figure*}[t]
  \centering
  \setlength{\tabcolsep}{0pt}
  \renewcommand{\arraystretch}{0}
  \begin{tabular}{@{}c@{\hspace{1pt}}cccccc@{}}
    & \small Sword & \small Running & \small Grasp 1 & \small Stick & \small Flip & \small Grasp 2 \\
    \rotatebox{90}{\small\hspace{4pt}Ground Truth} &
    \includegraphics[width=0.155\linewidth]{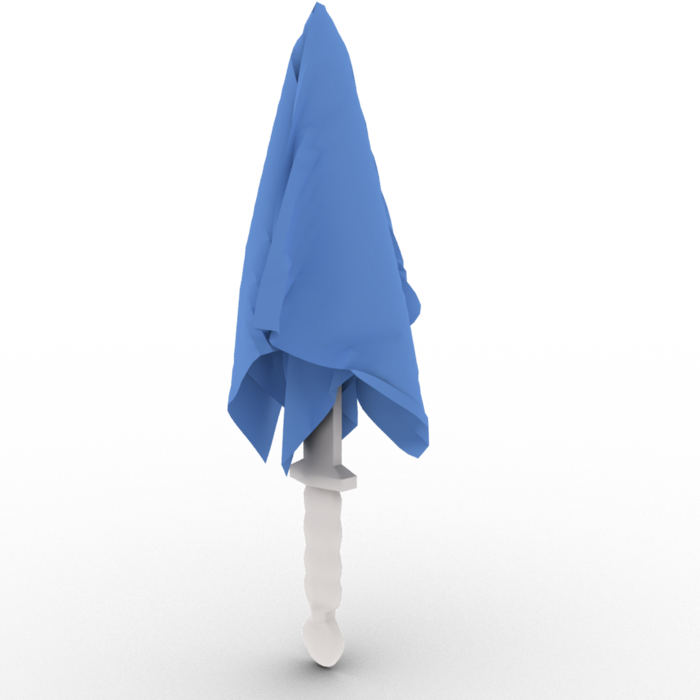} &
    \includegraphics[width=0.155\linewidth]{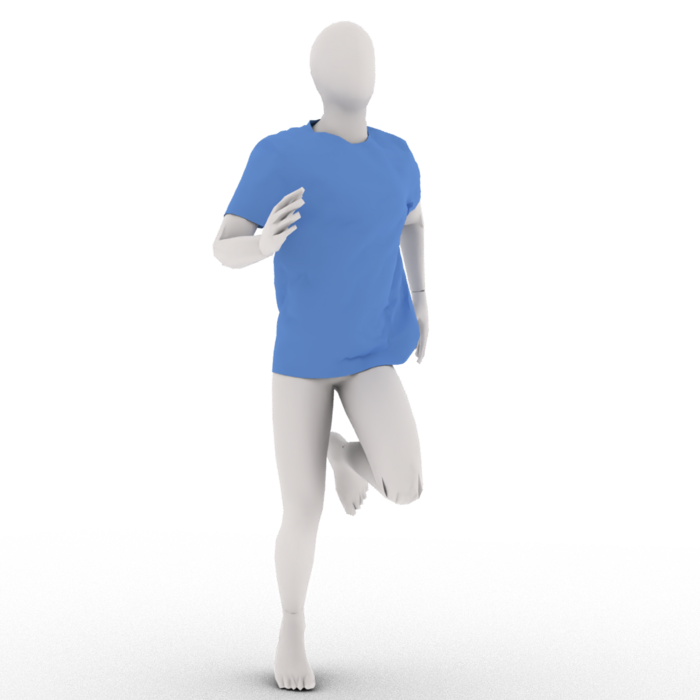} &
    \includegraphics[width=0.155\linewidth]{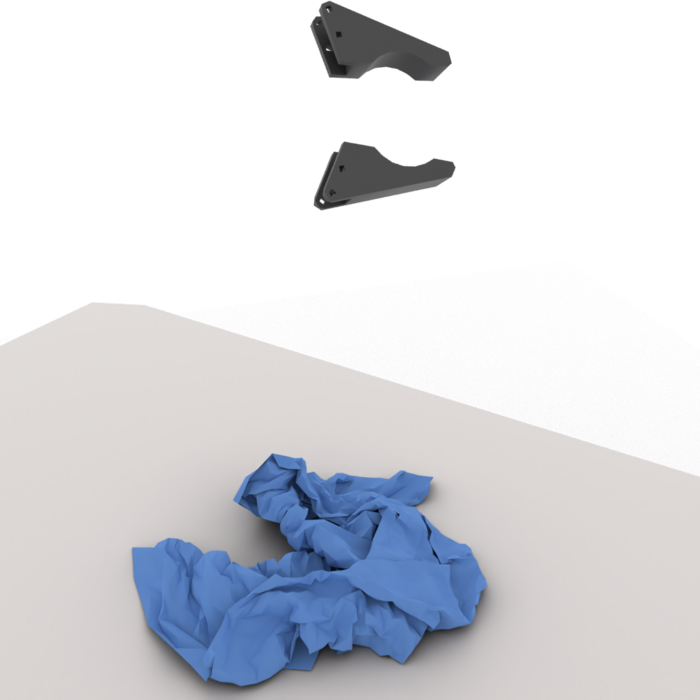} &
    \includegraphics[width=0.155\linewidth]{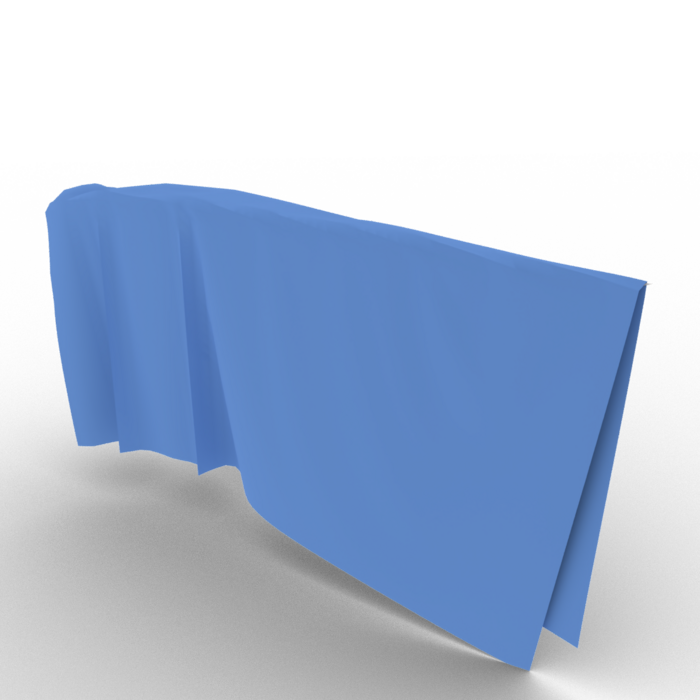} &
    \includegraphics[width=0.155\linewidth]{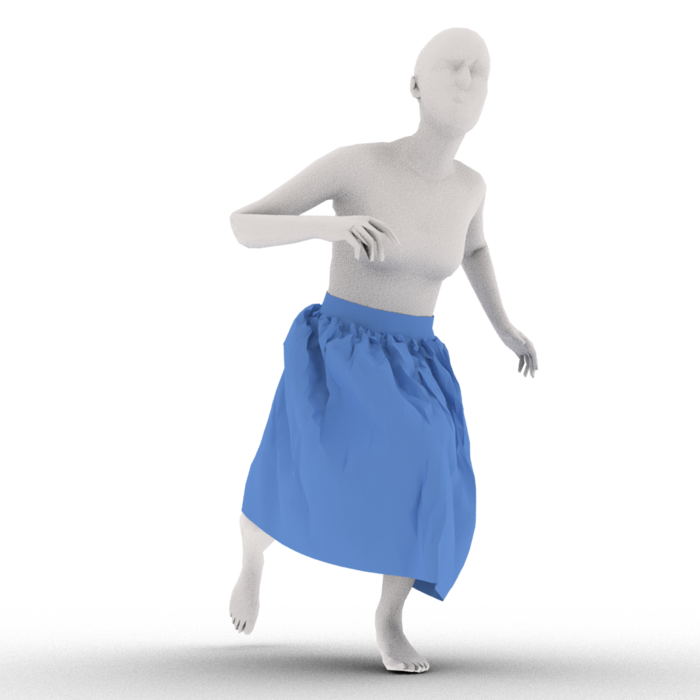} &
    \includegraphics[width=0.155\linewidth]{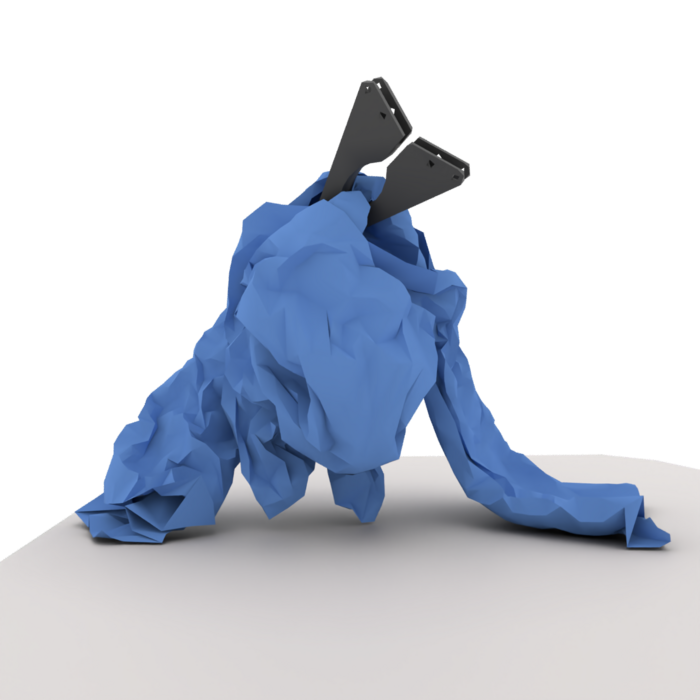} \\
    \rotatebox{90}{\small\hspace{4pt}SOTA GNN} &
    \includegraphics[width=0.155\linewidth]{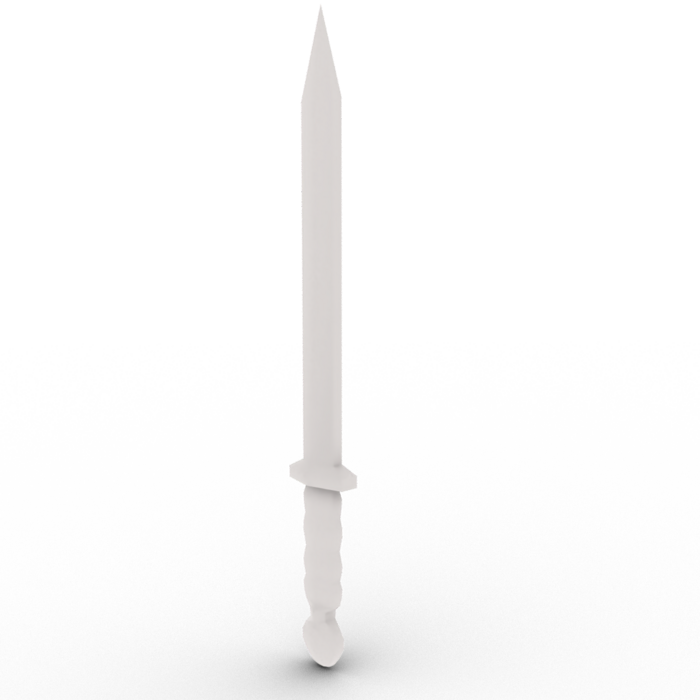} &
    \includegraphics[width=0.155\linewidth]{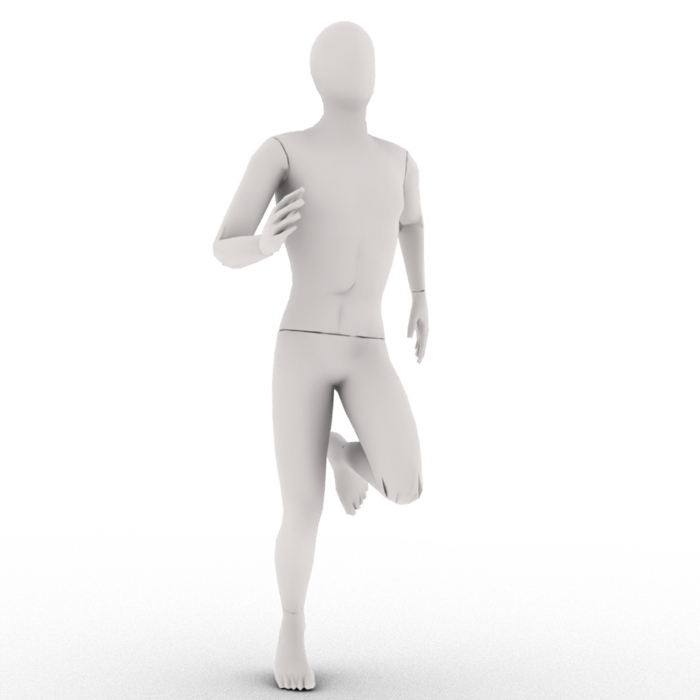} &
    \includegraphics[width=0.155\linewidth]{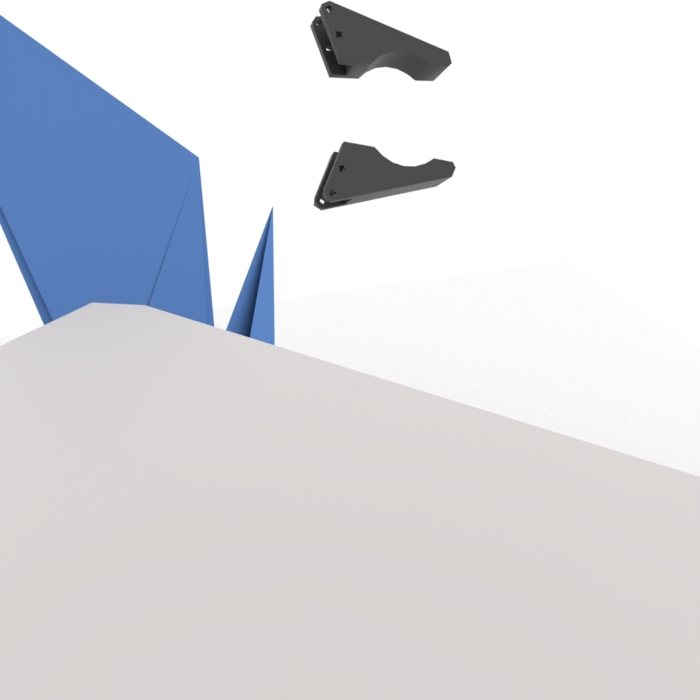} &
    \includegraphics[width=0.155\linewidth]{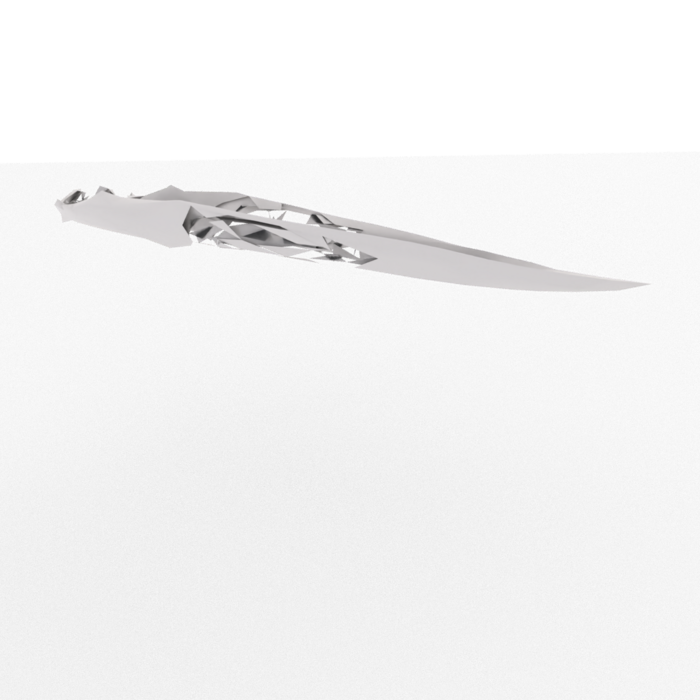} &
    \includegraphics[width=0.155\linewidth]{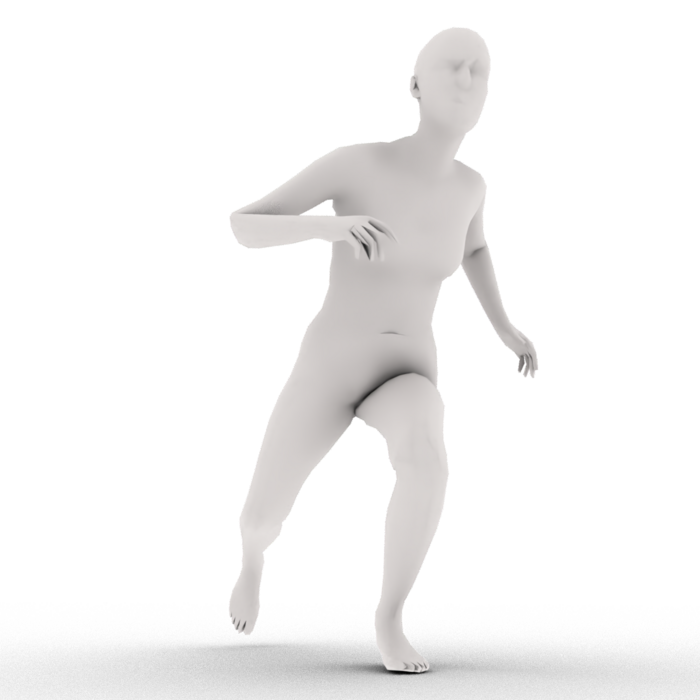} &
    \includegraphics[width=0.155\linewidth]{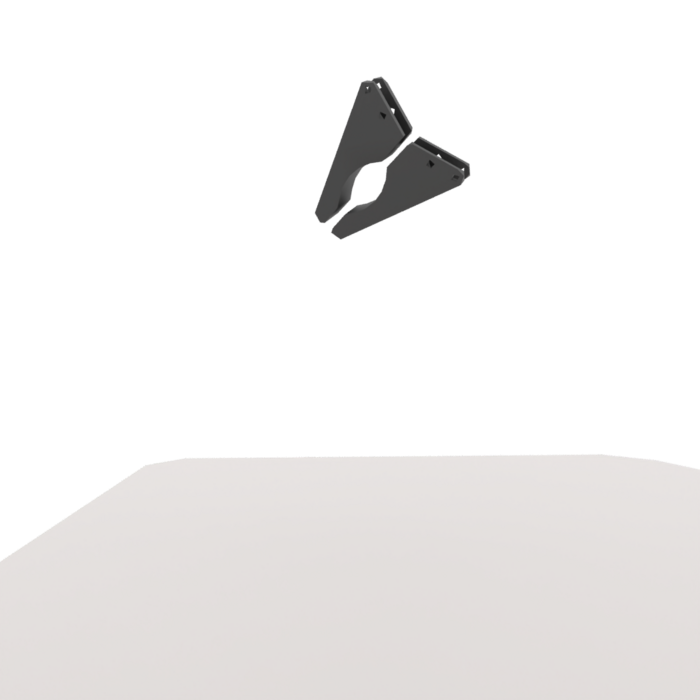} \\
    \rotatebox{90}{\small\hspace{4pt}MAT} &
    \includegraphics[width=0.155\linewidth]{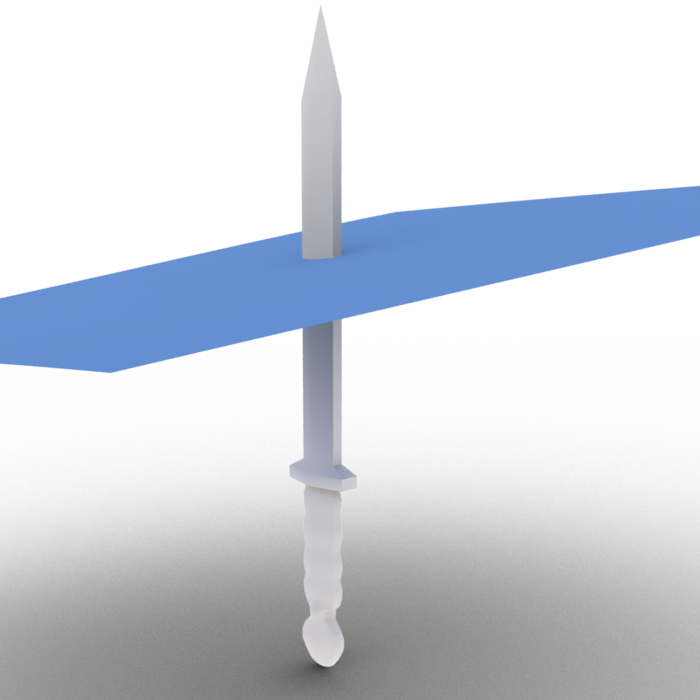} &
    \includegraphics[width=0.155\linewidth]{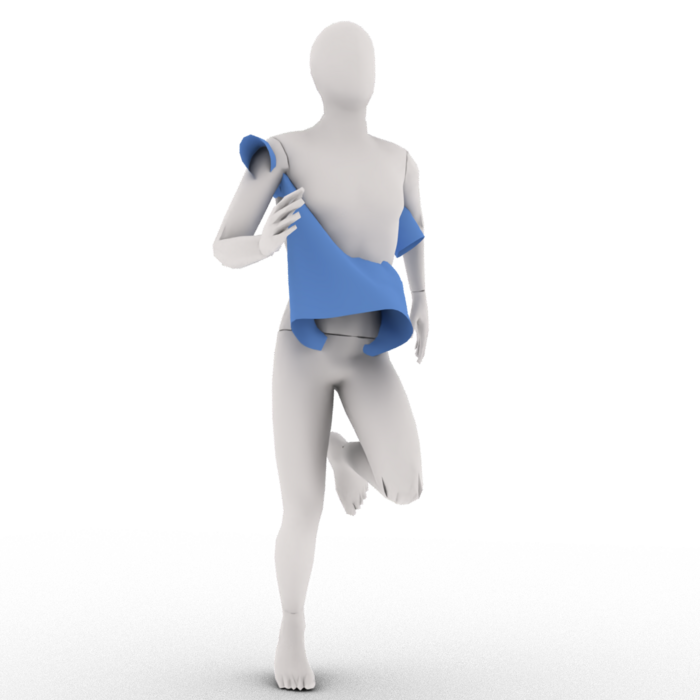} &
    \includegraphics[width=0.155\linewidth]{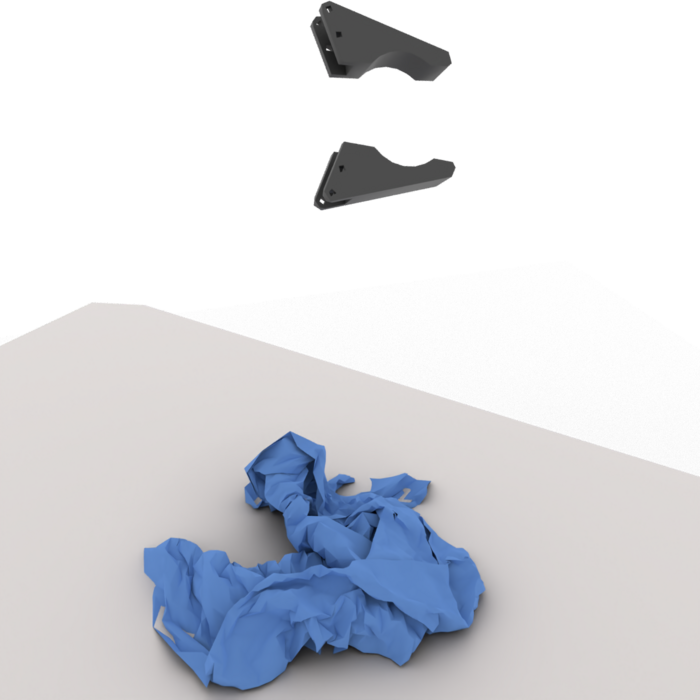} &
    \includegraphics[width=0.155\linewidth]{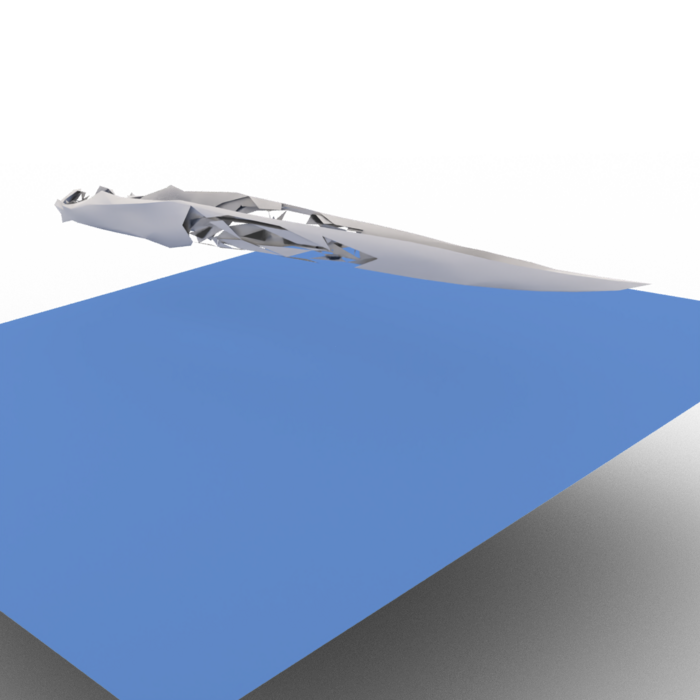} &
    \includegraphics[width=0.155\linewidth]{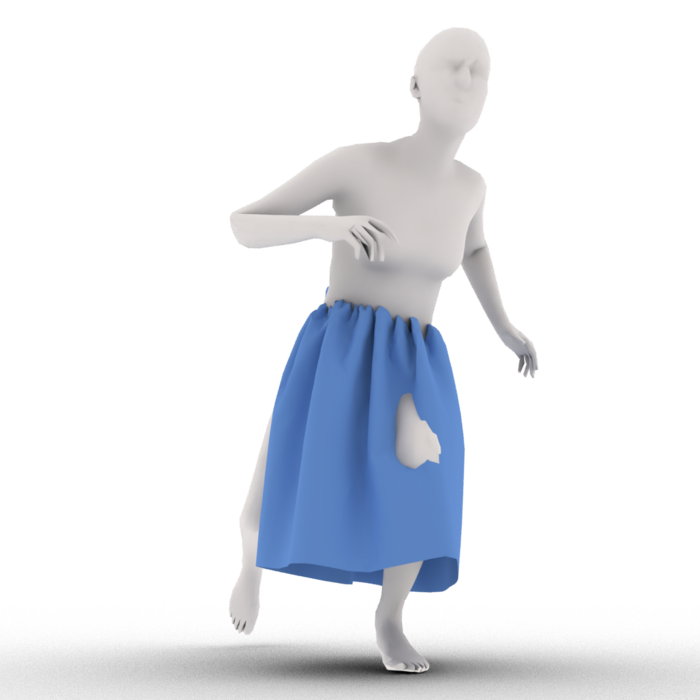} &
    \includegraphics[width=0.155\linewidth]{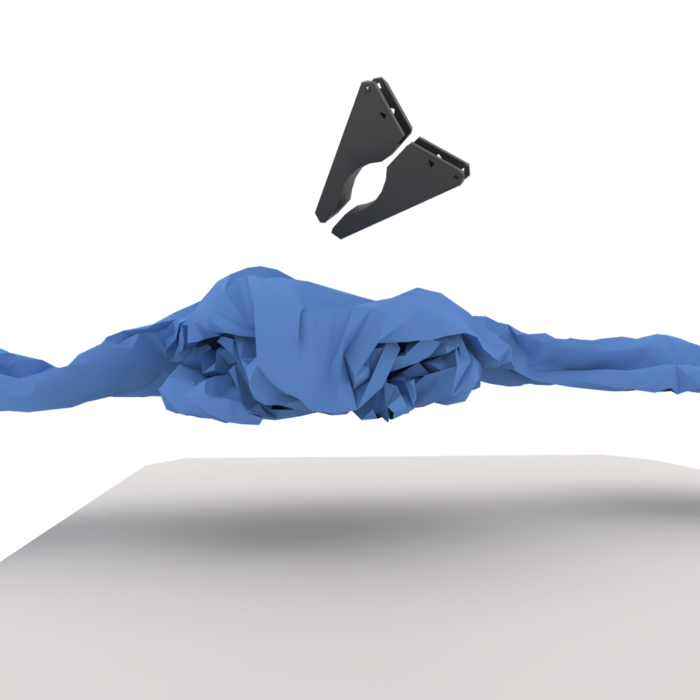} \\
    \rotatebox{90}{\small\hspace{4pt}LayersNet} &
    \includegraphics[width=0.155\linewidth]{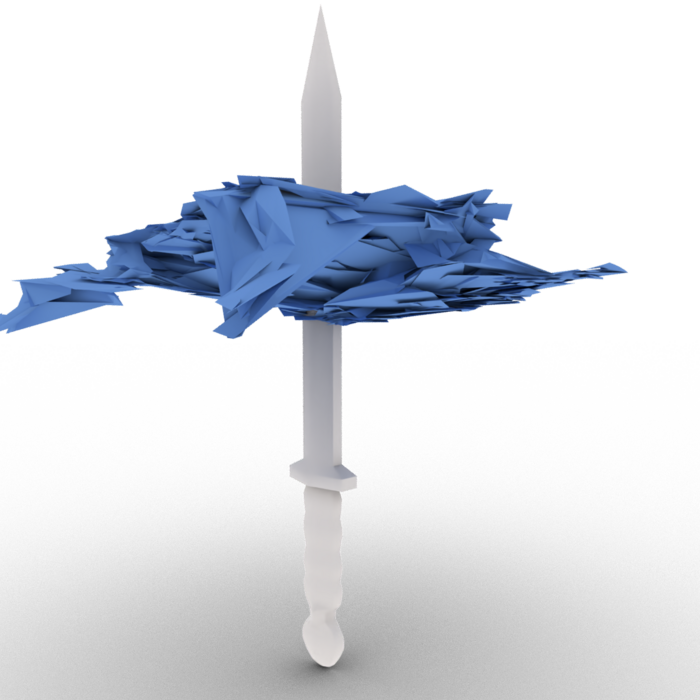} &
    \includegraphics[width=0.155\linewidth]{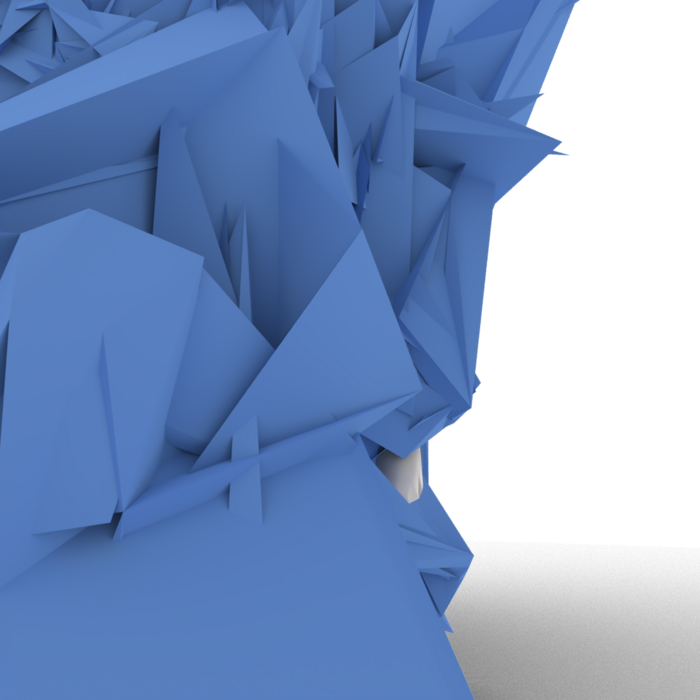} &
    \includegraphics[width=0.155\linewidth]{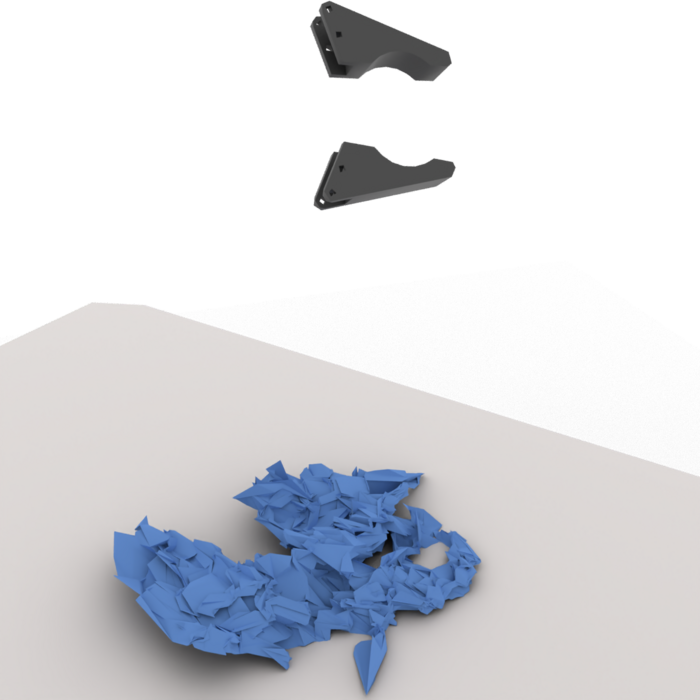} &
    \includegraphics[width=0.155\linewidth]{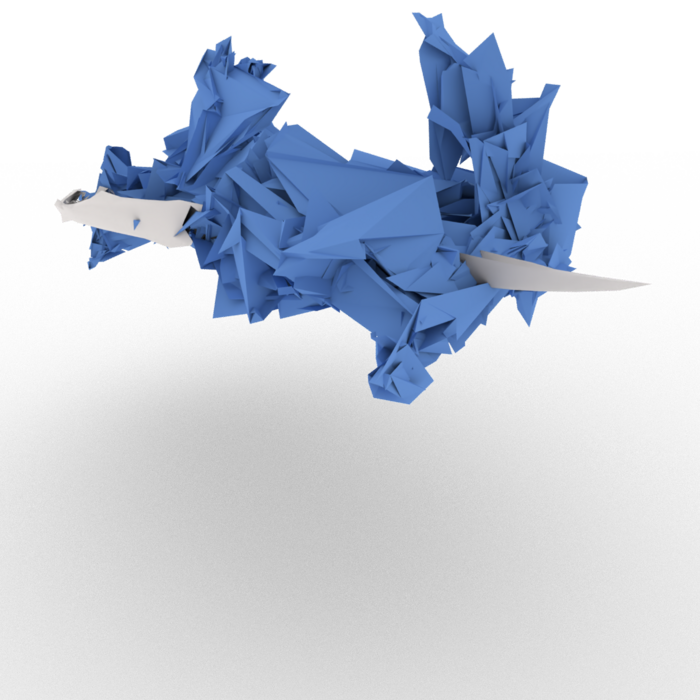} &
    \includegraphics[width=0.155\linewidth]{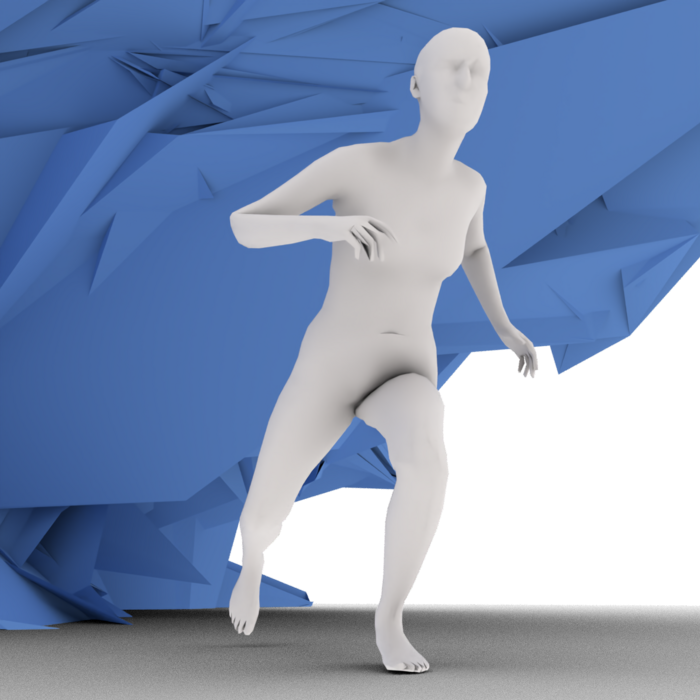} &
    \includegraphics[width=0.155\linewidth]{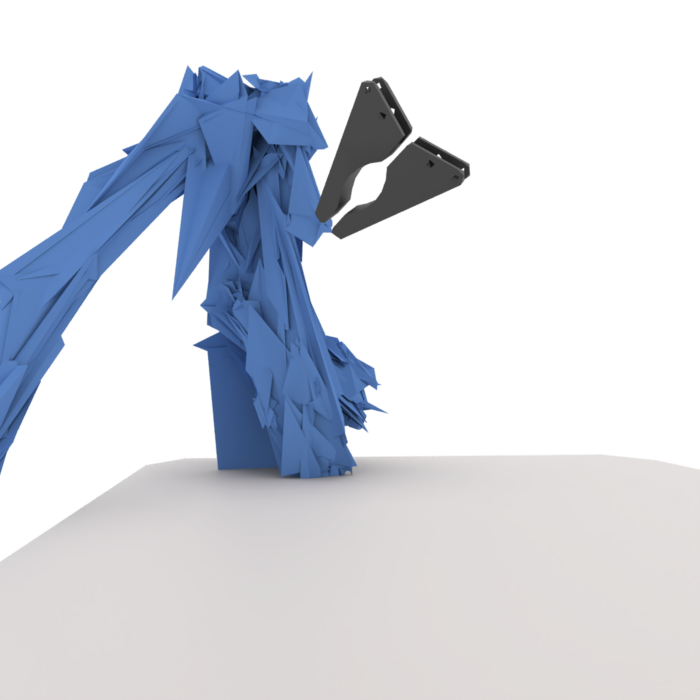} \\
    \rotatebox{90}{\small\hspace{4pt}\textbf{Ours}} &
    \includegraphics[width=0.155\linewidth]{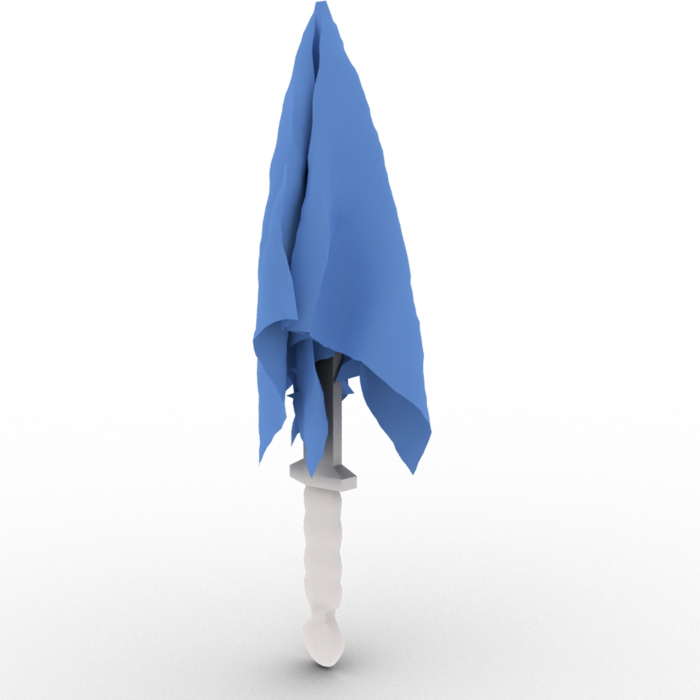} &
    \includegraphics[width=0.155\linewidth]{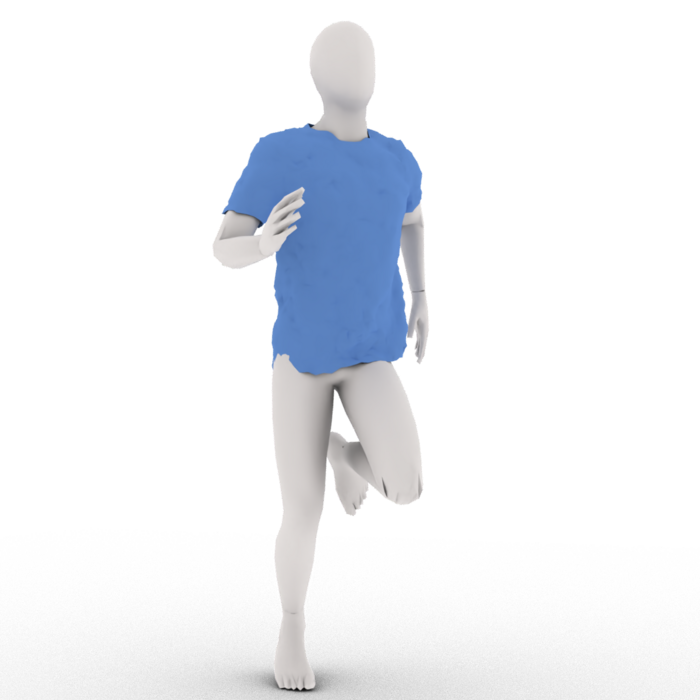} &
    \includegraphics[width=0.155\linewidth]{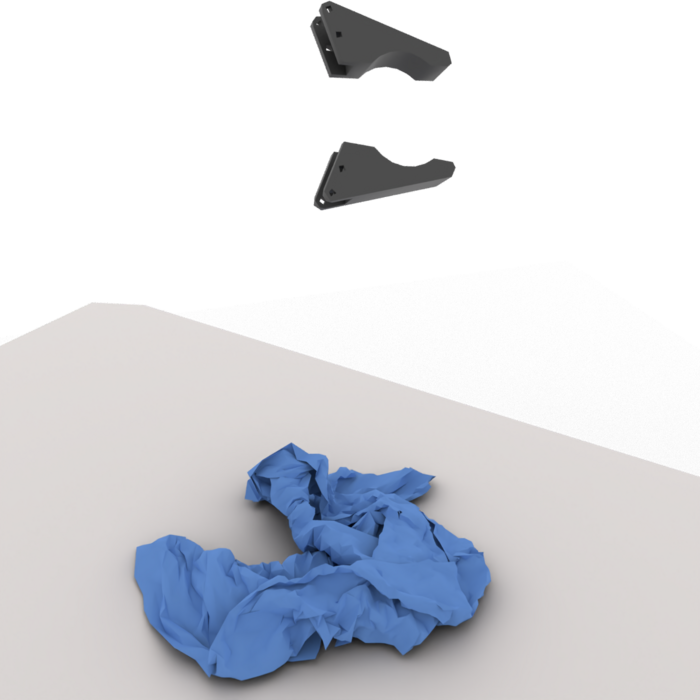} &
    \includegraphics[width=0.155\linewidth]{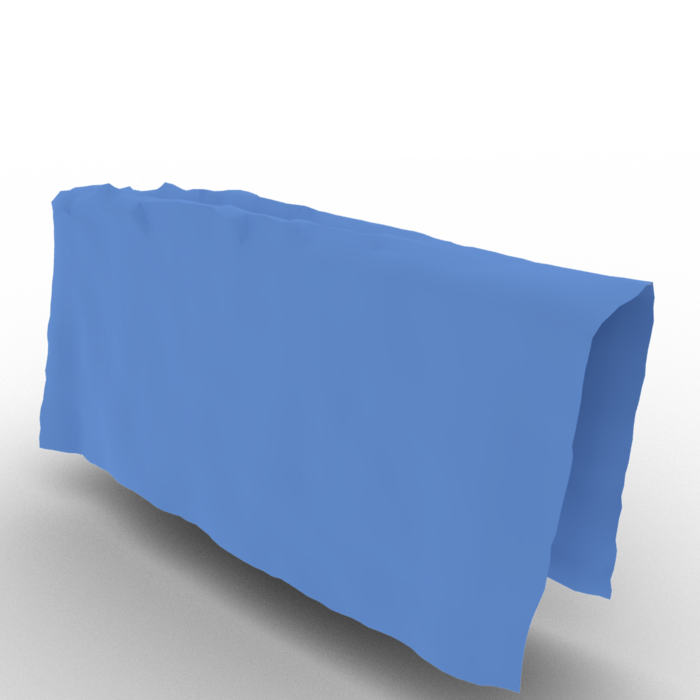} &
    \includegraphics[width=0.155\linewidth]{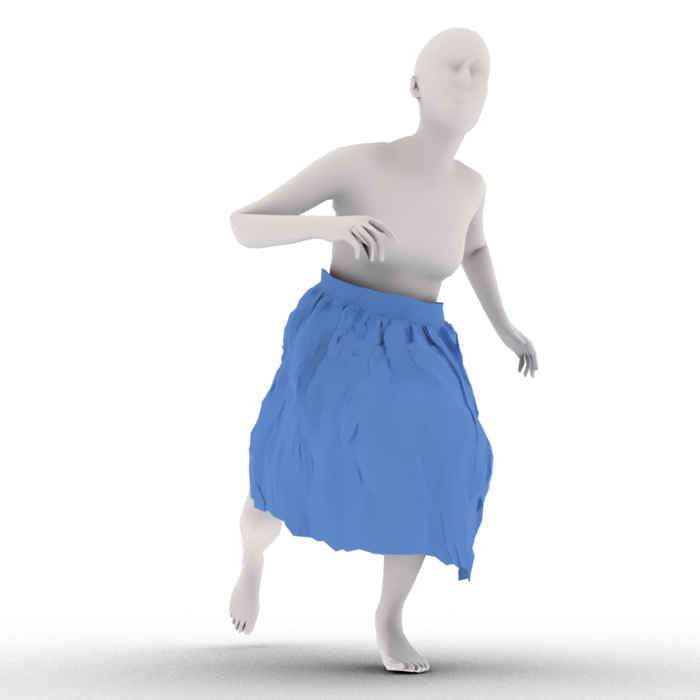} &
    \includegraphics[width=0.155\linewidth]{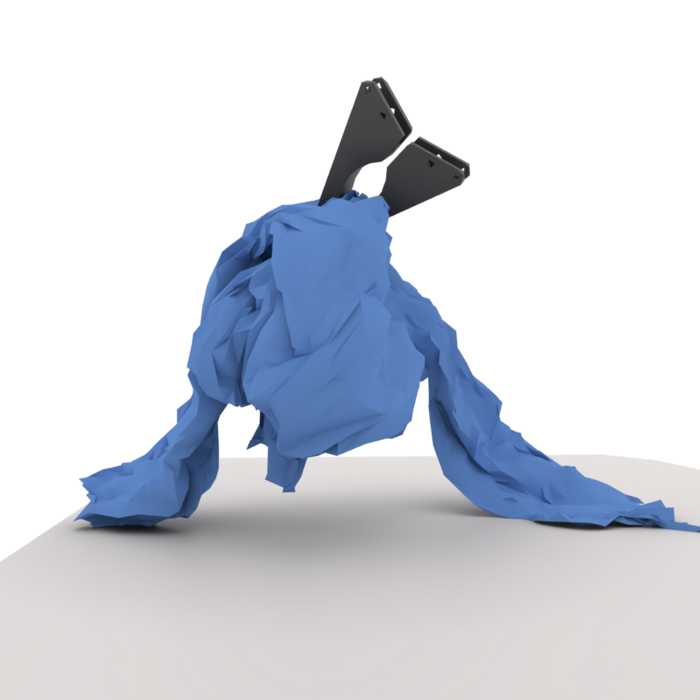} \\
  \end{tabular}
  \caption{\textbf{Qualitative comparison on unseen test sequences.} Columns 1, 4: Diverse Object Collision (sword, stick). Columns 2, 5: Human Garment (running, front-flip). Columns 3, 6: Robotic Manipulation (static resting, grasping).}
  \label{fig:qualitative}
\end{figure*}

We compare against three SOTA learning-based baselines: \textbf{SOTA GNN}~\cite{DBLP:conf/cvpr/GrigorevBH23, DBLP:conf/siggraph/0002BBHT24} (the hierarchical GNN backbone of HOOD/ContourCraft), \textbf{MAT}~\cite{DBLP:journals/cgf/LiWKCS24} (mesh-face tokenization with manifold-aware attention), and \textbf{LayersNet}~\cite{DBLP:conf/iccv/ShaoL023} (UV-patch tokenization). All methods are trained on the same training split and evaluated on unseen test sequences. We report three metrics: \textbf{MVE} (mean vertex error, cm), \textbf{Collision Rate} (\%, cloth--object penetration), and \textbf{Self-Collision Rate} (\%, self-intersections detected via CCD); formal definitions are in the supplementary. We adopt complementary CCD-post-processing settings for the two views: Table~\ref{tab:main_results} reports raw predictions (no post-processing) to isolate each model's underlying capability, whereas Figure~\ref{fig:qualitative} applies 10 CCD post-processing iterations uniformly to all methods so the visual comparison focuses on shape fidelity rather than residual penetration artifacts.

\begin{table}[t]
    \centering
    \caption{Quantitative comparison on unseen test sequences. MVE: Mean Vertex Error (cm). Coll.: collision rate with collision objects (\%). Self-C.: self-collision vertex rate (\%). \textit{Ours} denotes our model trained with the pretraining loss only; \textit{Ours (CCD Loss)} additionally finetunes with the differentiable CCD loss.}
    \label{tab:main_results}
    \resizebox{\linewidth}{!}{%
    \begin{tabular}{l|ccc|ccc|ccc}
    \toprule
    & \multicolumn{3}{c|}{\textbf{Human Garment}} & \multicolumn{3}{c|}{\textbf{Robotic Manip.}} & \multicolumn{3}{c}{\textbf{Diverse Object Collision}} \\
    Method & MVE $\downarrow$ & Coll. $\downarrow$ & Self-C. $\downarrow$ & MVE $\downarrow$ & Coll. $\downarrow$ & Self-C. $\downarrow$ & MVE $\downarrow$ & Coll. $\downarrow$ & Self-C. $\downarrow$ \\
    \midrule
    SOTA GNN & 59.13 & 27.14 & 57.43 & 59.42 & 17.42 & 50.02 & 142.67 & 16.92 & 9.67 \\
    MAT & 31.13 & 22.0 & \cellcolor{best}1.0 & 66.63 & \cellcolor{second}16.63 & \cellcolor{best}1.4 & 77.15 & \cellcolor{best}10.15 & \cellcolor{second}2.18 \\
    LayersNet & 149.09 & \cellcolor{best}13.07 & 79.12 & 69.54 & \cellcolor{best}13.01 & 77.91 & 154.02 & 18.87 & 71.31 \\
    \midrule
    \textbf{Ours (CCD Loss)} & \cellcolor{best}\textbf{6.53} & \textbf{16.32} & \cellcolor{second}\textbf{9.12} & \cellcolor{second}\textbf{15.03} & \textbf{17.32} & \cellcolor{second}\textbf{10.65} & \cellcolor{best}\textbf{8.91} & \textbf{17.44} & \cellcolor{best}\textbf{2.01} \\
    \textbf{Ours} & \cellcolor{second}\textbf{6.92} & \cellcolor{second}\textbf{14.12} & \textbf{9.79} & \cellcolor{best}\textbf{14.91} & \textbf{16.93} & \textbf{11.43} & \cellcolor{second}\textbf{9.03} & \cellcolor{second}\textbf{16.74} & \textbf{2.96} \\
    \bottomrule
    \end{tabular}%
    }
\end{table}

\textbf{Quantitative Comparison.}
Our method achieves the best MVE across all three scenarios, with approximately $4$--$9{\times}$ lower error than the strongest learning-based baseline on each scenario (and up to ${\sim}16{\times}$ lower than the SOTA GNN on Diverse Object Collision).
We caution that low collision/self-collision rates do not always indicate high quality: SOTA GNN's and LayersNet's clothes drift entirely away from the collision object (Figure~\ref{fig:qualitative}), and MAT degenerates into a near-rigid body with minimal local deformation---both trivially achieve low contact rates but yield poor MVE. Our method instead delivers both accurate vertex predictions \emph{and} the lowest Self-Collision Rate among non-degenerate methods.


\textbf{Qualitative Comparison.}
Figure~\ref{fig:qualitative} compares visual results across the three scenarios on unseen test cases. The visual failure modes mirror the quantitative caveats above: SOTA GNN and LayersNet both diverge in long-horizon rollouts, while MAT yields near-rigid garments; our method remains visually plausible and less penetrated across all scenarios.

\textbf{Generalization Analysis.}
The performance gap reflects a fundamental architectural distinction: our latent-space formulation uniformly encodes cloth vertices and collision triangles into a fixed-size token set, while the SOTA GNN's mesh-edge message passing, MAT's manifold-constrained attention, and LayersNet's UV-based parameterization each impose scenario-specific structural priors that limit generalization.

\textit{Isolating the architectural factor.}
Since the SOTA GNN backbone (HOOD, ContourCraft) is originally trained with a self-supervised physics-based loss, one might worry that retraining it under our supervised setup (Table~\ref{tab:main_results}) understates its true capability. To rule this out, we re-evaluate it under its \emph{native} self-supervised paradigm in two configurations: (i)~trained on \emph{only} Human Garment, matching its original deployment, and (ii)~trained jointly on all three scenarios, matching our unified setting; both are compared against (iii)~our unified model. Configuration~(i) produces reasonable visual quality on Human Garment but still deviates noticeably from the ground truth, while~(ii) degrades sharply across all scenarios, indicating that the SOTA GNN backbone struggles to absorb the increased data diversity. Our model outperforms~(ii) everywhere and even surpasses the scenario-specialized~(i) on its own distribution, confirming that the gap in Table~\ref{tab:main_results} stems from the SOTA GNN architecture itself, not the supervised setup. Full numbers and qualitative results are in Appendix~\ref{supp:unified} (Table~\ref{tab:unified}, Figure~\ref{fig:unified}).

\subsection{Ablation Studies}
\label{subsec:ablation}

\begin{figure*}[t]
  \centering
  \setlength{\tabcolsep}{0pt}
  \renewcommand{\arraystretch}{0}
  \begin{tabular}{@{}ccc@{\hspace{6pt}}ccc@{}}
    \small w/ DCD & \small + CCD Loss & \small + CCD Post. & \small w/ DCD & \small + CCD Loss & \small + CCD Post. \\
    \includegraphics[width=0.163\linewidth]{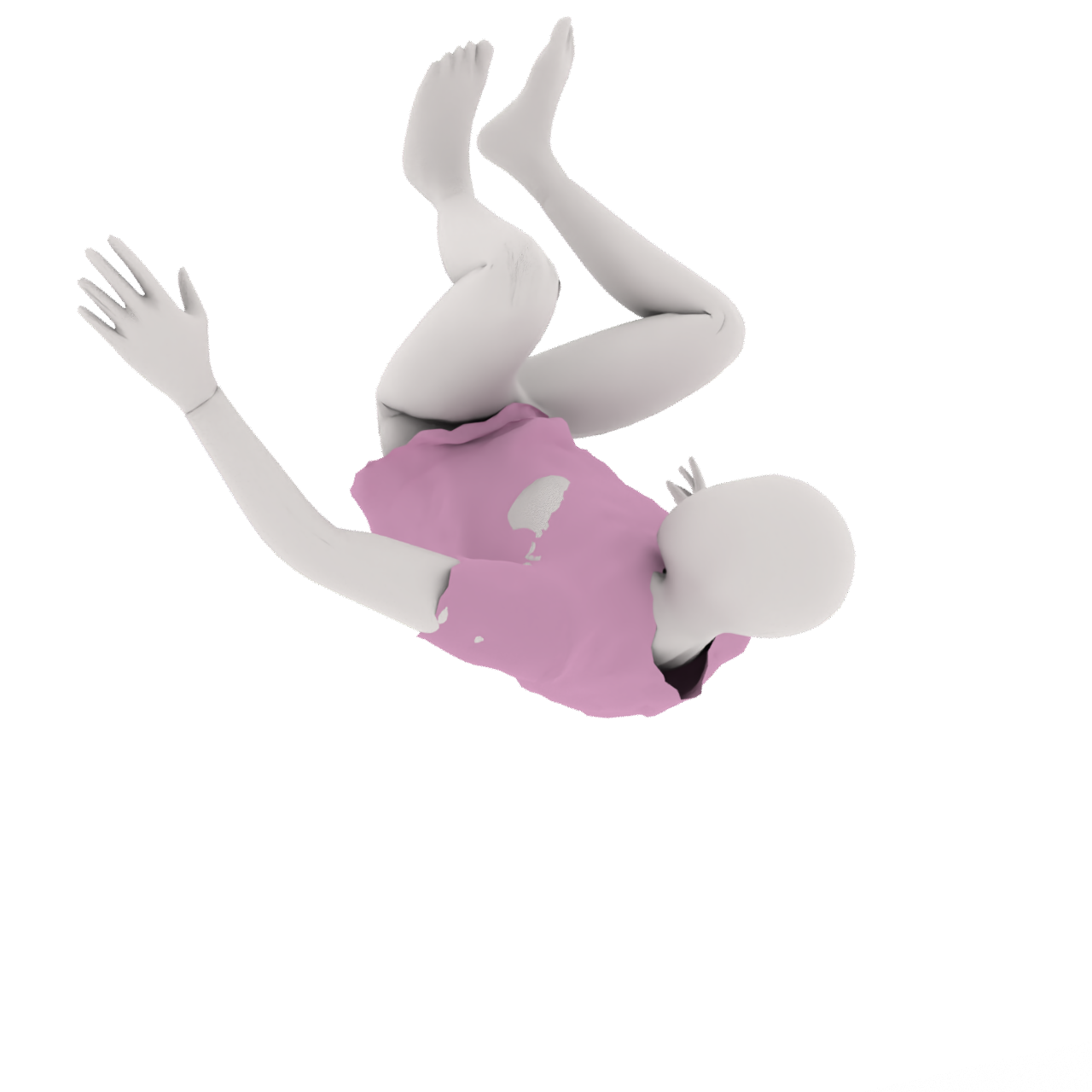} &
    \includegraphics[width=0.163\linewidth]{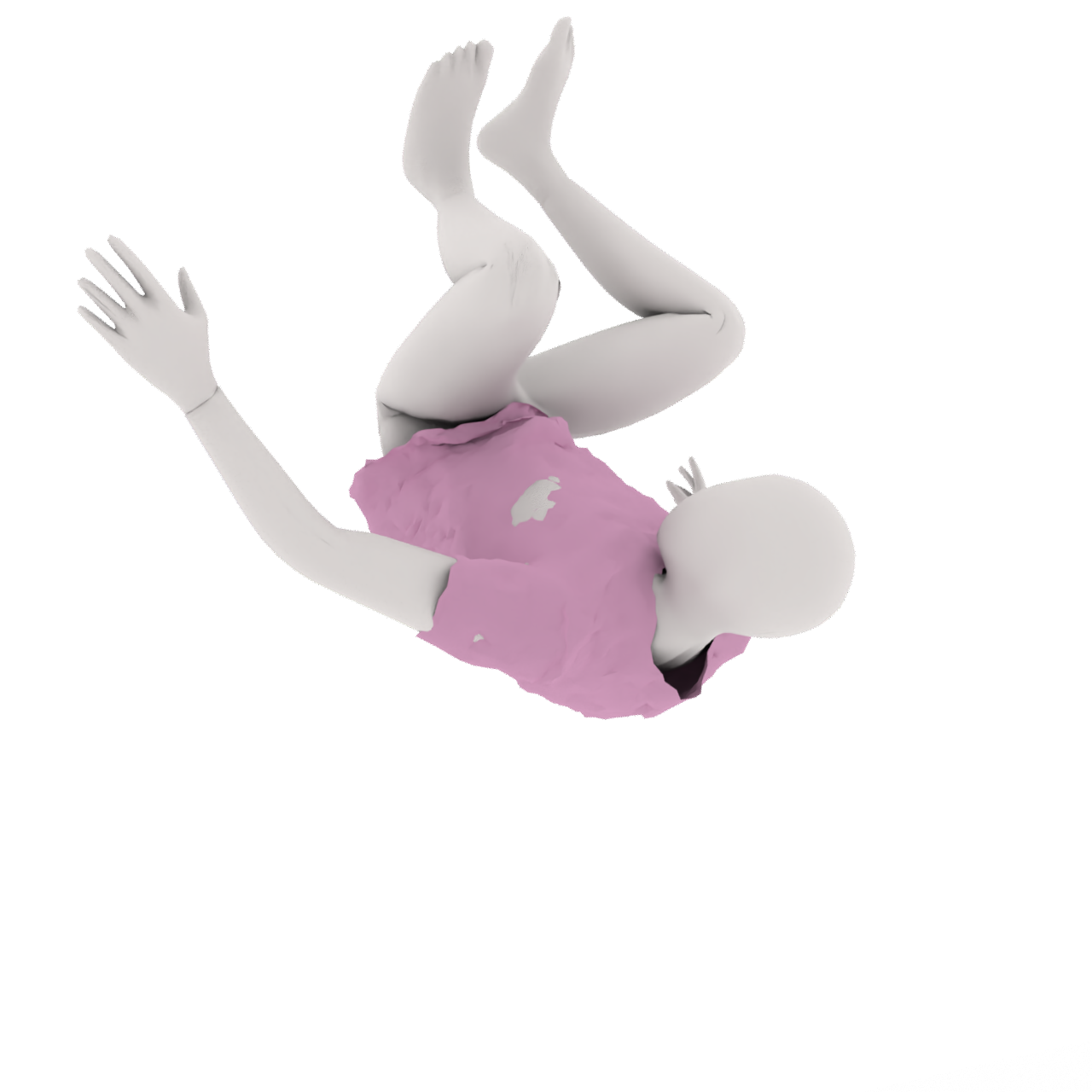} &
    \includegraphics[width=0.163\linewidth]{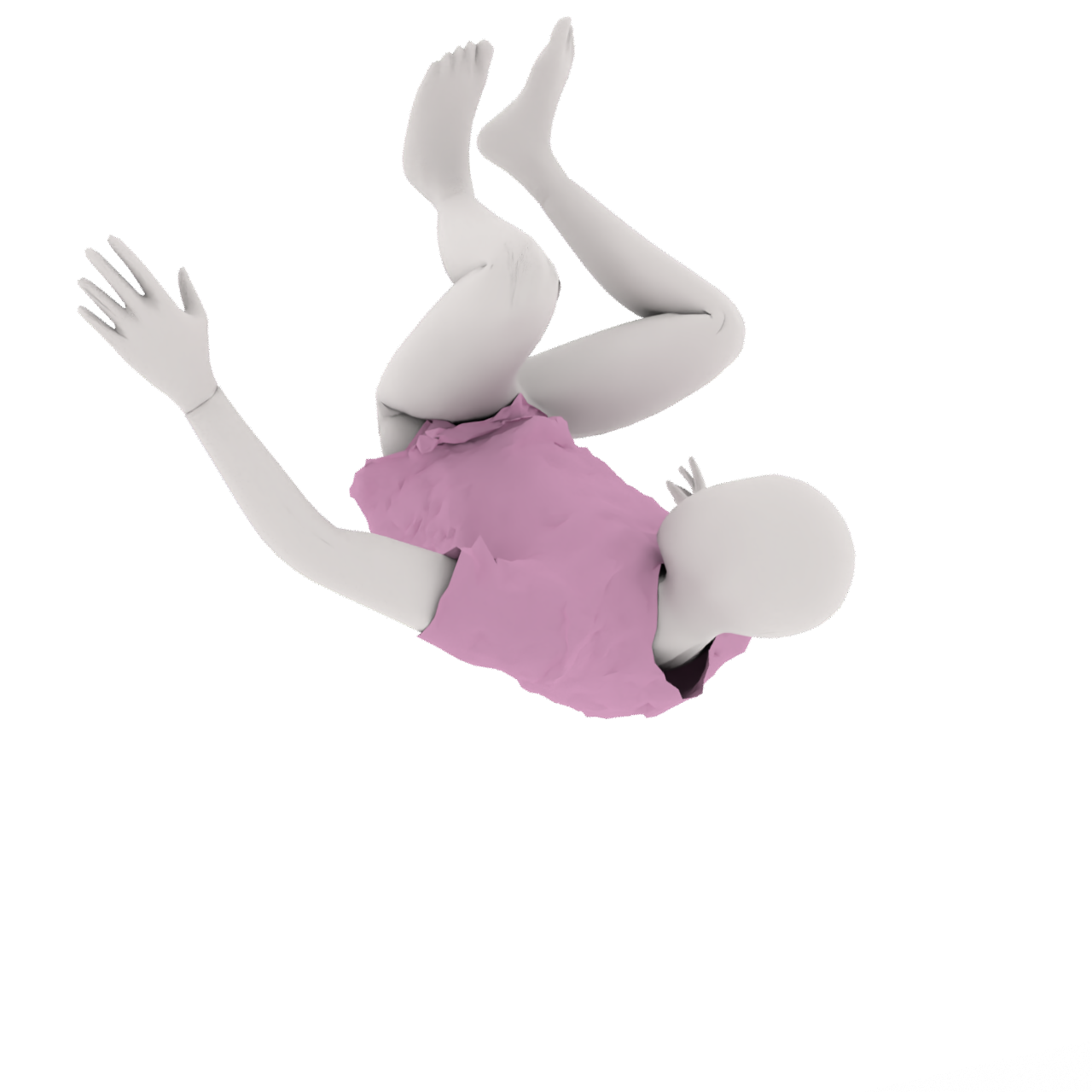} &
    \includegraphics[width=0.163\linewidth]{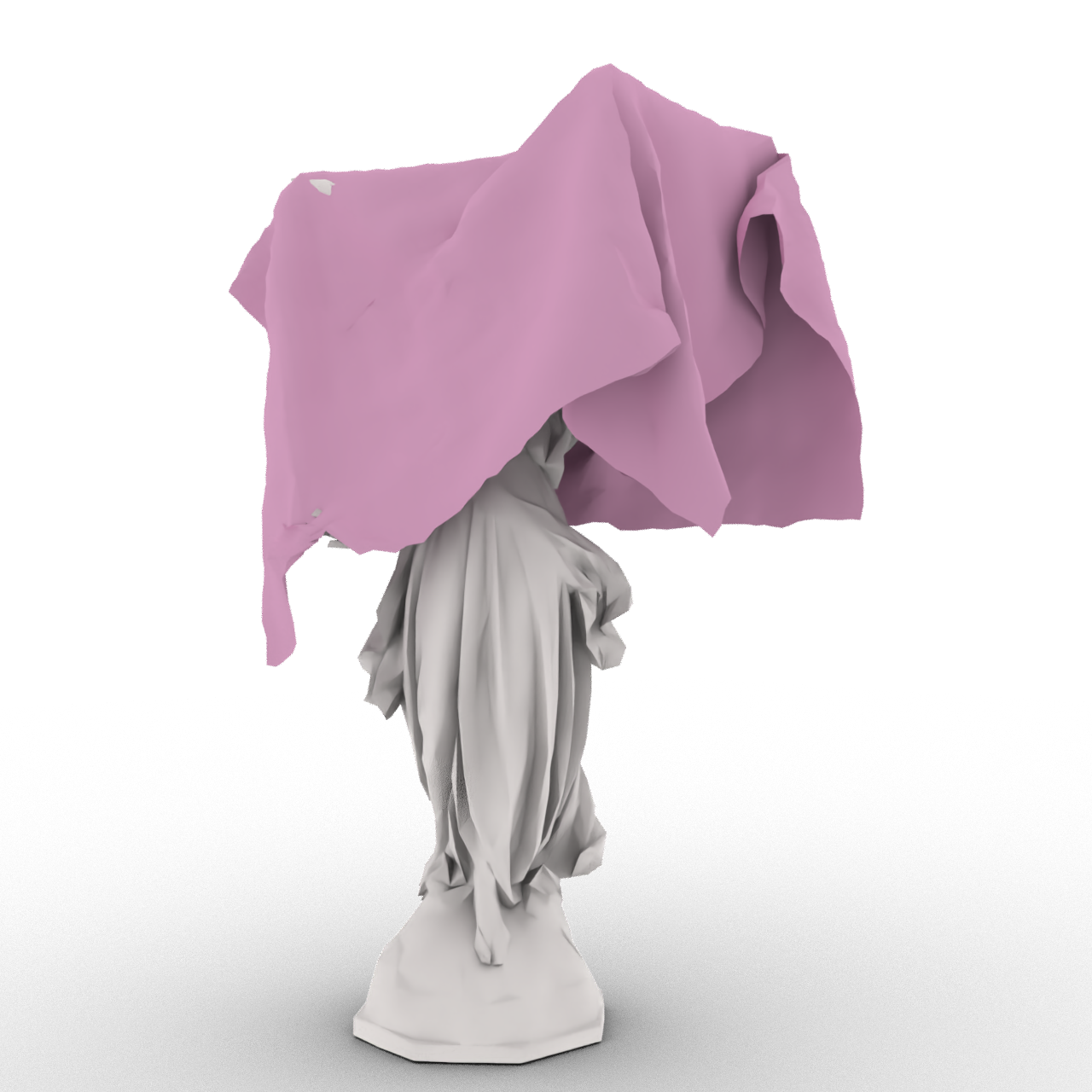} &
    \includegraphics[width=0.163\linewidth]{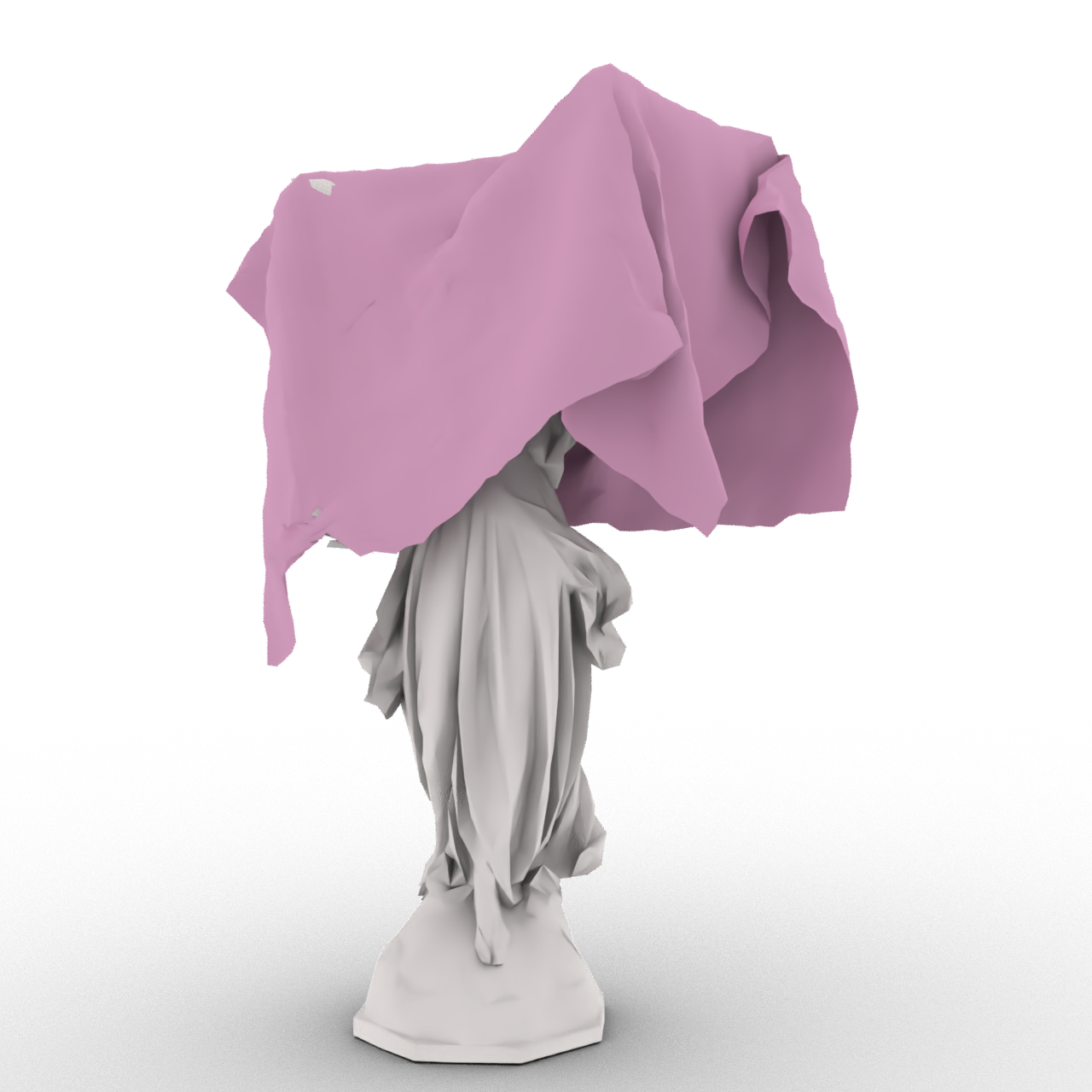} &
    \includegraphics[width=0.163\linewidth]{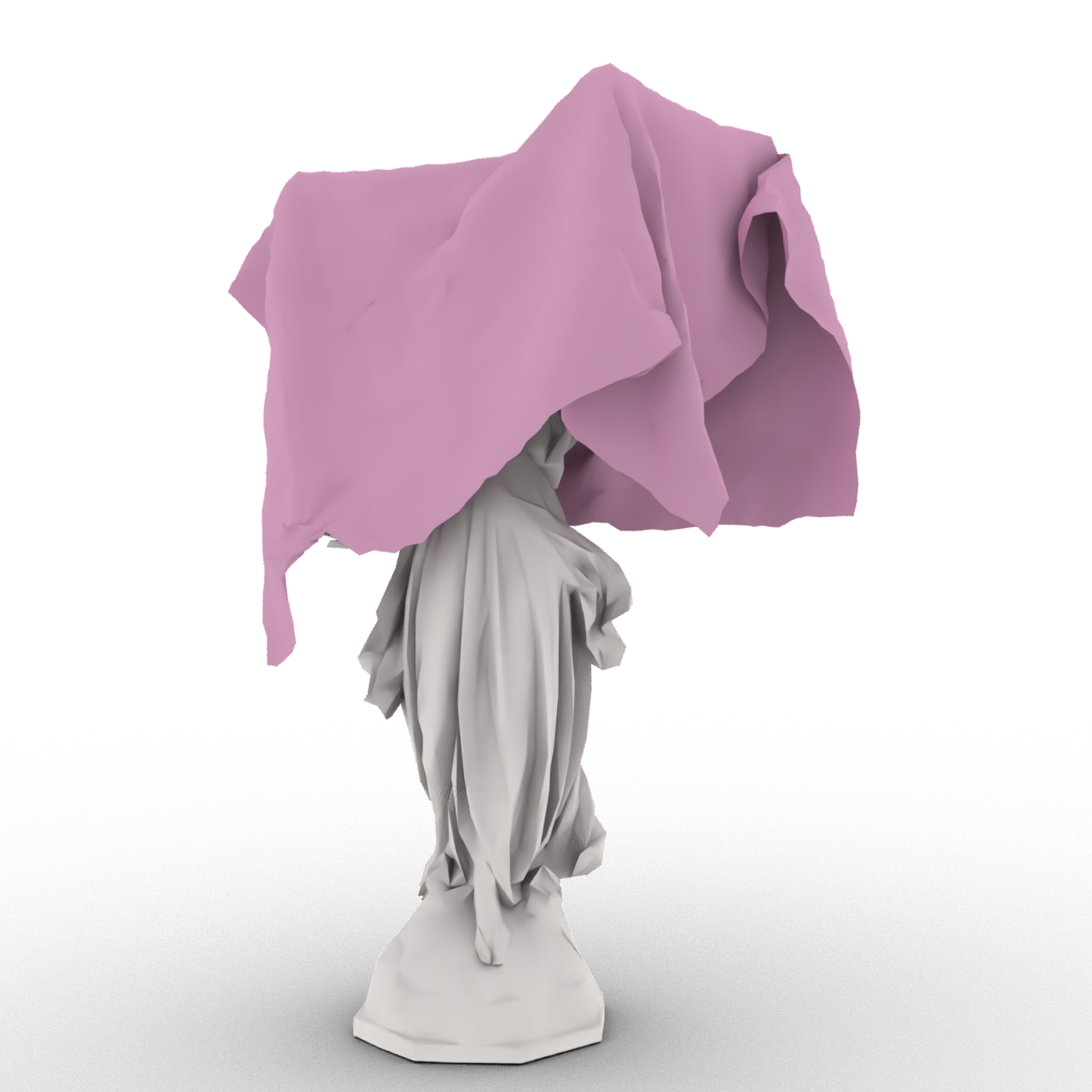} \\
    \includegraphics[width=0.163\linewidth]{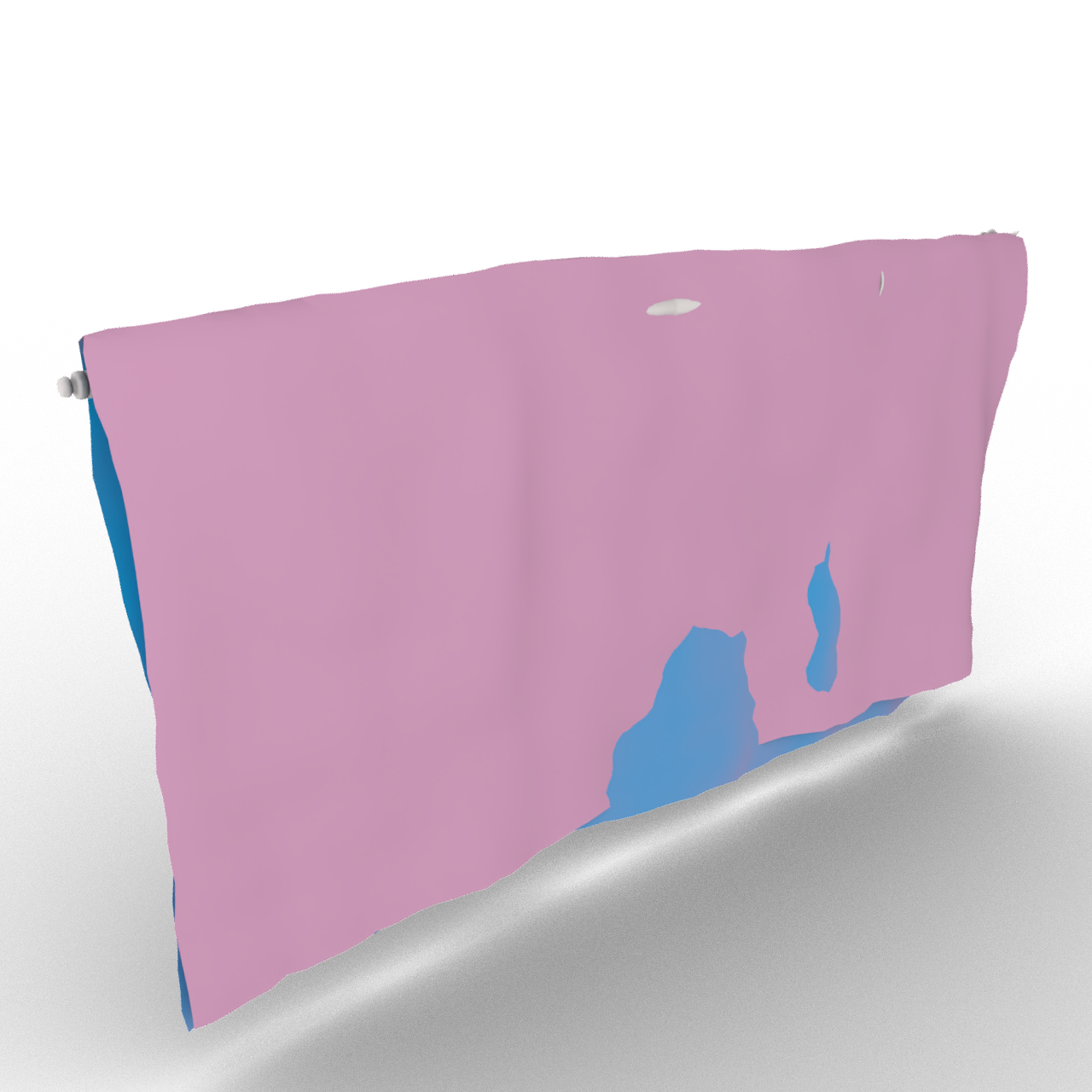} &
    \includegraphics[width=0.163\linewidth]{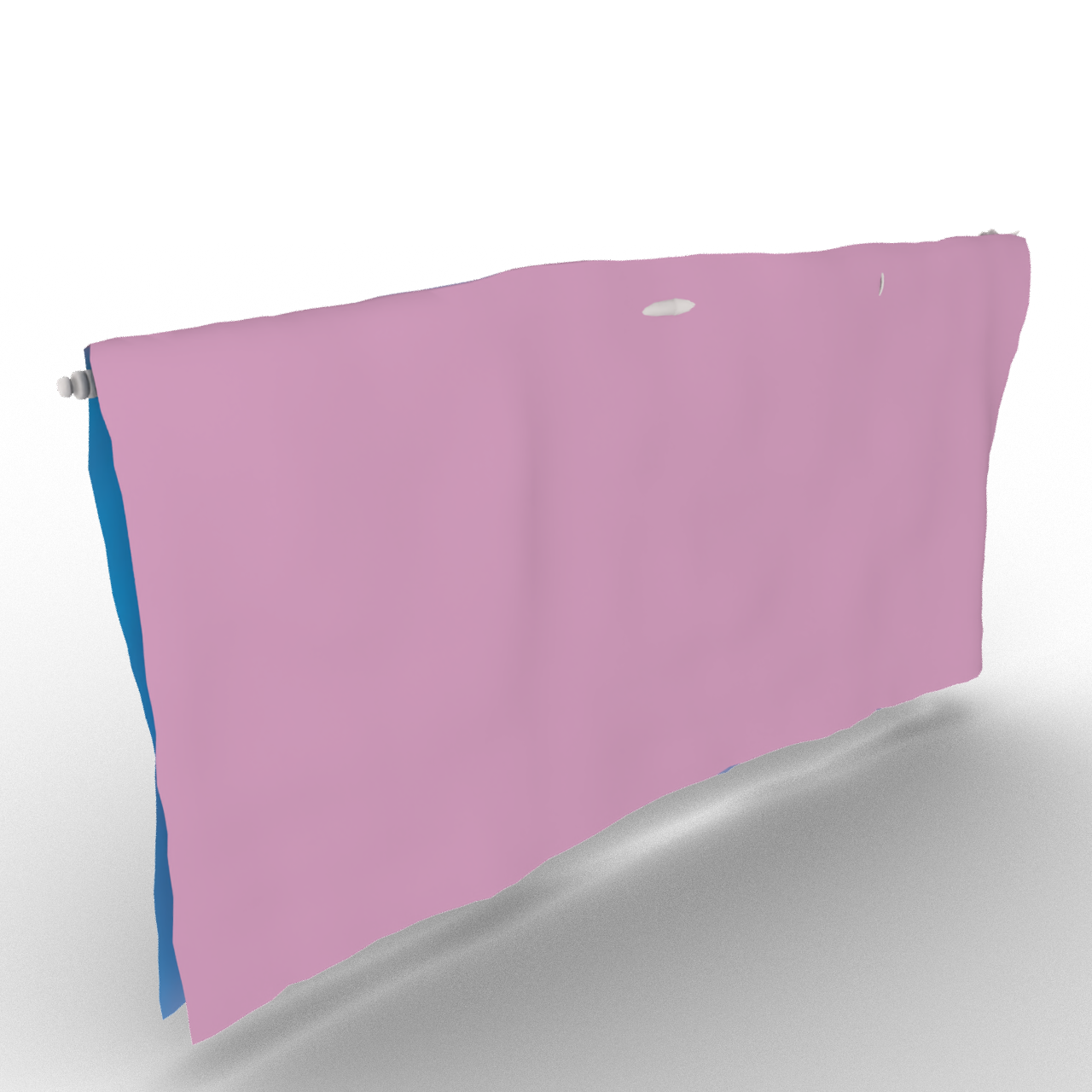} &
    \includegraphics[width=0.163\linewidth]{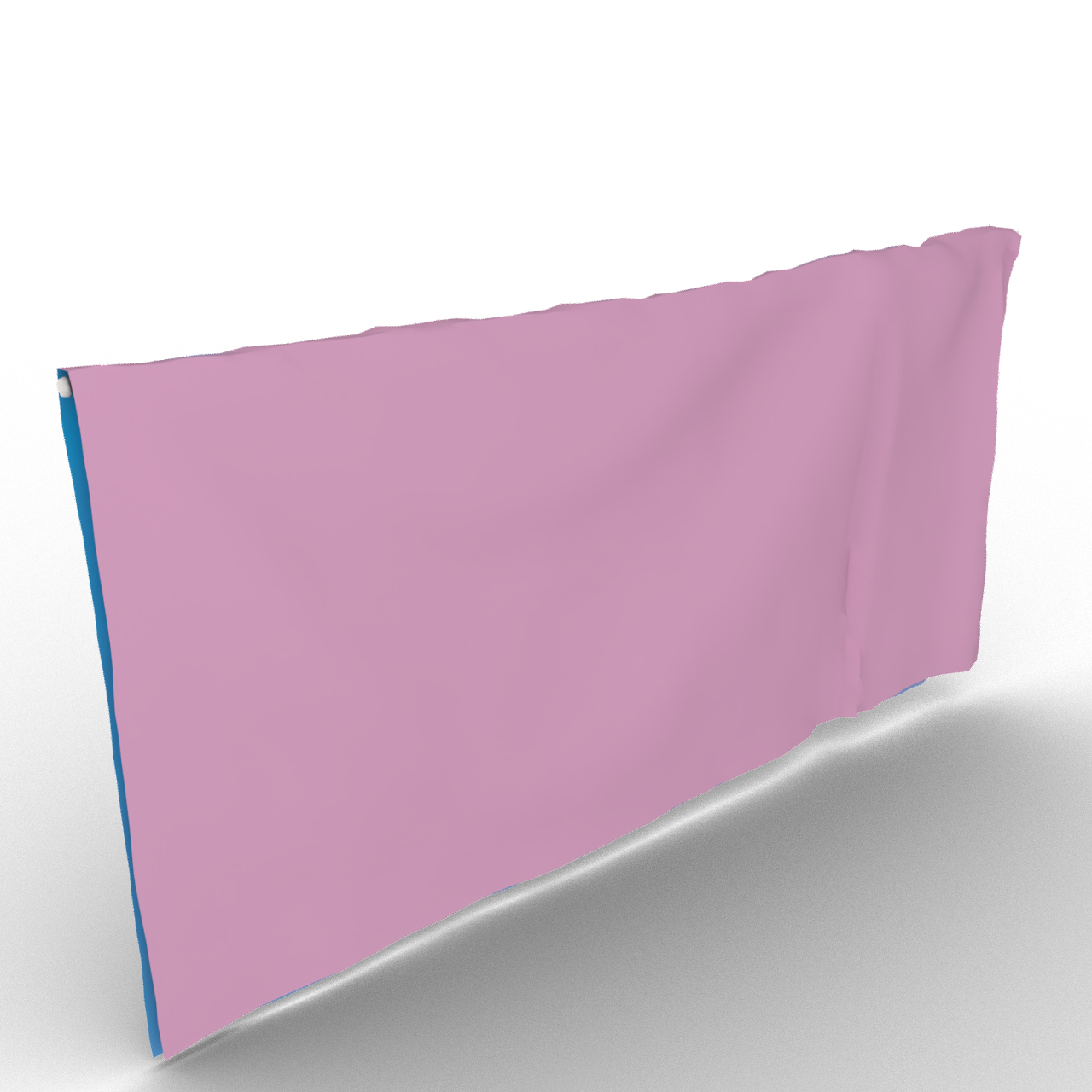} &
    \includegraphics[width=0.163\linewidth]{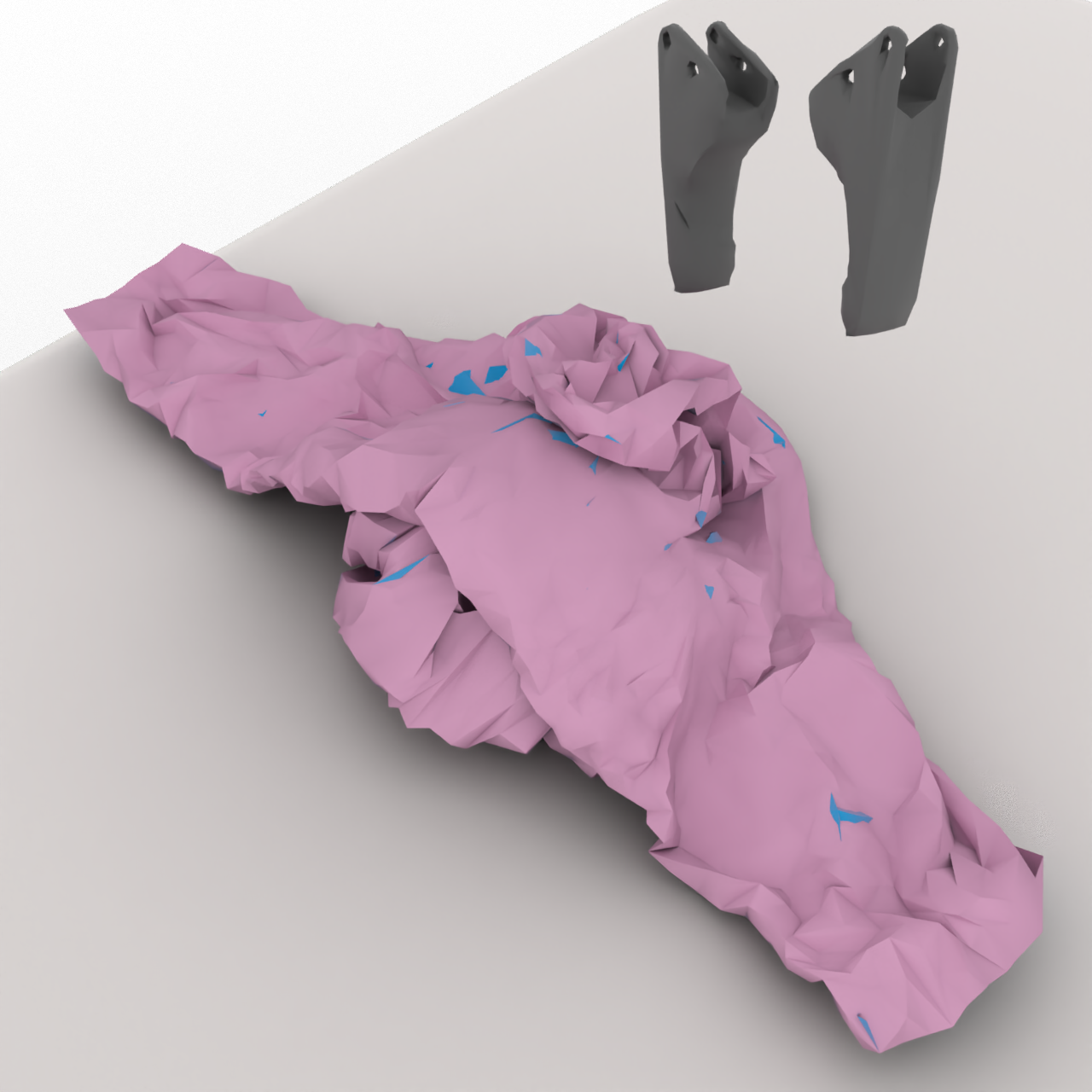} &
    \includegraphics[width=0.163\linewidth]{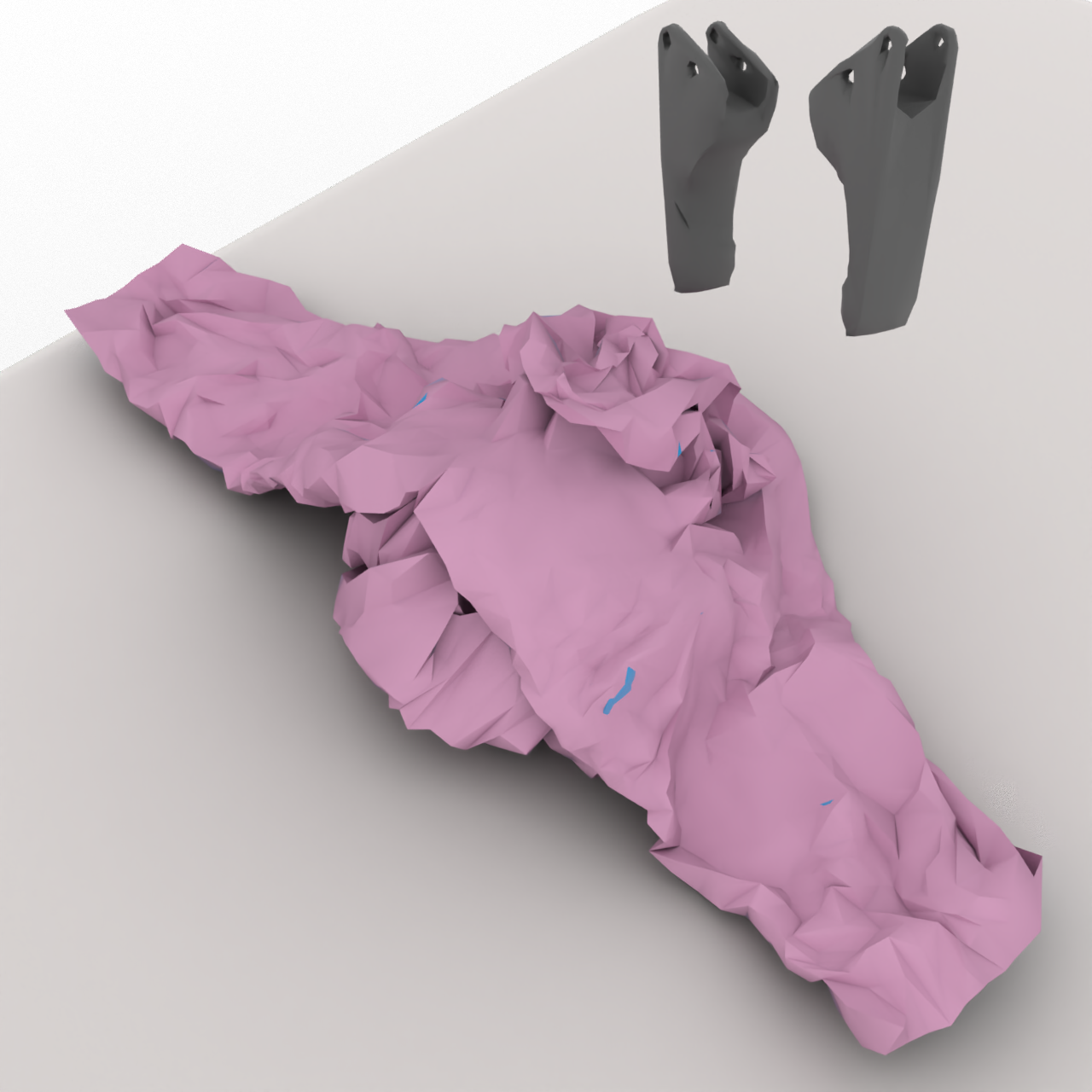} &
    \includegraphics[width=0.163\linewidth]{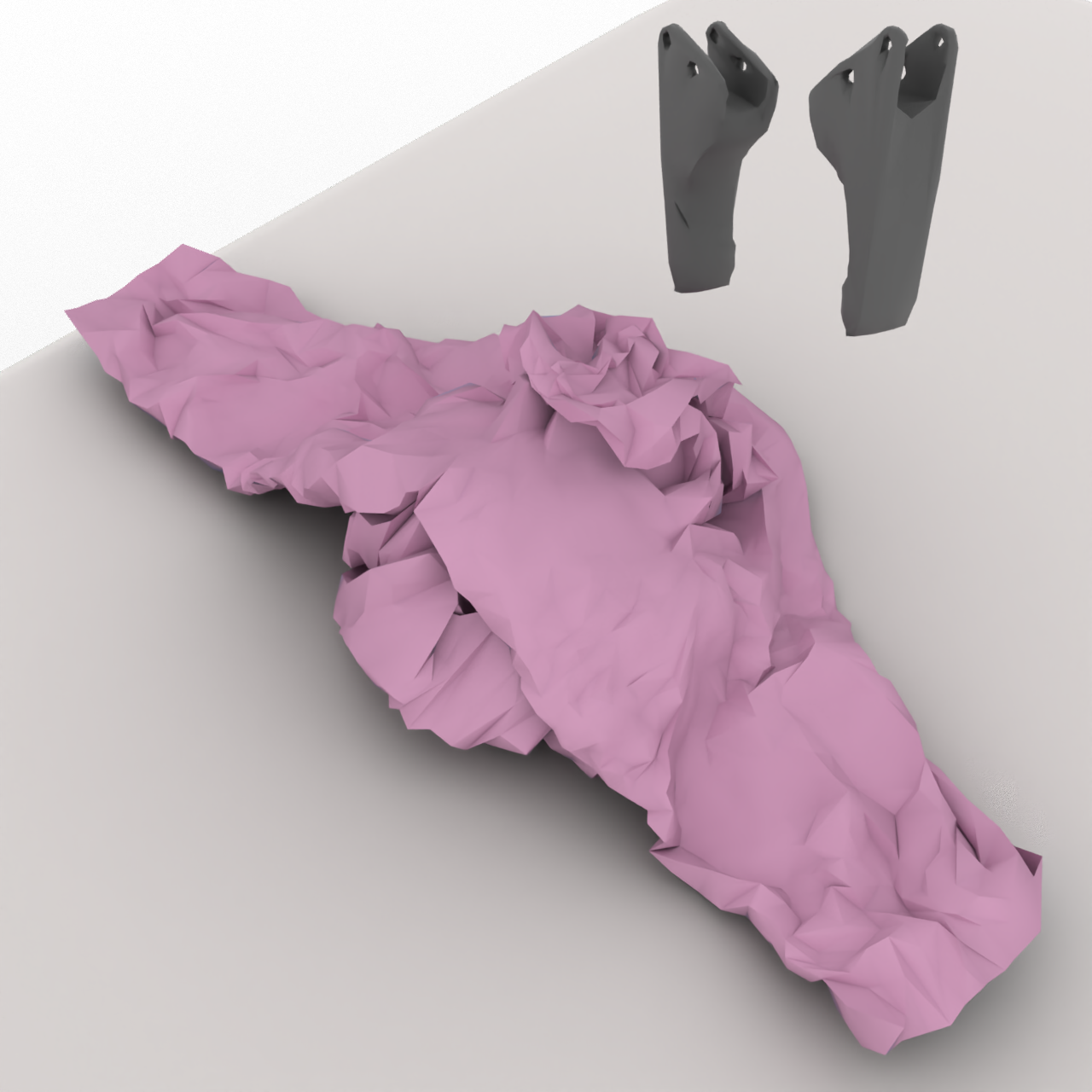} \\
  \end{tabular}
  \vspace{-2mm}
  \caption{\textbf{Effect of differentiable CCD.} Close-up views of collision-prone regions. \textit{Top row} (front flip, cloth on angel): cloth--collision penetrations that DCD loss cannot fully resolve; our CCD loss targets self-collision and does not improve these cases, but CCD post-processing eliminates them. \textit{Bottom row} (cloth on stick, robotic grasping): cloth self-collision artifacts, rendered with different colors for front (pink) and back (blue) faces for clarity; the CCD loss significantly reduces self-penetrations, and CCD post-processing resolves the remainder.}
  \label{fig:ablation_ccd}
  \vspace{-3mm}
\end{figure*}

\textbf{Impact of the CCD Module.}
Our high-fidelity penetration-free dataset additionally enables CCD-based operations, including the differentiable CCD loss during training and CCD post-processing at inference. We ablate the CCD module under three progressive settings: (1)~\emph{w/ DCD Loss} only, (2)~\emph{+ CCD Loss} during training, and (3)~\emph{+ CCD Post.} at inference. As shown in Figure~\ref{fig:ablation_ccd}, DCD loss alone leaves both cloth--collision and self-collision artifacts. Adding the CCD loss significantly reduces self-penetrations, while CCD post-processing eliminates remaining artifacts of both types. Furthermore, a direct comparison with ContourCraft~\cite{DBLP:conf/siggraph/0002BBHT24}, a state-of-the-art DCD-based approach is reported in Appendix~\ref{supp:ccd_vs_dcd}.

\textbf{Latent Compression Rate.}
\label{subsubsec:ablation_latent}
We vary the number of latent tokens $N_{latents}$ among 512, 1024, 2048, and no compression. As shown in Table~\ref{tab:ablation_latent}, three key findings emerge: (1)~MVE improves steadily from $N{=}512$ to $N{=}2048$, yet the uncompressed variant performs \emph{worse} due to insufficient training convergence, highlighting the importance of latent compression; (2)~$N{=}1024$ infers at ${\sim}$4.9\,ms/frame, well beyond real-time speed and ${\sim}18{\times}$ faster than the uncompressed variant; (3)~$N{=}1024$ offers the best accuracy--efficiency trade-off. A visual comparison and detailed analysis are provided in Appendix~\ref{supp:latent}.

\begin{table*}[t]
    \centering
    \begin{minipage}[t]{0.6\linewidth}
        \centering
        \caption{Ablation on latent compression rate (Human Garment, 50k steps). ``No Comp.'' operates directly on all mesh vertices.}
        \label{tab:ablation_latent}
        \resizebox{\linewidth}{!}{%
        \begin{tabular}{l|cc|c|c}
        \toprule
        $N_{latents}$ & MVE $\downarrow$ & Coll. $\downarrow$ & Infer. (ms) $\downarrow$ & Train (h) $\downarrow$ \\
        \midrule
        512            & 9.79 & 22.55 & \cellcolor{best}4.88 & \cellcolor{best}9.78 \\
        1024 (default) & \cellcolor{second}7.14 & 19.98 & \cellcolor{second}4.90 & \cellcolor{second}10.87 \\
        2048           & \cellcolor{best}6.01 & \cellcolor{best}15.19 & 10.75 & 15.68 \\
        No Comp.       & 10.01 & \cellcolor{second}18.56 & 90.07 & 18.01 \\
        \bottomrule
        \end{tabular}%
        }
    \end{minipage}%
    \hfill
    \begin{minipage}[t]{0.36\linewidth}
        \centering
        \caption{Ablation on spatial GNN (Human Garment, $N_{latents}{=}2048$, 50k steps).}
        \label{tab:ablation_gnn}
        \begin{tabular}{l|cc}
        \toprule
        Variant & MVE $\downarrow$ & Coll. $\downarrow$ \\
        \midrule
        w/o GNN & 7.18 & 22.98 \\
        \textbf{Ours} & \cellcolor{best}\textbf{6.01} & \cellcolor{best}\textbf{15.19} \\
        \bottomrule
        \end{tabular}%
    \end{minipage}
\end{table*}

\textbf{Spatial GNN.}
\label{subsubsec:ablation_gnn}
We replace the GNN encoder/decoder with simple MLPs (\textit{w/o Local GNN}), keeping the latent dimension fixed at $N_{latents}{=}2048$. As shown in Table~\ref{tab:ablation_gnn}, removing the GNN increases MVE from 6.01\,cm to 7.18\,cm (+19\%) and raises the collision rate from 15.19\% to 22.98\%. Without the GNN's explicit message passing along mesh edges, the MLP-only encoder must infer local surface geometry purely from per-vertex features, losing the topological connectivity that is critical for accurate spatial encoding and decoding.

\subsection{Scalability Analysis}
\label{subsec:scalability}

A key advantage of the latent-space formulation is scalable dynamics. We validate this by training all methods on the Human Garment subset (${\sim}$3.6k vertices) and evaluating both accuracy (MVE) and inference speed (ms/frame) at four test-time mesh resolutions: 5k, 10k, 20k, and 40k vertices. All timings are measured on a single NVIDIA RTX 4090 (24\,GB). Results are shown in Table~\ref{tab:scalability}.

\textbf{Cross-Resolution Accuracy.}
Our method achieves the best MVE across all four resolutions, demonstrating strong cross-resolution generalization. Even at 40k vertices---roughly $11{\times}$ the training resolution---our method maintains reasonable predictions and outperforms all baselines.

\begin{table}[t]
    \centering
    \caption{Scalability analysis across mesh resolutions on a single NVIDIA RTX 4090 (24\,GB). MVE (cm) $\downarrow$ / inference time (ms/frame) $\downarrow$. LayersNet runs out of memory at 40k vertices.}
    \label{tab:scalability}
    {\small
    \begin{tabular}{l|cc|cc|cc|cc}
    \toprule
    & \multicolumn{2}{c|}{\textbf{5k vertices}} & \multicolumn{2}{c|}{\textbf{10k vertices}} & \multicolumn{2}{c|}{\textbf{20k vertices}} & \multicolumn{2}{c}{\textbf{40k vertices}} \\
    Method & MVE $\downarrow$ & Time $\downarrow$ & MVE $\downarrow$ & Time $\downarrow$ & MVE $\downarrow$ & Time $\downarrow$ & MVE $\downarrow$ & Time $\downarrow$ \\
    \midrule
    SOTA GNN  & 55.34 & 130.01 & 74.24 & 147.12 & 73.98 & 231.57 & 201.46 & \cellcolor{second}471.78 \\
    MAT       & \cellcolor{second}32.68 & 66.1 & \cellcolor{second}54.12 & 294 & \cellcolor{second}67.1 & 634.32 & \cellcolor{second}72.36 & 1449.27 \\
    LayersNet & 149.27 & \cellcolor{second}59.67 & 79.15 & \cellcolor{second}90.64 & 69.24 & \cellcolor{second}138.08 & \multicolumn{2}{c}{OOM} \\
    \textbf{Ours} & \cellcolor{best}\textbf{7.21} & \cellcolor{best}\textbf{22.24} & \cellcolor{best}\textbf{5.81} & \cellcolor{best}\textbf{43.03} & \cellcolor{best}\textbf{34.91} & \cellcolor{best}\textbf{90.87} & \cellcolor{best}\textbf{57.32} & \cellcolor{best}\textbf{275.27} \\
    \bottomrule
    \end{tabular}%
    }
\end{table}

\textbf{Inference Speed.}
Our method is consistently the fastest across all resolutions, since the core Temporal Transformer cost stays fixed at $O(N_{latents}^2)$ regardless of mesh size. Notably, at 40k vertices our method is still ${\sim}1.7{\times}$ faster than the second-fastest one (SOTA GNN). Per-method timings and end-to-end pipeline timing details are provided in Appendix~\ref{supp:scalability_details}.

\subsection{Limitation of AR Rollout in Extreme Cases}
\label{subsec:ar_drift}

We observe that our model performs well in the vast majority of cases; however, on extreme cases the autoregressive (AR) rollout can still lead to considerable error accumulation. On an extreme Diverse Object Collision test case, the vertex error rapidly accumulates to ${\sim}37$\,cm within the first 60 frames and never recovers, despite velocity-based inputs and the rollout curriculum. This suggests that for highly nonlinear, contact-rich dynamics, the error accumulation of single-step AR prediction is difficult to eliminate through training strategies alone, motivating one-shot or chunked multi-step prediction paradigms. Detailed analysis is provided in Appendix~\ref{supp:ar_drift}.

\section{Conclusion}
\label{sec:conclusion}

We presented \textbf{ClothTransformer}, a unified Transformer framework that reformulates cloth simulation as autoregressive sequence modeling in a learned latent space. A single model handles diverse scenarios---body-driven garments, robotic manipulation, and free-fall collisions---and achieves approximately $4$--$9{\times}$ lower error than prior state-of-the-art methods across all scenarios. The latent-space formulation compresses arbitrary-resolution meshes into a fixed-size set of tokens, making temporal dynamics computation independent of mesh resolution. To support physically plausible training, we construct a diverse-scenario penetration-free dataset spanning all three settings, which enables our differentiable CCD module to suppress penetration artifacts.


\textbf{Limitations and Future Work.} Currently, our model infers material properties implicitly; future iterations could incorporate explicit physical parameters (\eg, stiffness) for greater artistic control. Furthermore, we observe that in extremely challenging scenarios, single-step autoregressive (AR) predictions of highly nonlinear physical dynamics can still lead to considerable error accumulation. This makes one-shot (or chunked, multi-step) prediction paradigms a potential alternative for future investigation. Additionally, extending the framework to handle topological changes (\eg, tearing) and integrating the framework with multimodal foundation models for text-guided physics generation are promising directions for future research.


\bibliographystyle{plainnat}
\bibliography{refs_full}


\clearpage
\appendix

\begin{center}
  {\LARGE\bfseries Appendix: Supplementary Material\par}
\end{center}
\vspace{1em}

\section{Architecture Details}
\label{supp:architecture}

Here we provide additional details of the three components introduced in Sec.~\ref{sec:method} of the main paper.

\textbf{Spatial Encoder.}
The encoder maps the physical state at frame $T$ into a fixed-size set of latent vectors $\mathbf{Z}_T \in \mathbb{R}^{K \times D}$, processing two inputs: the cloth mesh at frame $T$ and the collision geometry at frame $T{+}1$.

\textit{Cloth Feature Extraction.}
Each cloth vertex is encoded with two embeddings:
(1)~a \textbf{Position Embedding} that applies sinusoidal positional encoding to the 3D coordinates $\mathbf{X}_T$, mapping them into a high-dimensional feature space;
(2)~a \textbf{Velocity Embedding} that projects the vertex velocities $\mathbf{V}_T$ into the same feature dimension to capture instantaneous motion.
In practice, directly using velocity as input leads to error accumulation during autoregressive rollout, even with data normalization. We therefore compute the velocity as the position difference between two consecutive frames, $\mathbf{V}_T = \mathbf{X}_T - \mathbf{X}_{T-1}$, and multiply it by a scaling coefficient to keep its magnitude comparable to the position embedding, which stabilizes long-horizon inference.
These embeddings are fused and processed by a 2-layer GNN that aggregates features along mesh edges $\mathcal{E}$, yielding topology-aware \textit{Cloth Vertex Tokens}.

\textit{Collision Triangle Embedding.}
Collision objects are represented as triangles from the lookahead frame $T{+}1$. Each triangle is encoded using:
(1)~\textbf{Vertex and Velocity Embedding} applied to its three vertices, ensuring the model is aware of the object's motion trajectory;
(2)~\textbf{Geometric Features} including the surface normal $\mathbf{n}$ and triangle area $A$.
The geometric and dynamic features are concatenated to produce \textit{Collision Triangle Tokens}.

\textit{Latent Compression.}
A set of $K$ learnable query tokens $\mathbf{Q}_{\text{learn}}$ attends to the concatenated Cloth Vertex Tokens and Collision Triangle Tokens via cross-attention, compressing the variable-sized input into a fixed set of $K$ latent tokens $\mathbf{Z}_T$. We set $K{=}1024$ by default.

\textbf{Temporal Transformer.}
The Transformer takes the latent state $\mathbf{Z}_T$ as input and evolves it forward in time. It uses block-causal masking across frames and self-attention within each frame to model inter-token dependencies. The architecture consists of 12 layers with 12 attention heads, an embedding dimension of 768, and a feed-forward dimension of 3072 with SwiGLU activation. The output is the predicted next-frame latent state $\mathbf{Z}_{T+1}$.

\textbf{Spatial Decoder.}
The decoder reconstructs vertex positions from the predicted latent tokens. Rest-shape vertices are encoded via sinusoidal Position Embedding into \textit{Rest Vertex Tokens}, which serve as queries in a cross-attention layer against the predicted latent tokens. This retrieves the dynamic state for each vertex based on its canonical position. A final 2-layer GNN refines the output to ensure local surface smoothness, followed by a projection layer mapping features to 3D coordinates.


\section{CCD Implementation Details}
\label{supp:ccd}

This section expands on the Continuous Collision Detection module and its differentiable loss introduced in the main paper (Sec.~\ref{subsec:ccd}), detailing the underlying cubic root finding and the inference-time post-processing.

\textbf{Cubic Root Finding.}
Both Point-Triangle (VF, FV, Self-VF) and Edge-Edge (EE, Self-EE) CCD tests reduce to finding the roots of a cubic polynomial
\begin{equation}
    P(t) = at^3 + bt^2 + ct + d = 0, \quad t \in [0, 1],
\end{equation}
where $t$ parameterizes the linear trajectory between consecutive frames. Standard iterative root solvers are computationally expensive and numerically unstable when applied to thousands of primitive pairs. We adopt the method of Yuksel~\cite{yuksel2022fast}, which analytically computes the critical points of $P(t)$ (roots of $P'(t)$) to decompose $[0, 1]$ into monotonic intervals, then checks for sign changes at interval boundaries and applies Newton-Raphson iteration only within intervals where a root is guaranteed. This approach enables efficient and robust CCD on large meshes.

\textbf{Iterative Post-Processing.}
At inference, the CCD post-processing resolves collisions iteratively. For each detected collision at time $t_c$, we compute $t_{\text{safe}} = \max(0,\, t_c - \epsilon)$ and reset the penetrating vertices to their positions at $t_{\text{safe}}$ via linear interpolation along the motion vector. Since resolving one collision may introduce secondary collisions, this process is repeated until convergence (i.e., no new collisions are detected) or a maximum iteration count is reached.

\section{Dataset Simulation Details}
\label{supp:dataset}

We simulate the Baraff and Witkin~\cite{DBLP:conf/siggraph/BaraffW98} cloth model using the GIPC solver~\cite{DBLP:journals/tog/HuangCLK24} with the following material parameters: stretching Young's modulus $E_s = 10^6$\,Pa, bending Young's modulus $E_b = 10^5$\,Pa, Poisson's ratio $\nu = 0.49$, shear stiffness $G = 5 \times 10^6$\,Pa, and cloth density $\rho = 200$\,g/m$^2$. The simulation uses a fixed time step of $\Delta t = 1/60$\,s, a friction coefficient of $\mu = 0.4$, and each sequence spans 240 frames (4 seconds).

\section{Latent Compression Analysis}
\label{supp:latent}

\begin{figure}[t]
  \centering
  \setlength{\tabcolsep}{1pt}
  \renewcommand{\arraystretch}{0}
  \begin{tabular}{@{}ccccc@{}}
    \small Ground Truth & \small $N{=}512$ & \small $N{=}1024$ & \small $N{=}2048$ & \small No Comp. \\
    \includegraphics[width=0.195\linewidth]{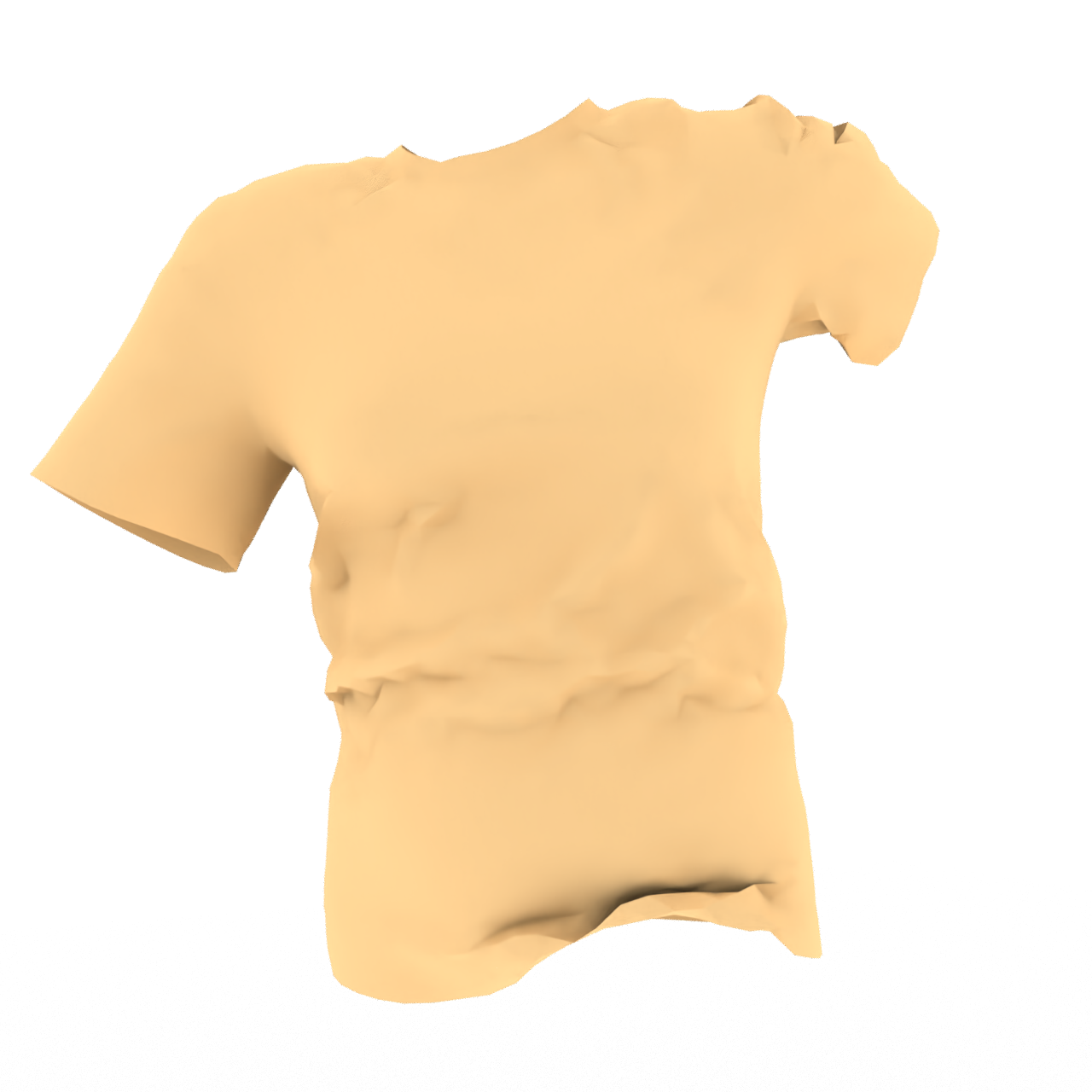} &
    \includegraphics[width=0.195\linewidth]{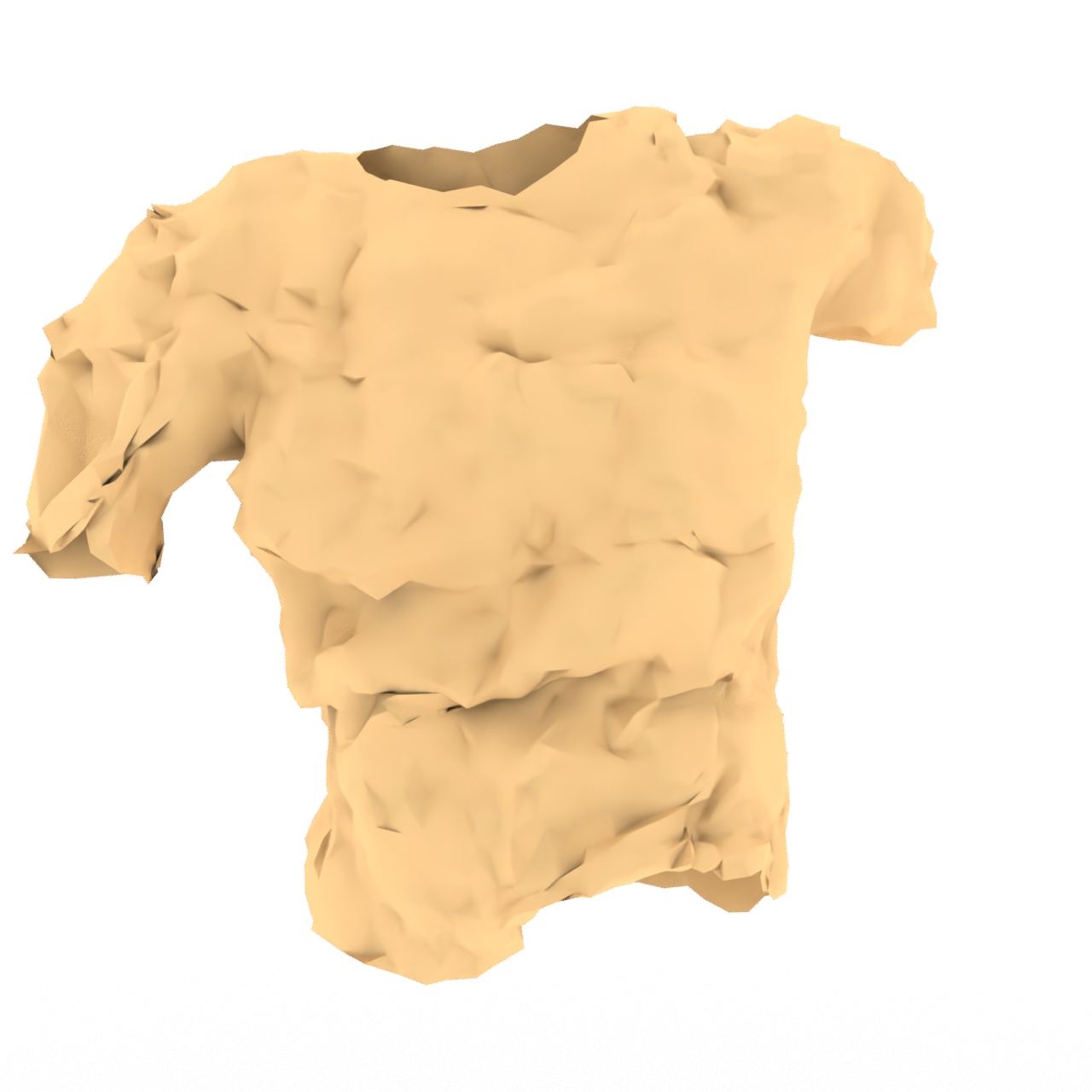} &
    \includegraphics[width=0.195\linewidth]{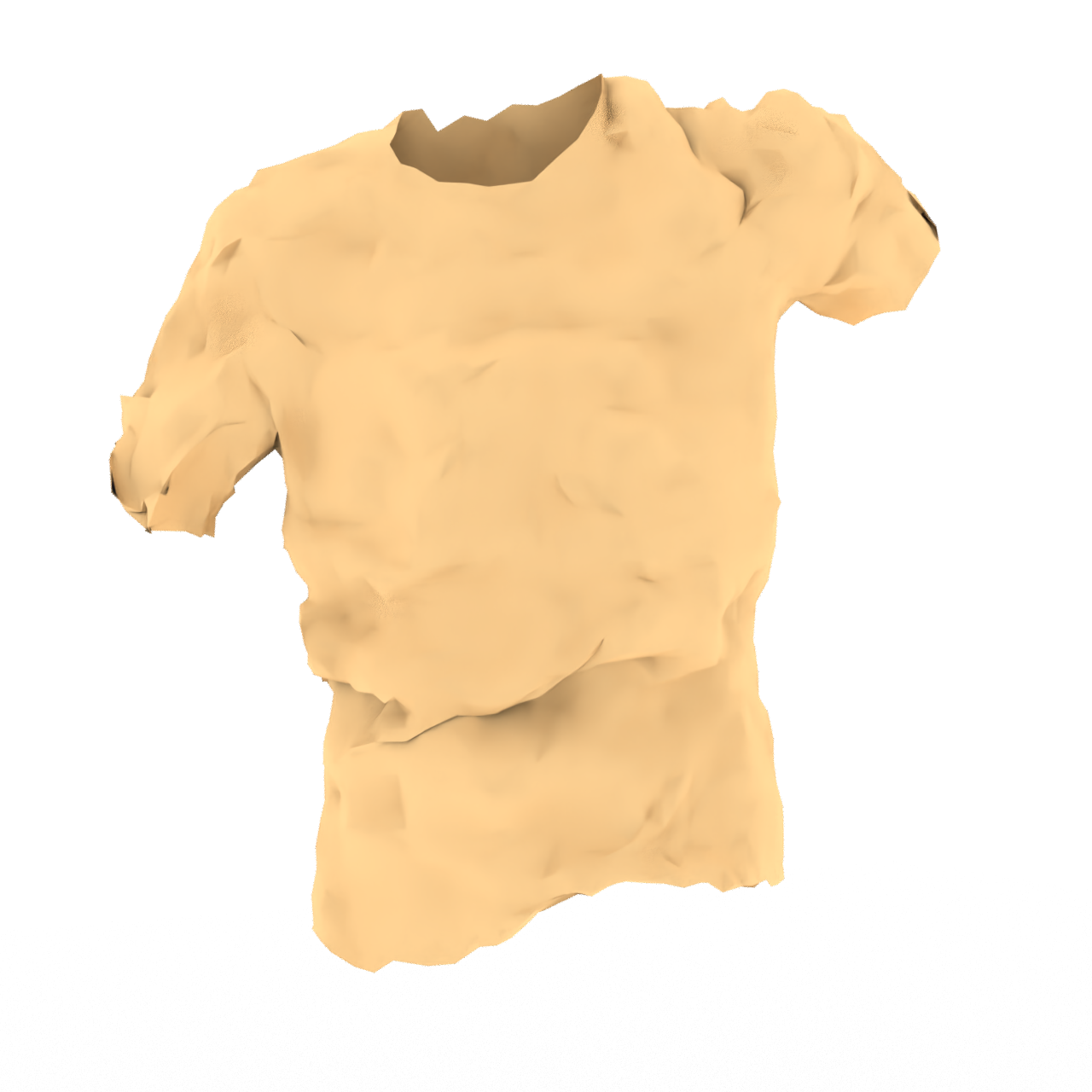} &
    \includegraphics[width=0.195\linewidth]{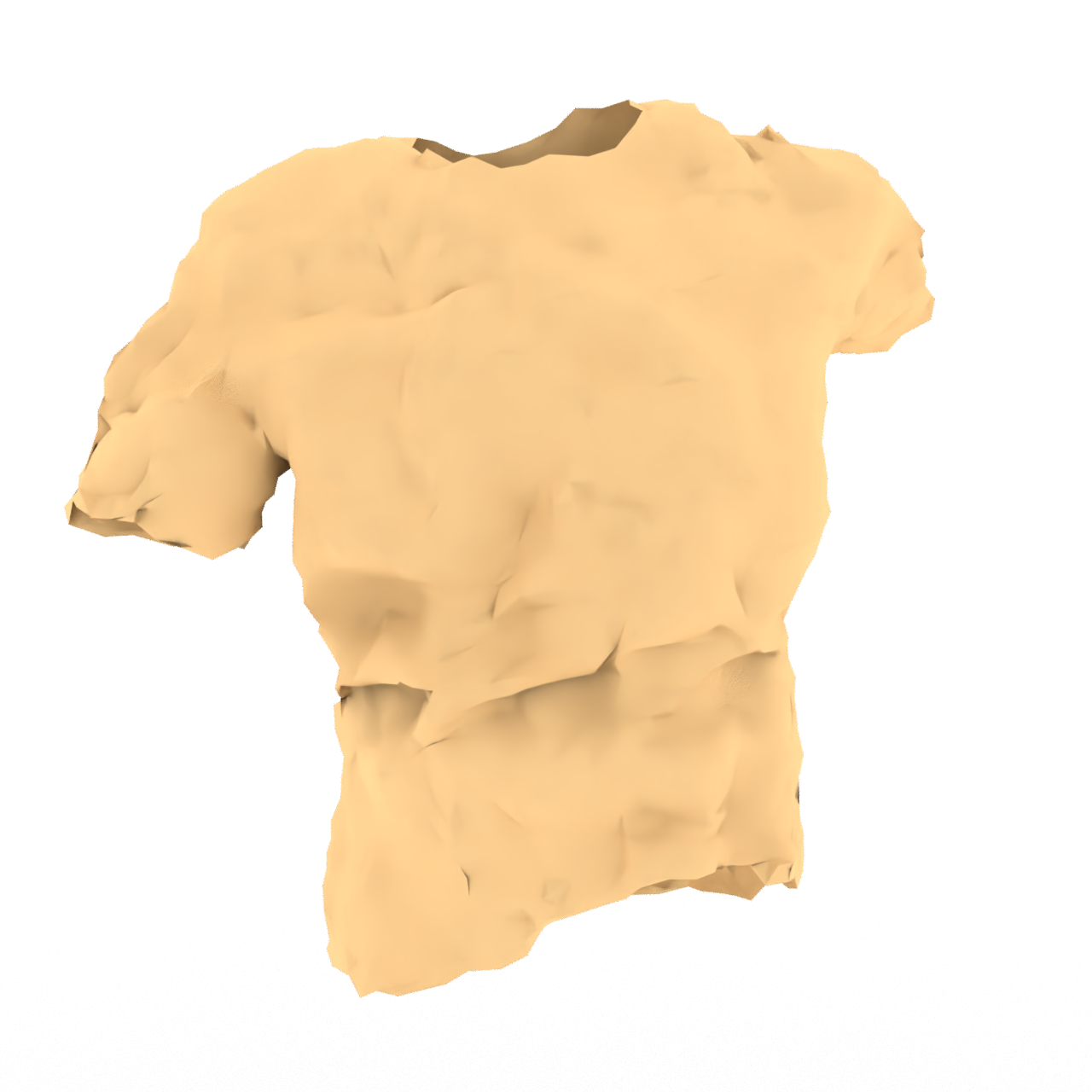} &
    \includegraphics[width=0.195\linewidth]{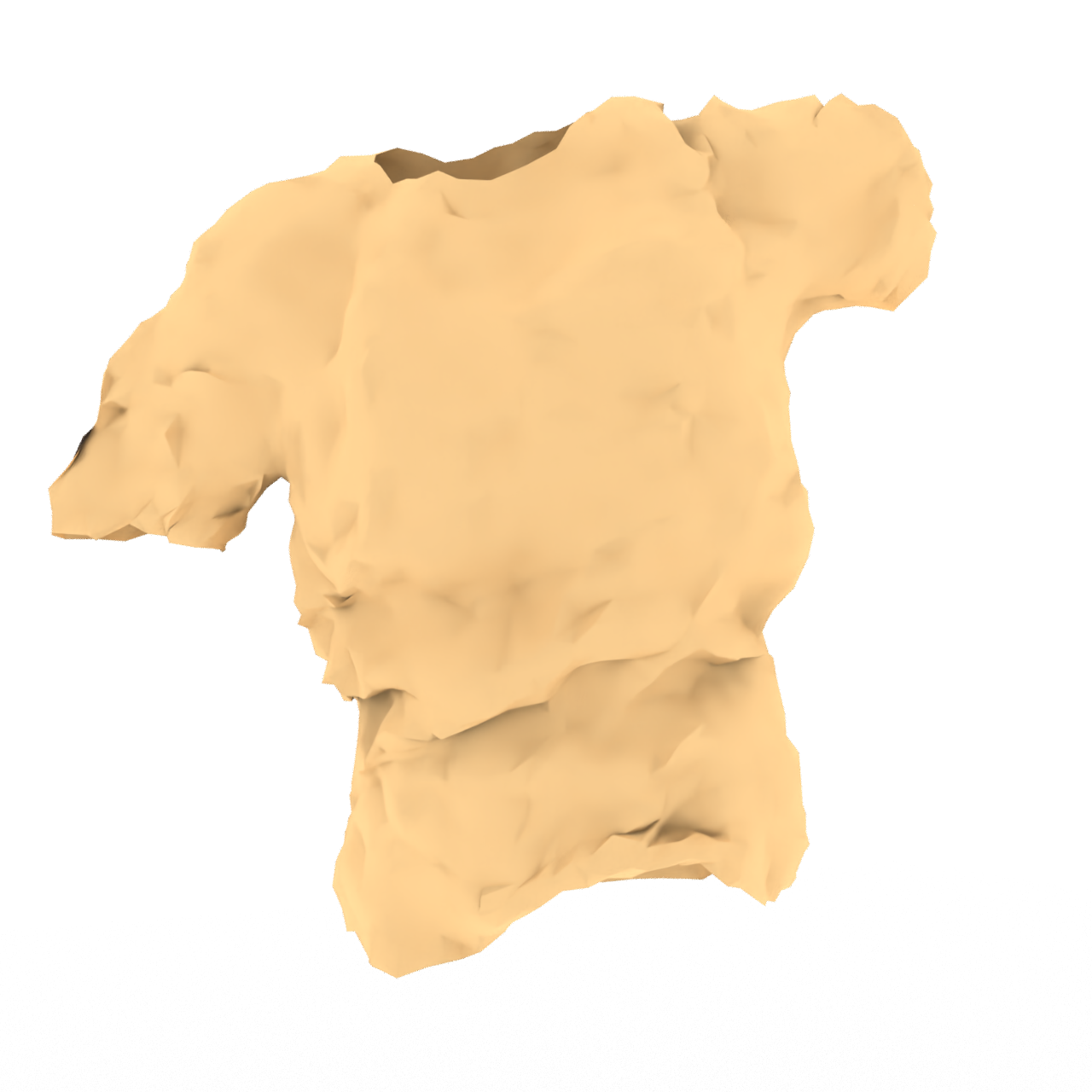} \\
    \includegraphics[width=0.195\linewidth]{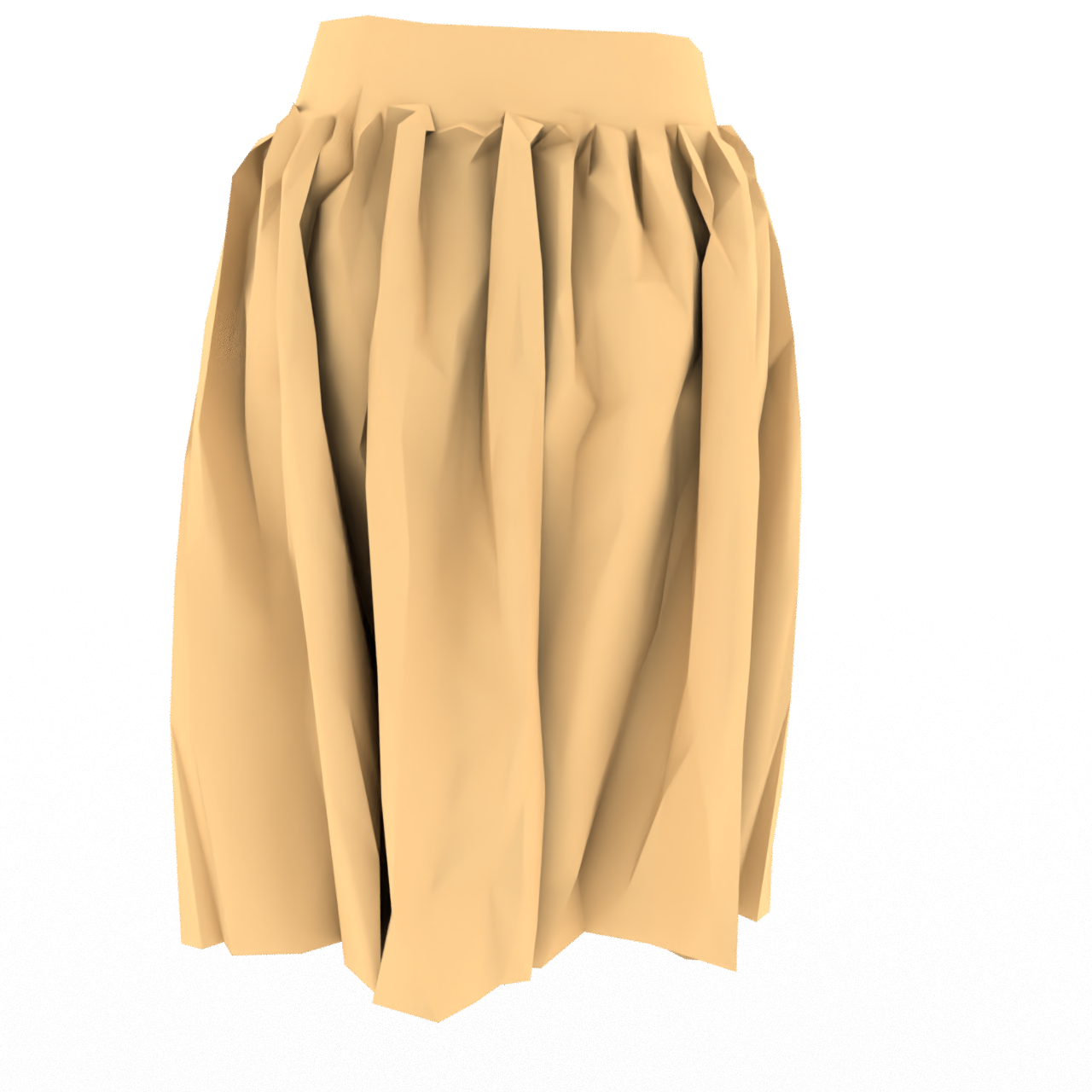} &
    \includegraphics[width=0.195\linewidth]{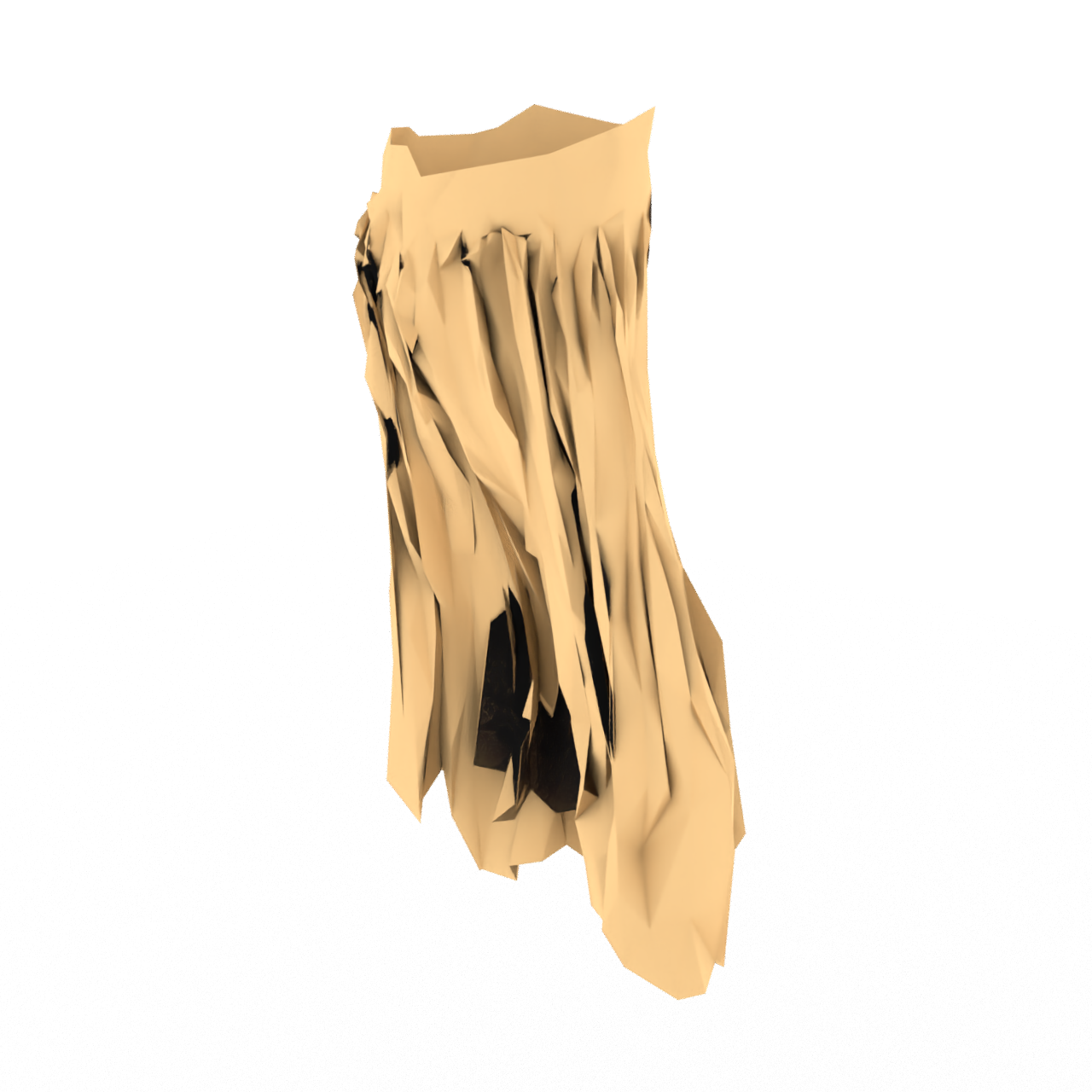} &
    \includegraphics[width=0.195\linewidth]{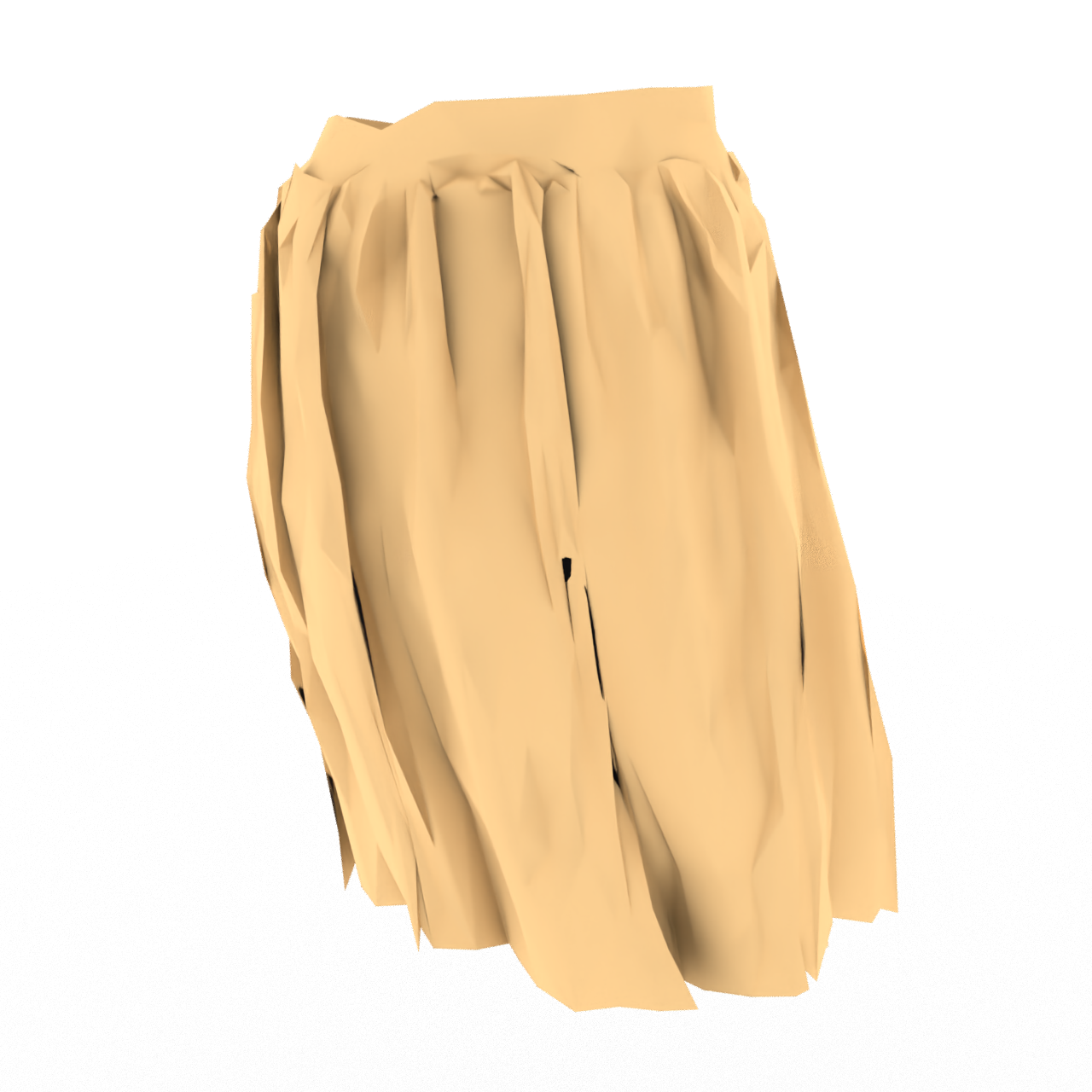} &
    \includegraphics[width=0.195\linewidth]{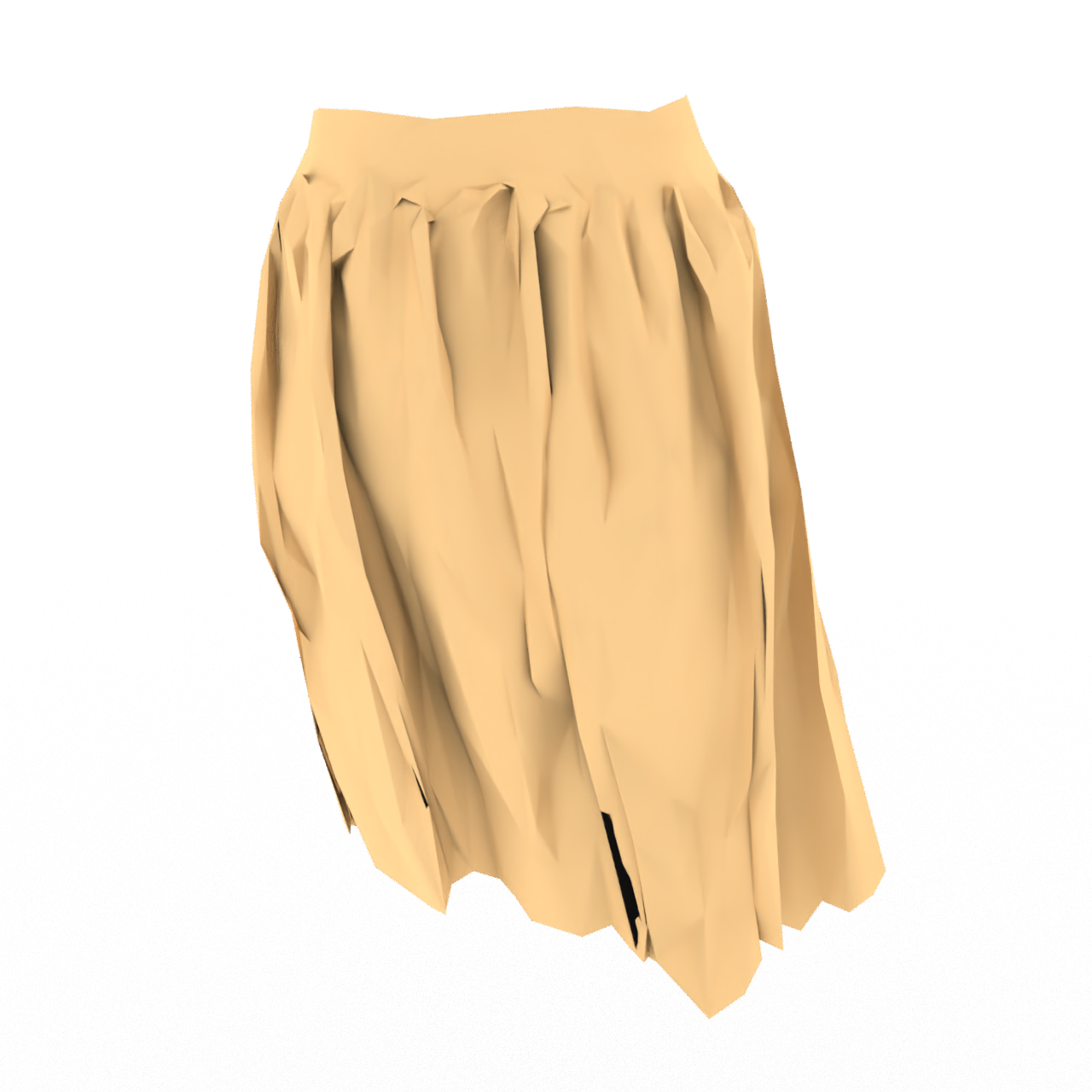} &
    \includegraphics[width=0.195\linewidth]{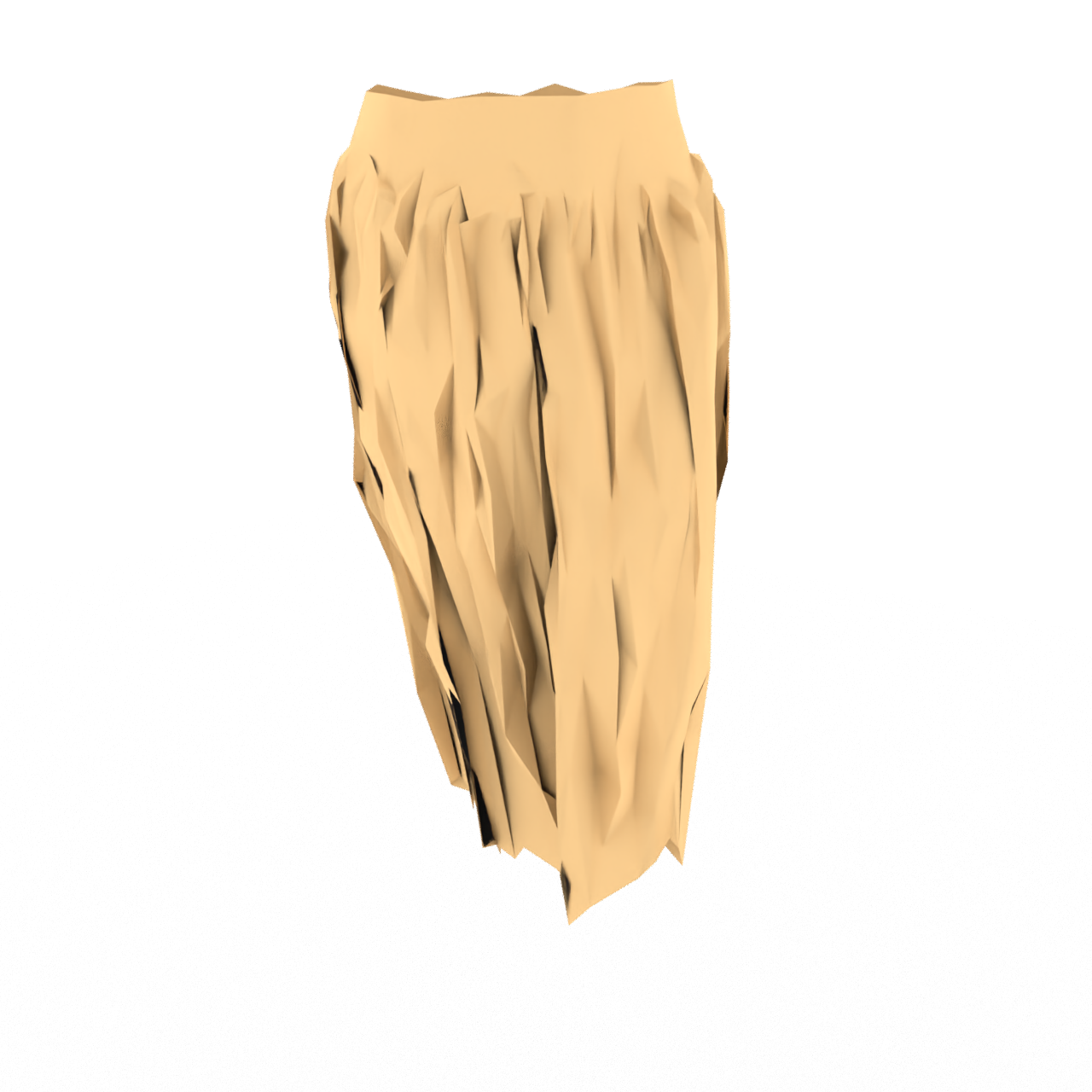} \\
  \end{tabular}
  \caption{\textbf{Visual comparison of different latent compression rates.} $N{=}512$ produces abnormal deformation artifacts due to excessive compression. The uncompressed variant (No Comp.) exhibits similar artifacts due to insufficient training convergence. Our default $N{=}1024$ achieves a good balance between quality and efficiency.}
  \label{fig:ablation_latent_vis}
\end{figure}

We provide a detailed analysis of the latent compression ablation (Table~\ref{tab:ablation_latent} in the main paper). Figure~\ref{fig:ablation_latent_vis} qualitatively illustrates the visual artifacts at different compression rates.

\textbf{Accuracy.}
MVE improves steadily as $N_{latents}$ increases from 512 to 2048, with our default $N{=}1024$ reducing MVE by 27\% relative to $N{=}512$ and $N{=}2048$ achieving a further 16\% reduction. Notably, the uncompressed variant (``No Comp.'') performs \emph{worse} than both $N{=}1024$ and $N{=}2048$: because the full-resolution Transformer has substantially more parameters, it requires far more training iterations to converge, despite already consuming 18.01 GPU hours---nearly $1.7{\times}$ the training cost of our default $N{=}1024$ (10.87\,h). This emphasizes the importance of latent compression.

\textbf{Efficiency.}
The latent bottleneck provides dramatic gains in both inference speed and training cost. Both $N{=}512$ and $N{=}1024$ run at ${\sim}4.9$\,ms per frame---well within real-time budgets---while $N{=}2048$ is ${\sim}2.2{\times}$ slower at 10.75\,ms due to the quadratic attention cost $O(N_{latents}^2)$. The uncompressed variant is ${\sim}18{\times}$ slower than our default at 90.07\,ms. Training time follows a similar trend: $N{=}512$ and $N{=}1024$ require only 9.78\,h and 10.87\,h respectively, while $N{=}2048$ takes 15.68\,h and the uncompressed variant 18.01\,h.

\section{Scalability Analysis Details}
\label{supp:scalability_details}

\textbf{Per-Method Inference Speed.}
We provide a per-method analysis of the scalability results in Table~\ref{tab:scalability} of the main paper. Our method scales from 22.24\,ms at 5k vertices to 275.27\,ms at 40k vertices. This growth is approximately linear and stems solely from the lightweight spatial encoder/decoder; the core Temporal Transformer cost remains fixed at $O(N_{latents}^2)$ regardless of mesh size. LayersNet is the second fastest at lower resolutions (59.67\,ms at 5k, 90.64\,ms at 10k) but exhausts the 24\,GB GPU memory at 40k vertices, as its UV-patch tokenization materializes full-resolution feature maps that grow quadratically with mesh size. The SOTA GNN scales moderately, from 130.01\,ms at 5k to 471.78\,ms at 40k, due to its multi-level message-passing graph. MAT scales the worst with mesh resolution, jumping from 66.1\,ms at 5k to 1449.27\,ms at 40k, as its per-face attention cost grows steeply with mesh size. At 5k vertices, our method is ${\sim}2.7{\times}$ faster than LayersNet and ${\sim}5.8{\times}$ faster than the SOTA GNN.

\textbf{End-to-End Pipeline Timing.}
The per-frame timings in Table~\ref{tab:scalability} of the main paper measure the neural network forward pass only. With CCD post-processing enabled (${\sim}$10 iterations on average), the full pipeline adds approximately 30\,ms on a single RTX 4090, yielding a total of ${\sim}$52\,ms/frame at 5k vertices and ${\sim}$305\,ms/frame at 40k vertices. For reference, the GIPC solver~\cite{DBLP:journals/tog/HuangCLK24} used to generate our ground truth requires approximately 10\,s per frame on this scenario. Our full pipeline thus achieves roughly a $200{\times}$ speedup over GIPC while maintaining competitive accuracy, making it practical for interactive applications.

\section{Ablation Study Settings}
\label{supp:ablation}

The ablation studies reported in the main paper use different configurations depending on the experiment:

\textbf{Impact of the CCD Module} (Figure~\ref{fig:ablation_ccd} in the main paper and Figure~\ref{fig:ccd_vs_dcd} in this supplementary). This ablation was conducted on a smaller subset of the full training data to reduce computational cost.

\textbf{Latent Compression Rate and Spatial GNN} (Section~\ref{subsec:ablation} in the main paper). These two ablations were conducted using a smaller network configuration (hidden dimension $D{=}256$, 6 layers, 8 attention heads) on the Human Garment scenario, trained for 50k steps. This lighter setup allows efficient exploration of the design space; the trends observed are consistent with the full-scale model.

\section{Evaluation Details}
\label{supp:eval}

\textbf{Baseline Adaptation.}
The SOTA GNN backbone~\cite{DBLP:conf/cvpr/GrigorevBH23, DBLP:conf/siggraph/0002BBHT24} was originally designed for unsupervised training with physics-based losses. For a fair comparison, we adapt it to a supervised setting by replacing its unsupervised objectives with the same loss used in our pretraining stage, and train it on our penetration-free dataset.

\textbf{Metric Definitions.}
\begin{itemize}
    \item \textbf{MVE (cm):} Mean Vertex Error, the average Euclidean distance between predicted and ground-truth vertex positions over all vertices and frames:
    \begin{equation}
        \text{MVE} = \frac{1}{T \cdot N_v} \sum_{t=1}^{T} \sum_{i=1}^{N_v} \| \hat{\mathbf{x}}_i^t - \mathbf{x}_i^t \|_2.
    \end{equation}
    \item \textbf{Collision Rate (\%):} The percentage of cloth vertices that penetrate the collision object, averaged over all frames.
    \item \textbf{Self-Collision Rate (\%):} The percentage of cloth vertices involved in self-intersections, detected via CCD between consecutive frames. Unlike discrete checks, this metric captures tunneling events where a vertex passes through and returns within a single time step. The involved vertices include the colliding vertex and the three triangle vertices (for Self-VF) or the four edge endpoints (for Self-EE).
\end{itemize}

\section{CCD vs.\ DCD-Based Self-Collision Handling}
\label{supp:ccd_vs_dcd}

\begin{figure}[t]
  \centering
  \setlength{\tabcolsep}{1pt}
  \renewcommand{\arraystretch}{0}
  \begin{tabular}{@{}cccc@{}}
    \small Contact Loss & \small ContourCraft & \small CCD Loss & \small CCD Post. \\
    \includegraphics[width=0.24\linewidth]{figs/ccd_grasp_contact.png} &
    \includegraphics[width=0.24\linewidth]{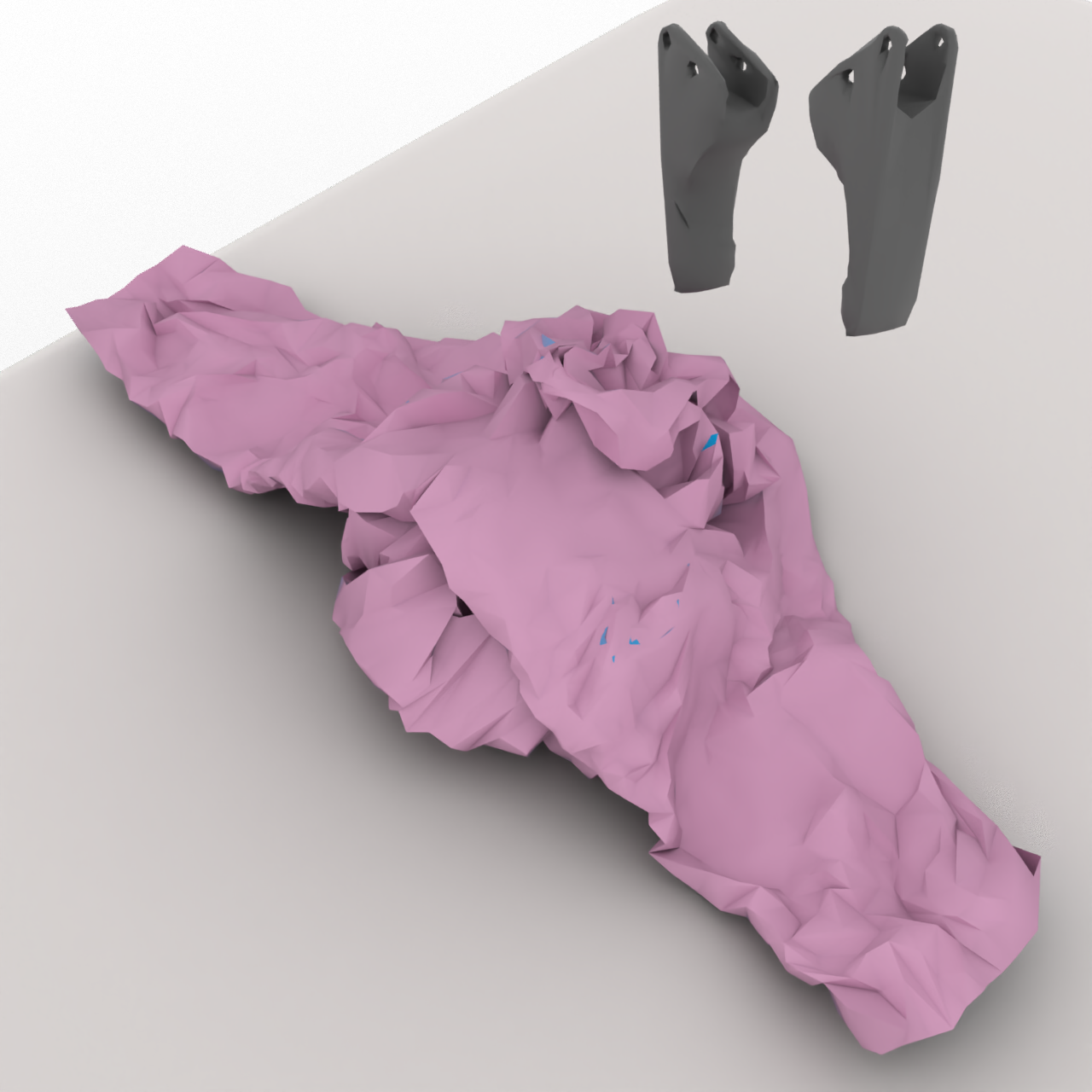} &
    \includegraphics[width=0.24\linewidth]{figs/ccd_grasp_ccd_loss.png} &
    \includegraphics[width=0.24\linewidth]{figs/ccd_grasp_ccd_post.png} \\
  \end{tabular}
  \caption{\textbf{CCD vs.\ DCD-based self-collision handling} on a challenging folded-cloth grasping scenario. ContourCraft~\cite{DBLP:conf/siggraph/0002BBHT24} and our CCD loss both reduce self-collisions but leave residual artifacts. Our CCD post-processing fully resolves remaining intersections.}
  \label{fig:ccd_vs_dcd}
\end{figure}

To further validate the advantage of CCD over DCD-based approaches, we compare against ContourCraft~\cite{DBLP:conf/siggraph/0002BBHT24}, a representative DCD method that detects self-intersecting contours and learns to resolve them. As shown in Figure~\ref{fig:ccd_vs_dcd}, on a challenging robotic grasping scenario where the cloth is put on a table (producing dense self-collisions), both ContourCraft and our CCD loss reduce self-collisions to some extent, but neither fully eliminates them. Our CCD post-processing, which iteratively resolves trajectory-level intersections, achieves clean, intersection-free results.

\section{Unified vs.\ Specialized Training}
\label{supp:unified}

\begin{figure}[t]
    \centering
    \setlength{\tabcolsep}{1pt}
    \renewcommand{\arraystretch}{0}
    \begin{tabular}{@{}c@{\hspace{2pt}}cccc@{}}
        & \small Frame 1 & \small Frame 2 & \small Frame 3 & \small Frame 4 \\
        \rotatebox{90}{\small\hspace{2pt}SOTA GNN (single)} &
        \includegraphics[width=0.22\linewidth]{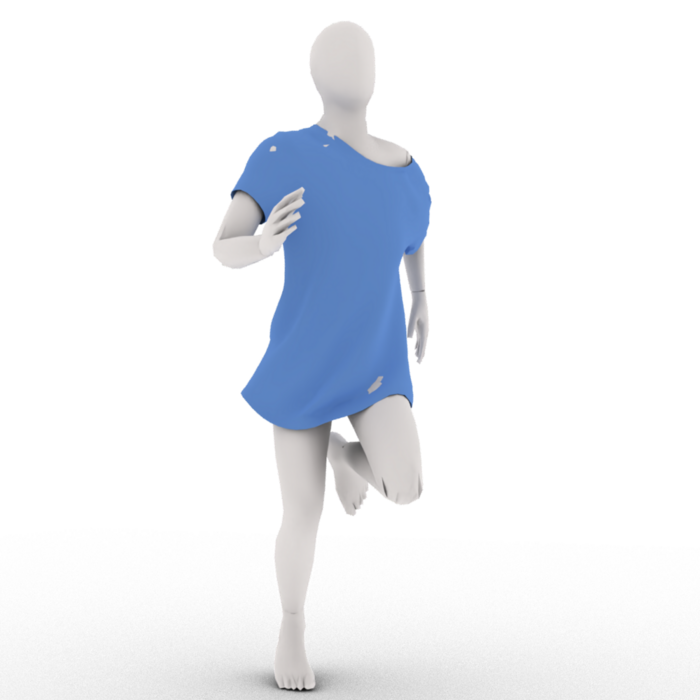} &
        \includegraphics[width=0.22\linewidth]{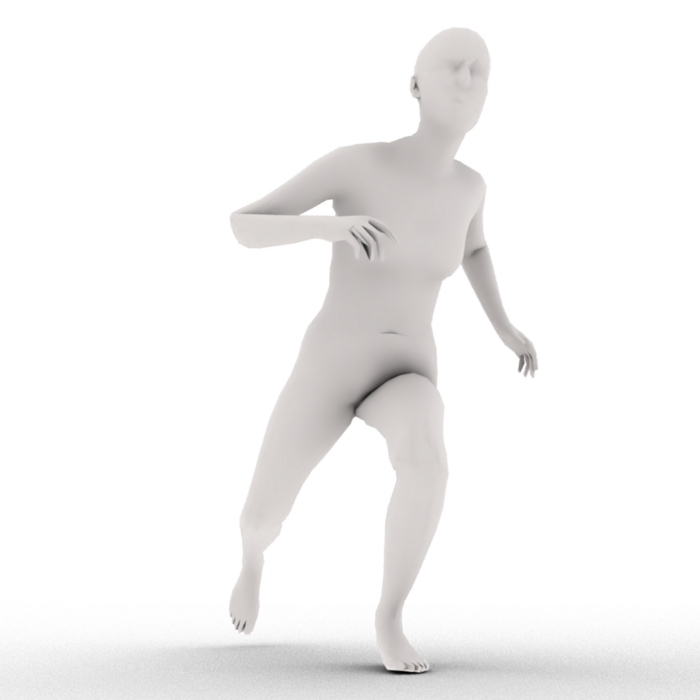} &
        \includegraphics[width=0.22\linewidth]{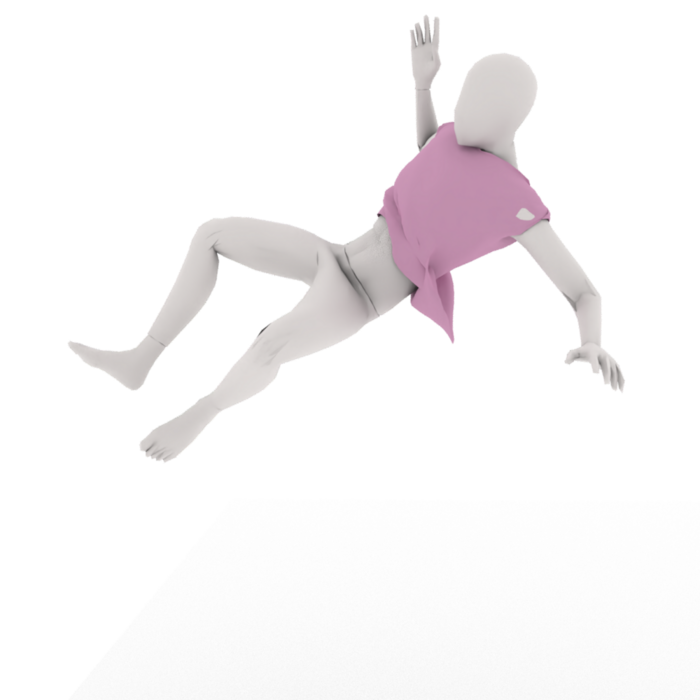} &
        \includegraphics[width=0.22\linewidth]{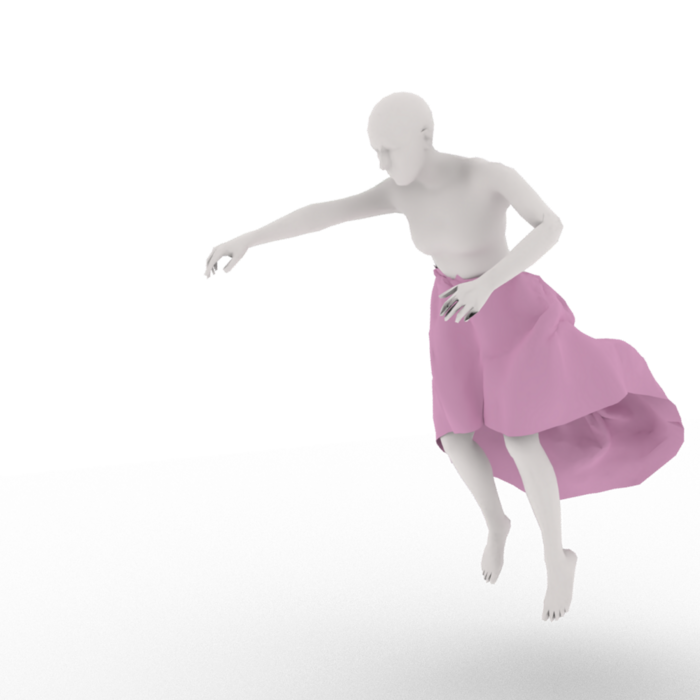} \\
        \rotatebox{90}{\small\hspace{2pt}SOTA GNN (unified)} &
        \includegraphics[width=0.22\linewidth]{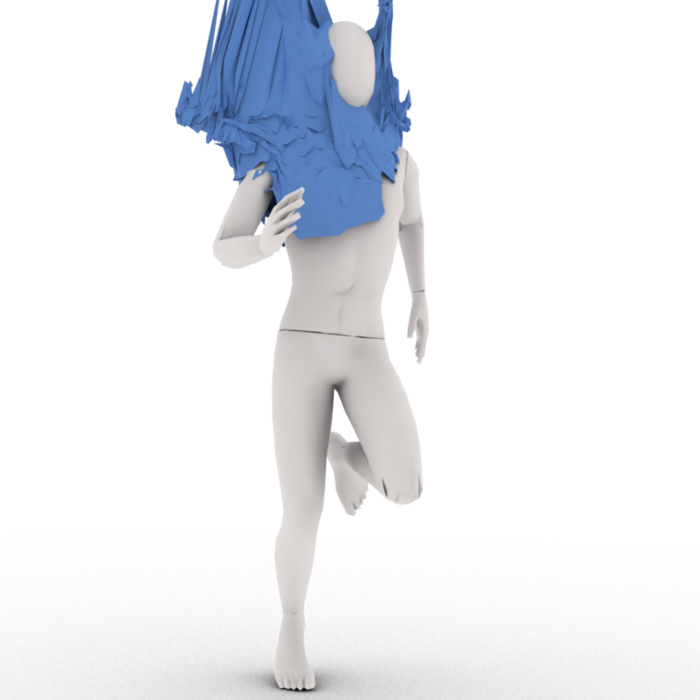} &
        \includegraphics[width=0.22\linewidth]{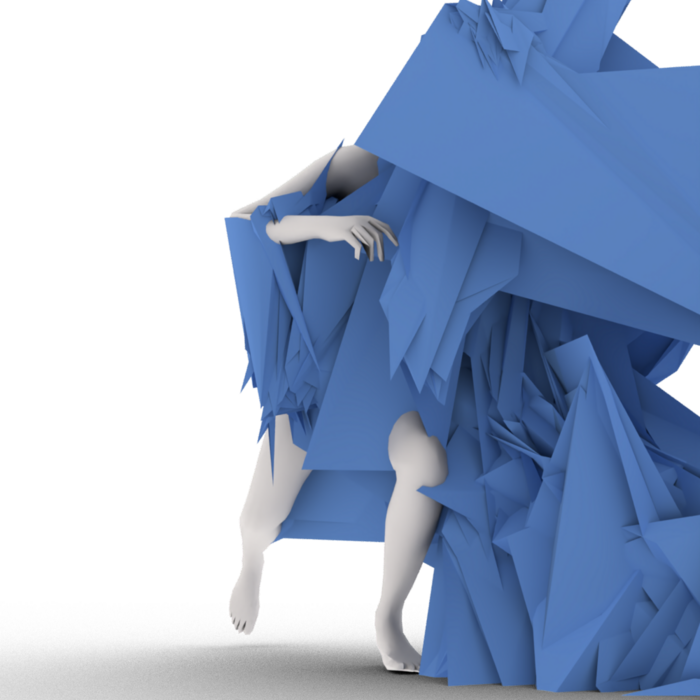} &
        \includegraphics[width=0.22\linewidth]{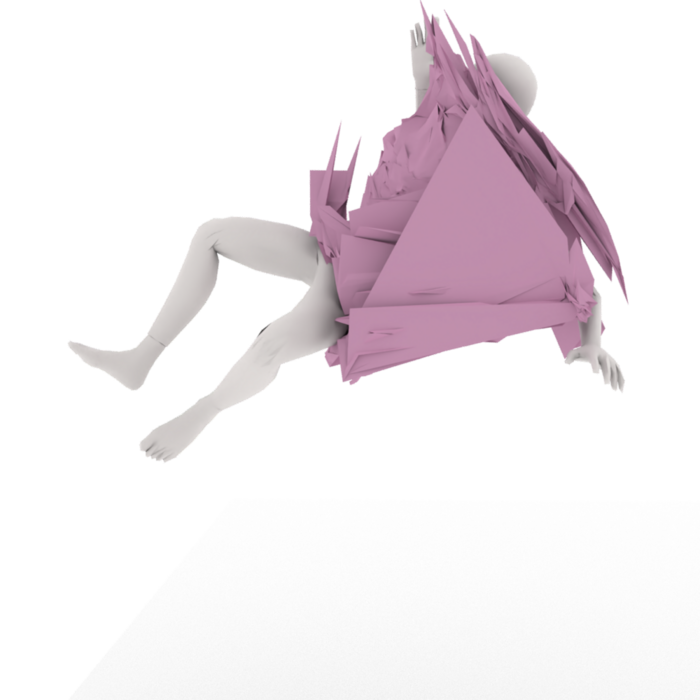} &
        \includegraphics[width=0.22\linewidth]{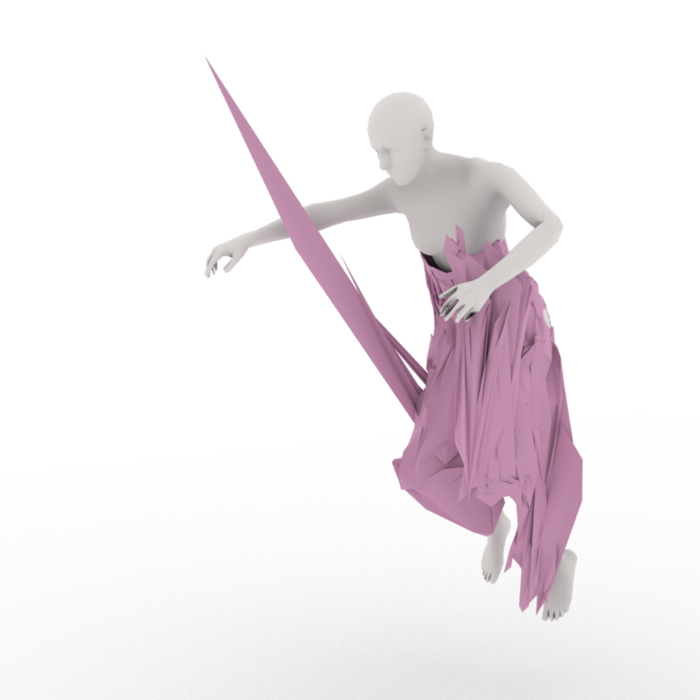} \\
        \rotatebox{90}{\small\hspace{2pt}\textbf{Ours (unified)}} &
        \includegraphics[width=0.22\linewidth]{figs/main_res/run_ours_small.png} &
        \includegraphics[width=0.22\linewidth]{figs/main_res/flip_ours_small.png} &
        \includegraphics[width=0.22\linewidth]{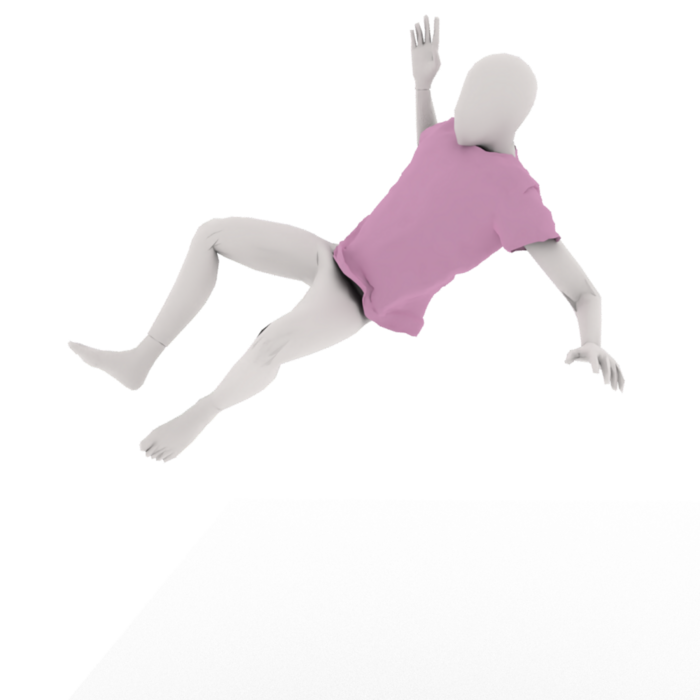} &
        \includegraphics[width=0.22\linewidth]{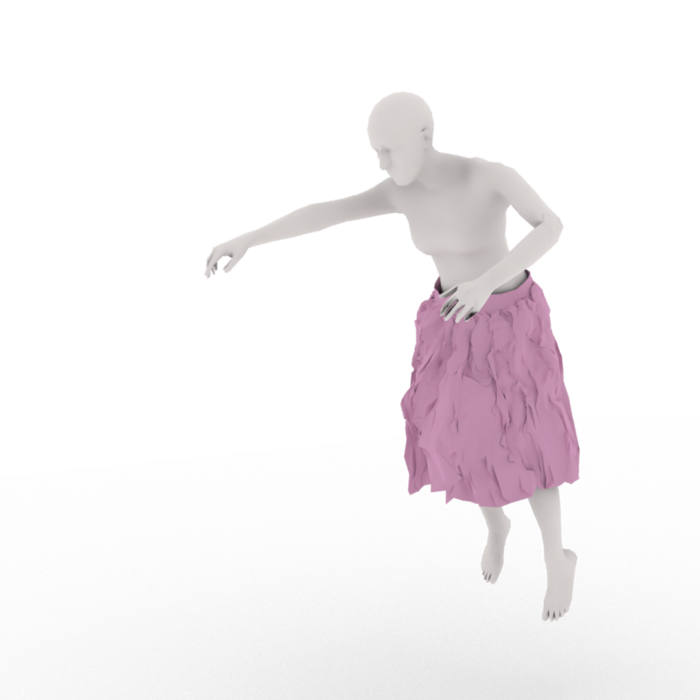} \\
    \end{tabular}
    \caption{\textbf{Unified vs.\ specialized training on the Human Garment scenario.} Each column shows a different frame. \textit{Top}: SOTA GNN trained unsupervised on Human Garment only. \textit{Middle}: SOTA GNN trained unsupervised on the unified dataset (all three scenarios). \textit{Bottom}: our unified model. The single-scenario SOTA GNN looks reasonable but still deviates from ground truth; the unified-training SOTA GNN degrades sharply; our unified model produces visually plausible results that match or exceed the specialized SOTA GNN.}
    \label{fig:unified}
\end{figure}

We provide quantitative and qualitative results for the comparison discussed in Section~\ref{subsec:comparison} of the main paper. SOTA GNN methods (e.g., HOOD~\cite{DBLP:conf/cvpr/GrigorevBH23}, ContourCraft~\cite{DBLP:conf/siggraph/0002BBHT24}) are typically trained with unsupervised physics-based losses. Table~\ref{tab:unified} reports MVE on the Human Garment scenario for the three settings, and Figure~\ref{fig:unified} shows the corresponding visual results. Training the SOTA GNN unsupervised on a single scenario (Human Garment) yields reasonable visual quality on that scenario but still deviates from the ground truth (16.95\,cm MVE). Training the same SOTA GNN unsupervised on the unified dataset (all three scenarios) degrades sharply, with severe stretching artifacts on Human Garment frames (76.92\,cm). Our unified model achieves the lowest MVE (6.92\,cm) and produces visually plausible results that match or exceed the single-scenario specialized SOTA GNN on its own training distribution.

\begin{table}[!ht]
    \centering
    \caption{Quantitative comparison of unified vs.\ specialized training on the Human Garment scenario. MVE (cm) $\downarrow$.}
    \label{tab:unified}
    \begin{tabular}{l|c}
    \toprule
    Method & MVE $\downarrow$ \\
    \midrule
    SOTA GNN (single, unsupervised) & \cellcolor{second}16.95 \\
    SOTA GNN (unified, unsupervised) & 76.92 \\
    \textbf{Ours (unified)} & \cellcolor{best}\textbf{6.92} \\
    \bottomrule
    \end{tabular}
\end{table}

\section{Extreme Cases Analysis Details}
\label{supp:ar_drift}

This section details the analysis summarized in Sec.~\ref{subsec:ar_drift}. We rank 100 Diverse Object Collision test sequences by AR rollout error (raw predictions, no CCD post-processing, same protocol as Table~\ref{tab:main_results}; split-mean MVE: 9.03\,cm) and select the 95th-percentile sequence.

\begin{figure}[t]
  \centering
  \includegraphics[width=0.72\linewidth]{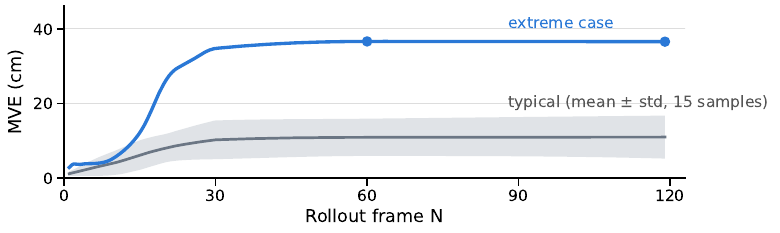}
  \caption{AR rollout error on the 95th-percentile Diverse Object Collision test sequence (blue), compared against 15 randomly sampled typical test sequences (gray: mean $\pm$ one standard deviation). Curves show the per-frame MVE over the first 30 frames and at horizons $N{=}60/120$ (markers), smoothly interpolated in between. Typical sequences quickly settle at a bounded error consistent with Table~\ref{tab:main_results}, whereas the extreme case accumulates roughly $4{\times}$ that level.}
  \label{fig:ar_drift_curve}
\end{figure}

Table~\ref{tab:ar_drift} reports the MVE of the AR rollout at increasing horizons, and Figure~\ref{fig:ar_drift_curve} plots the per-frame MVE over the full rollout. The error accumulates rapidly in the early rollout---2.9\,cm at frame 1, 5.2\,cm at frame 10, and 27.1\,cm at frame 20---and then plateaus at ${\sim}37$\,cm, roughly $4{\times}$ the split-mean level: early deviations feed back into subsequent inputs and drive the cloth toward an incorrect yet self-consistent resting state, in which later predictions are locally stable but globally wrong, and from which the rollout does not recover. This drift persists despite velocity-based inputs and the 5-step rollout curriculum (Sec.~\ref{subsec:implementation}).

\begin{table}[t]
  \centering
  \caption{AR rollout error (MVE, cm) at rollout horizon $N$ on the 95th-percentile Diverse Object Collision test sequence, compared with the average over 15 randomly sampled typical test sequences. The values in parentheses denote the cumulative mean MVE over the first $N$ frames.}
  \label{tab:ar_drift}
  {\small
  \begin{tabular}{l|ccc}
  \toprule
  & $N{=}60$ & $N{=}120$ & Full (240) \\
  \midrule
  Extreme case & 36.6 (26.3) & 36.5 (31.4) & 36.7 (34.0) \\
  Typical (15 samples) & 10.9 (8.5) & 11.0 (9.7) & 10.9 (10.3) \\
  \bottomrule
  \end{tabular}%
  }
\end{table}

\section{More Results}
\label{supp:more_results}

\begin{figure}[t]
    \centering
    \setlength{\tabcolsep}{0pt}
    \begin{tabular}{cccccc}
        & Ground Truth & SOTA GNN & MAT & LayersNet & \textbf{Ours} \\
        \rotatebox{90}{\small Jump Forward Kick} &
        \includegraphics[width=0.18\linewidth]{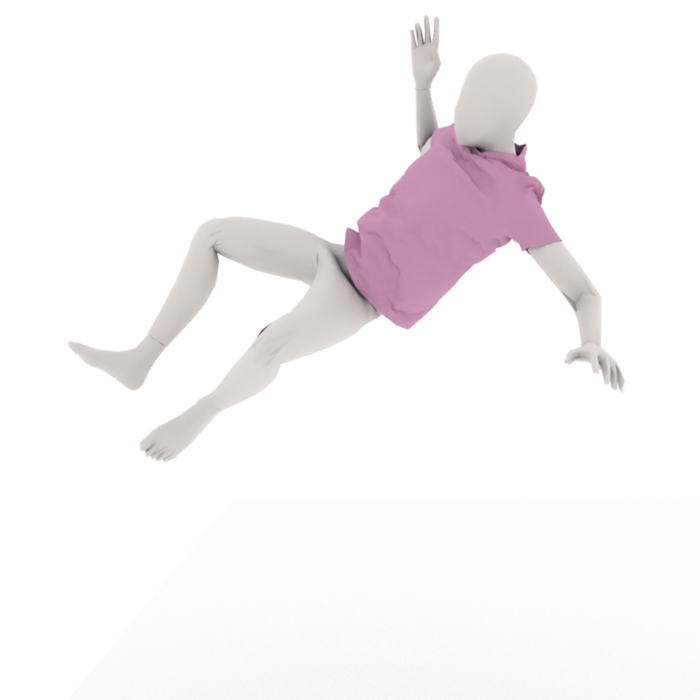} &
        \includegraphics[width=0.18\linewidth]{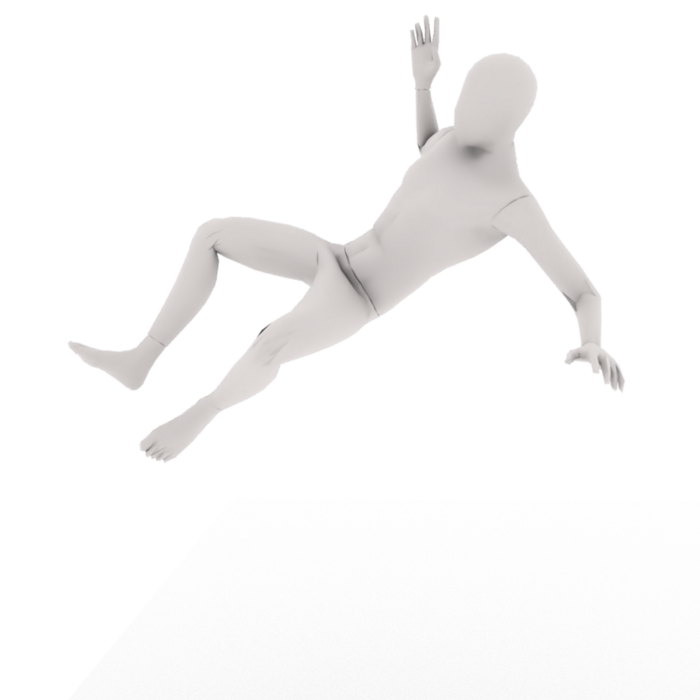} &
        \includegraphics[width=0.18\linewidth]{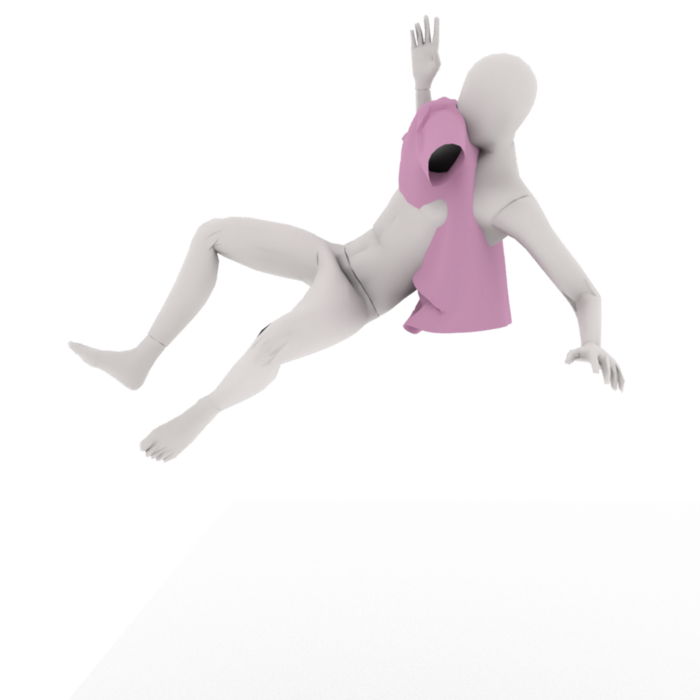} &
        \includegraphics[width=0.18\linewidth]{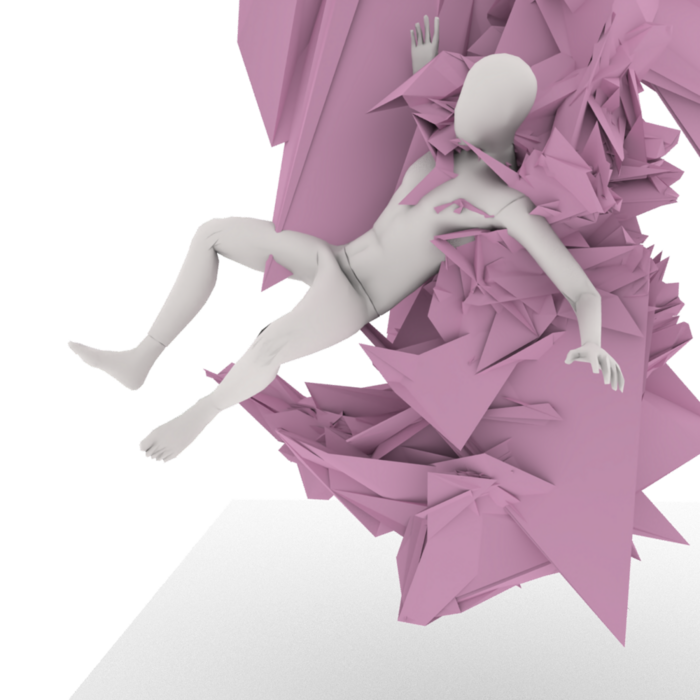} &
        \includegraphics[width=0.18\linewidth]{figs/supp/more_0_ours_small.png} \\
        \rotatebox{90}{\small Front Flip} &
        \includegraphics[width=0.18\linewidth]{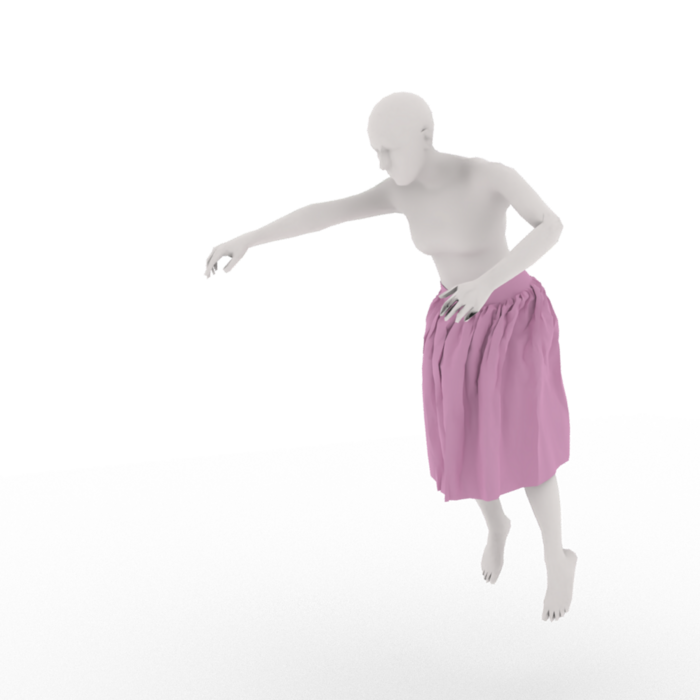} &
        \includegraphics[width=0.18\linewidth]{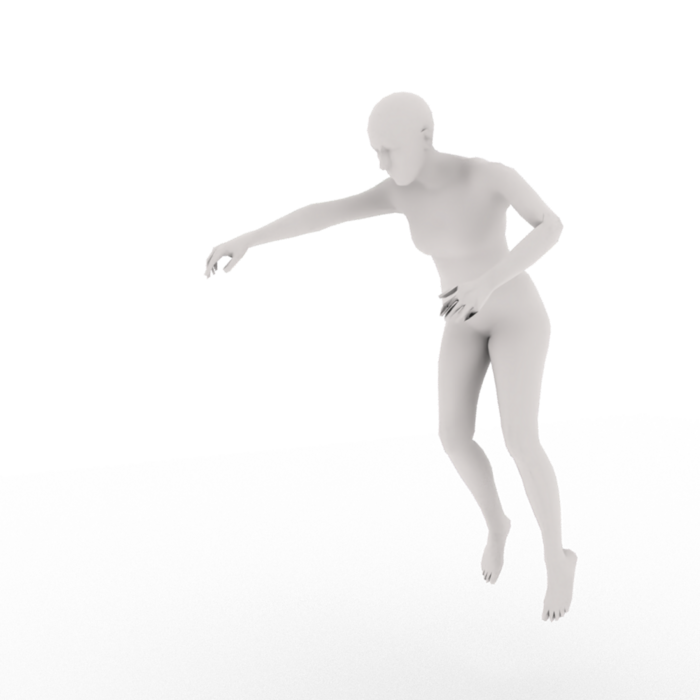} &
        \includegraphics[width=0.18\linewidth]{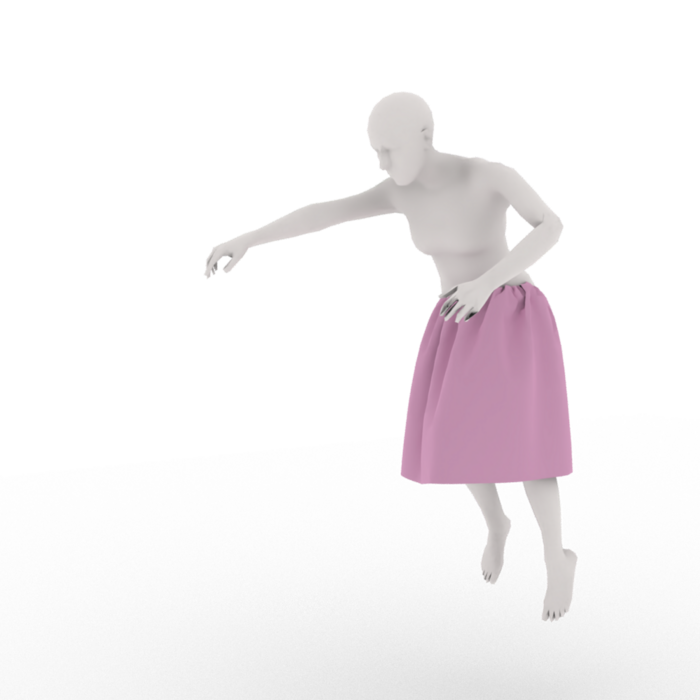} &
        \includegraphics[width=0.18\linewidth]{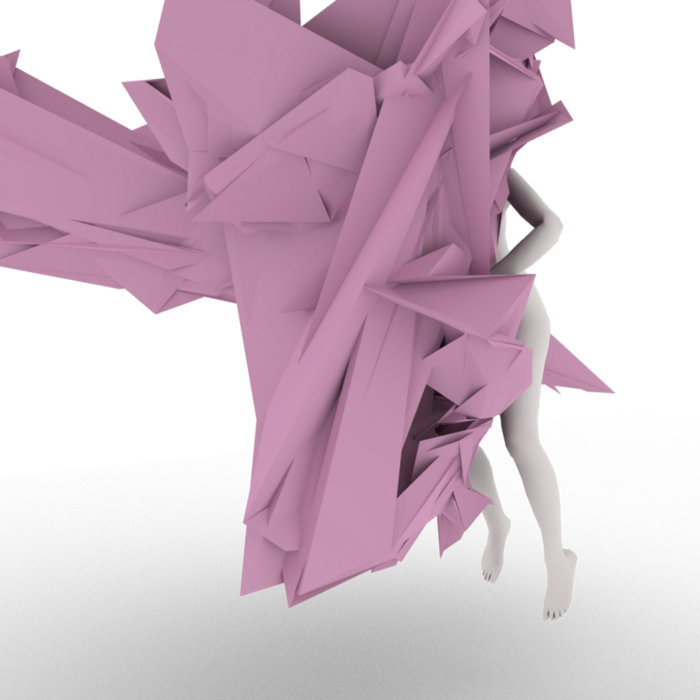} &
        \includegraphics[width=0.18\linewidth]{figs/supp/more_1_ours_small.png} \\
        \rotatebox{90}{\small Guan Dao} &
        \includegraphics[width=0.18\linewidth]{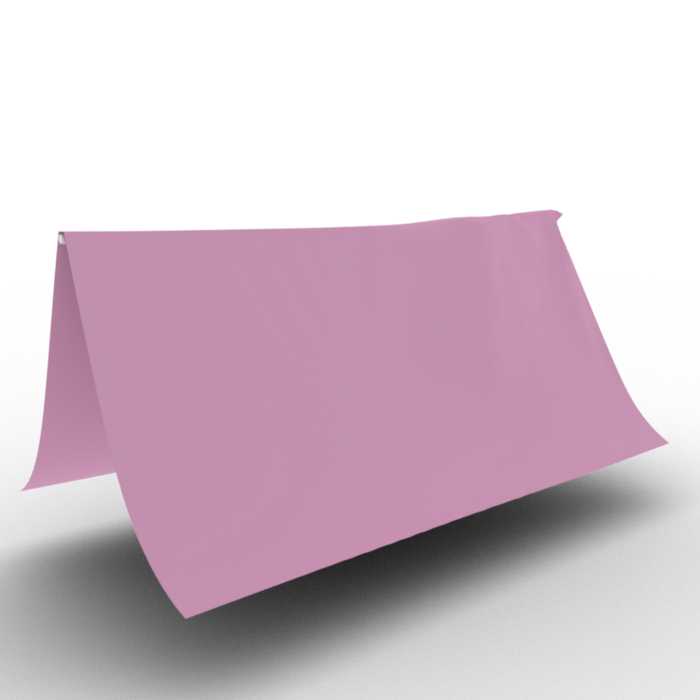} &
        \includegraphics[width=0.18\linewidth]{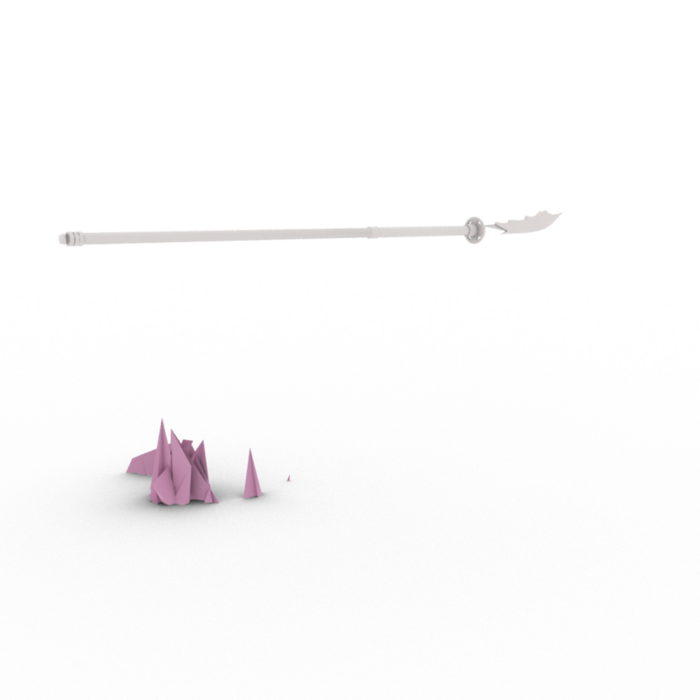} &
        \includegraphics[width=0.18\linewidth]{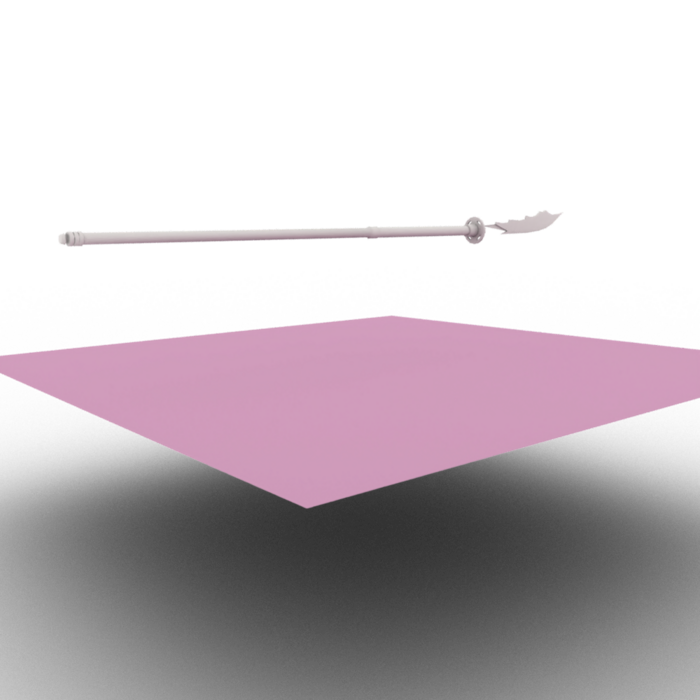} &
        \includegraphics[width=0.18\linewidth]{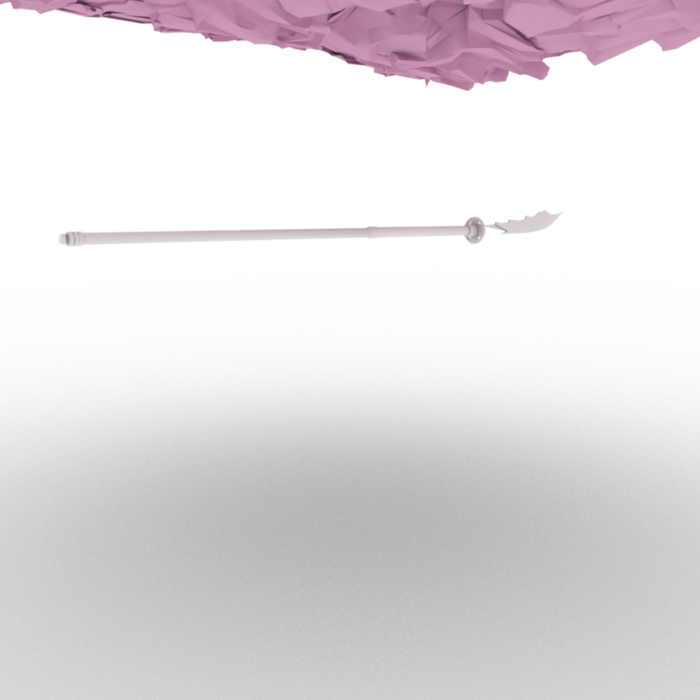} &
        \includegraphics[width=0.18\linewidth]{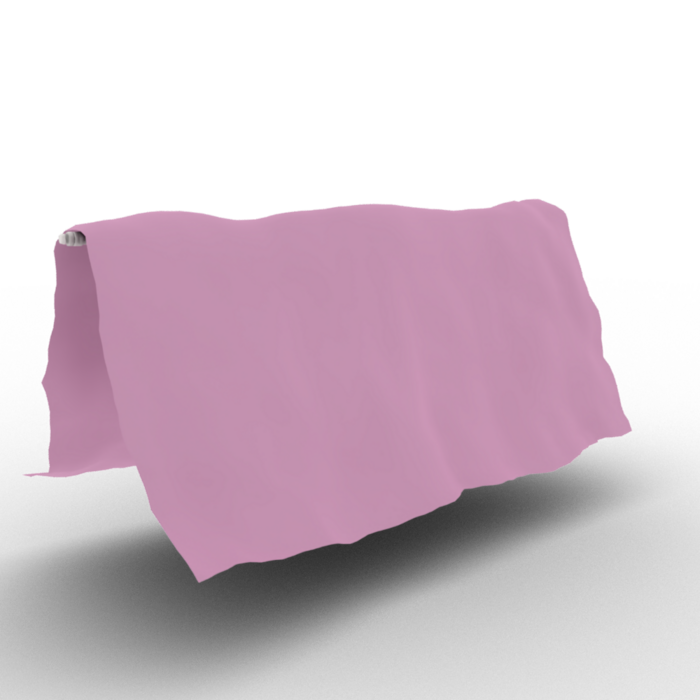} \\
        \rotatebox{90}{\small Character} &
        \includegraphics[width=0.18\linewidth]{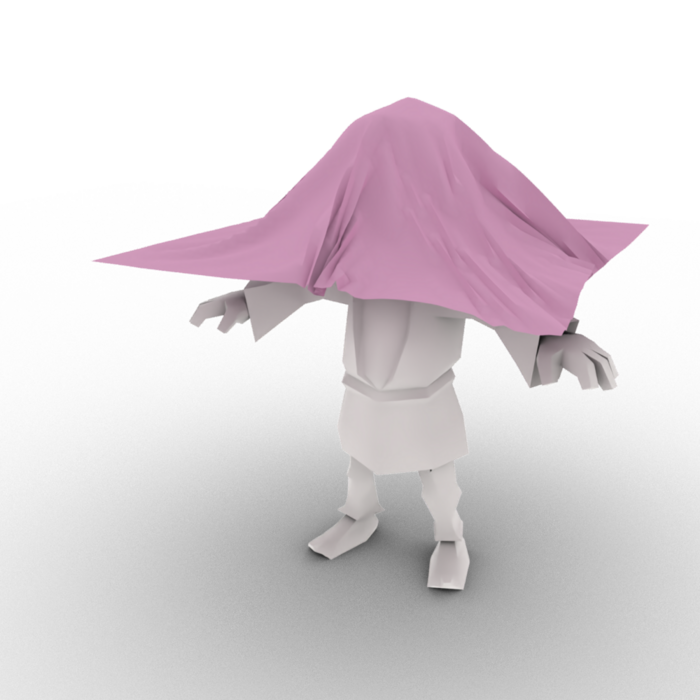} &
        \includegraphics[width=0.18\linewidth]{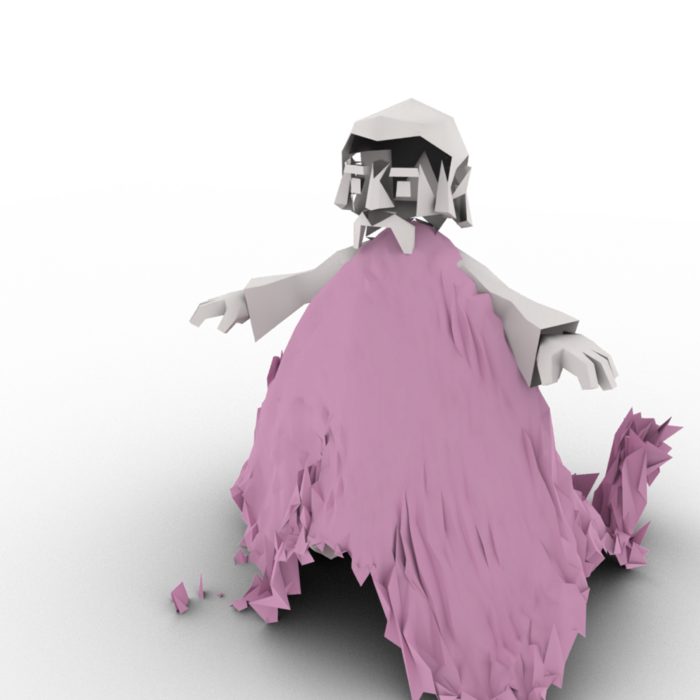} &
        \includegraphics[width=0.18\linewidth]{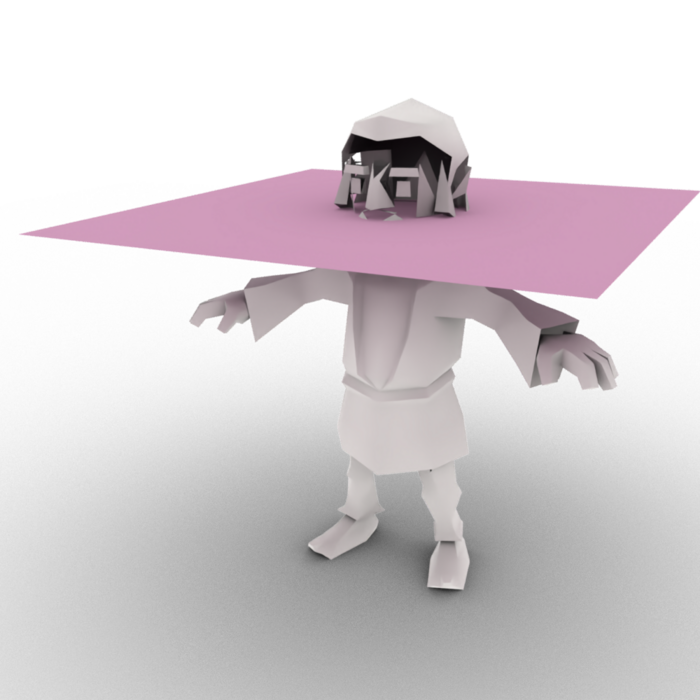} &
        \includegraphics[width=0.18\linewidth]{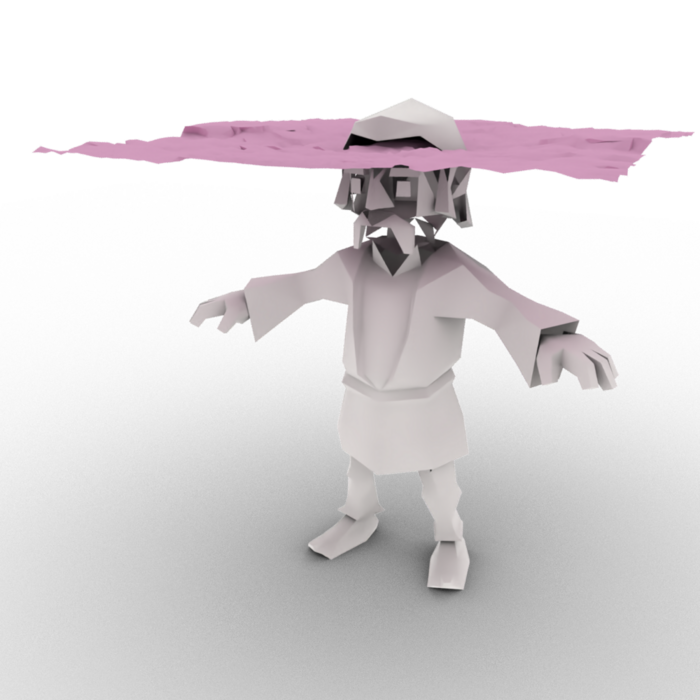} &
        \includegraphics[width=0.18\linewidth]{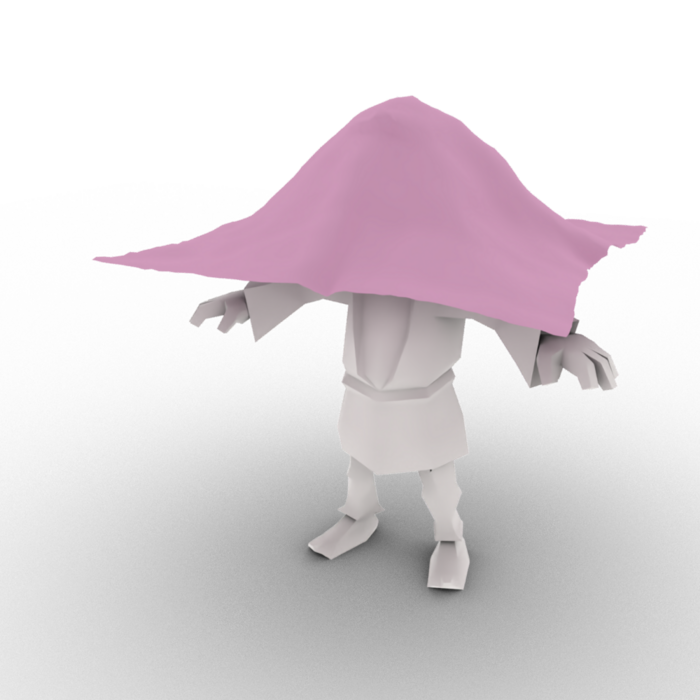} \\
        \rotatebox{90}{\small Grasp 1} &
        \includegraphics[width=0.18\linewidth]{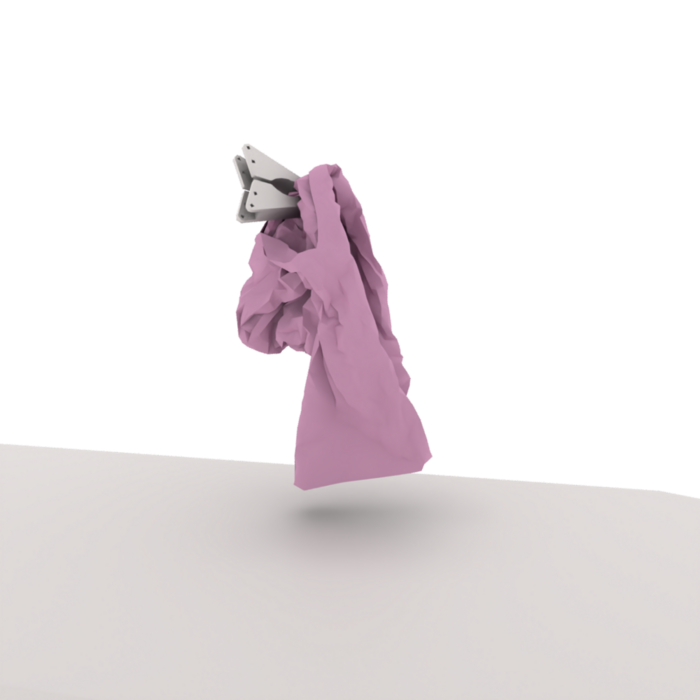} &
        \includegraphics[width=0.18\linewidth]{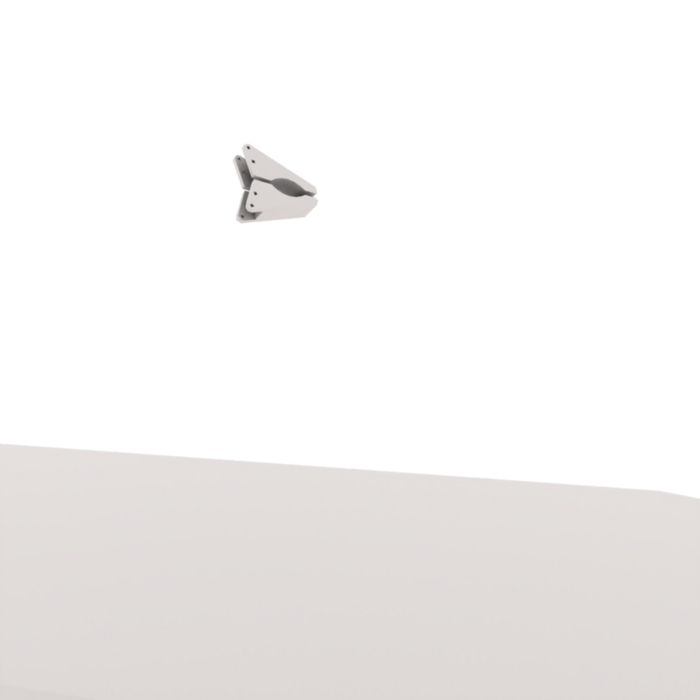} &
        \includegraphics[width=0.18\linewidth]{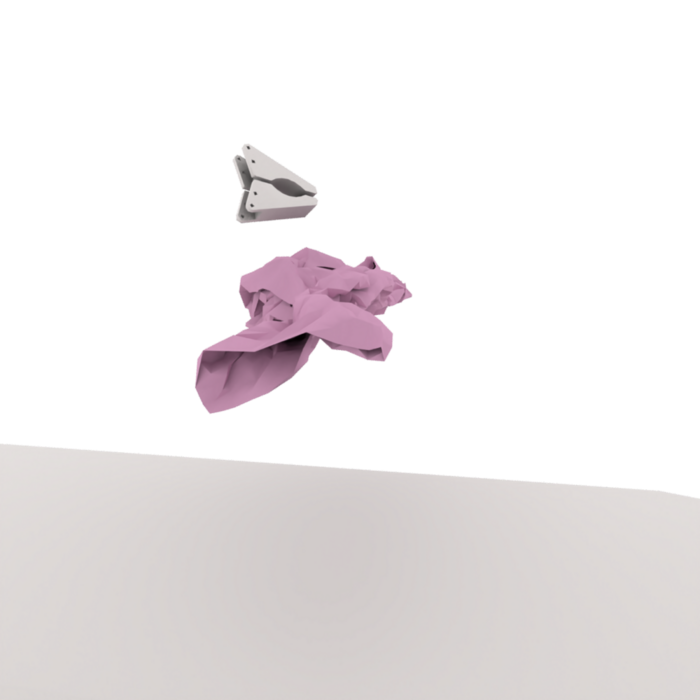} &
        \includegraphics[width=0.18\linewidth]{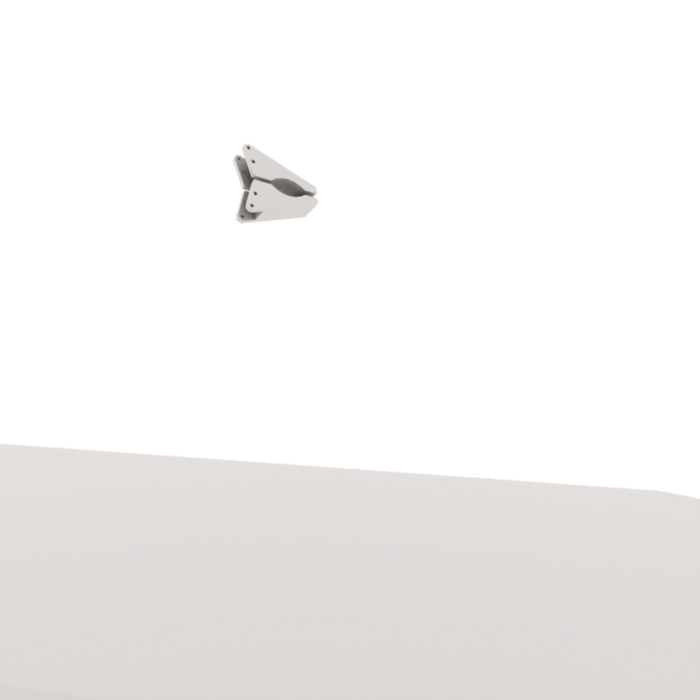} &
        \includegraphics[width=0.18\linewidth]{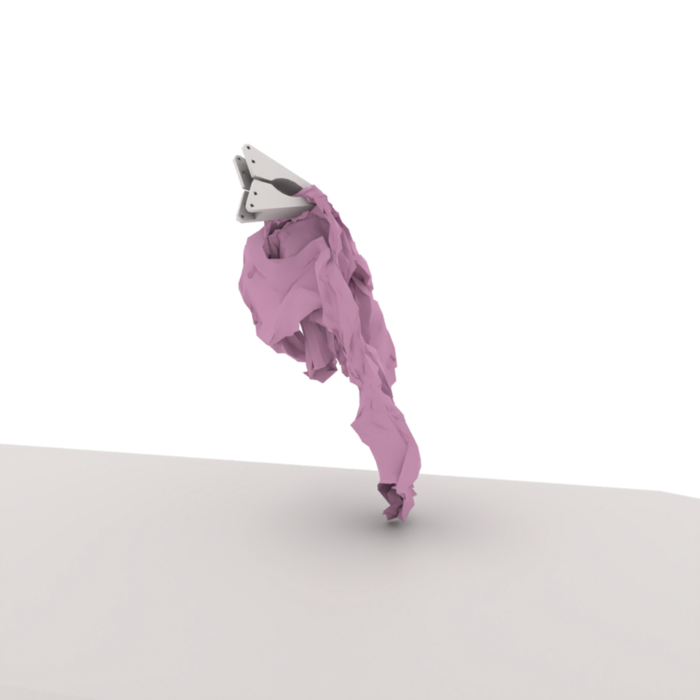} \\
        \rotatebox{90}{\small Grasp 2} &
        \includegraphics[width=0.18\linewidth]{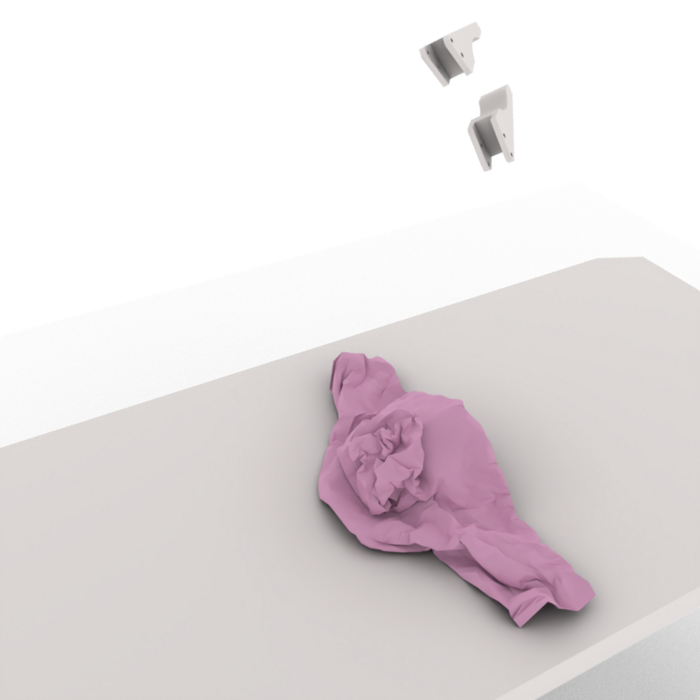} &
        \includegraphics[width=0.18\linewidth]{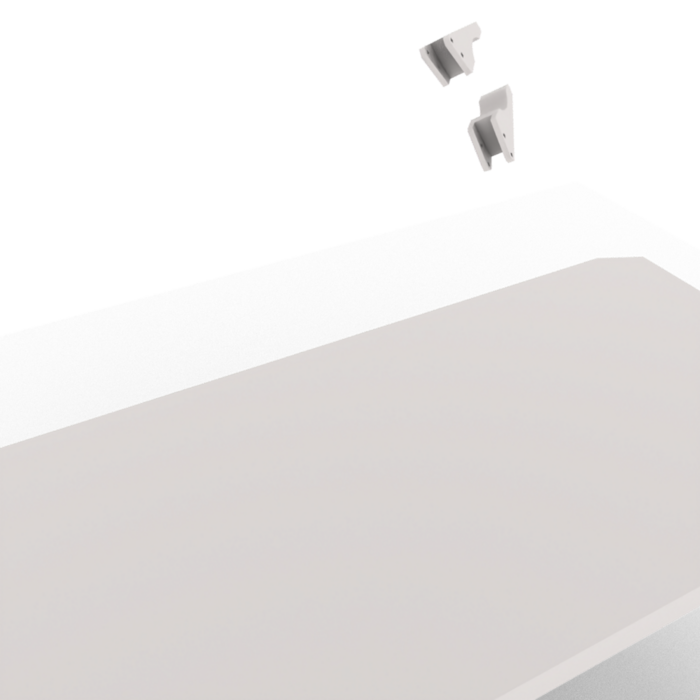} &
        \includegraphics[width=0.18\linewidth]{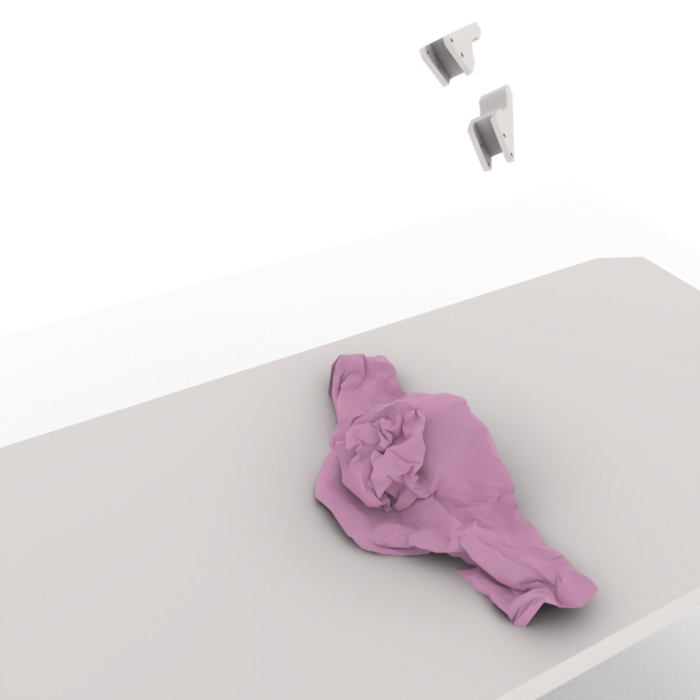} &
        \includegraphics[width=0.18\linewidth]{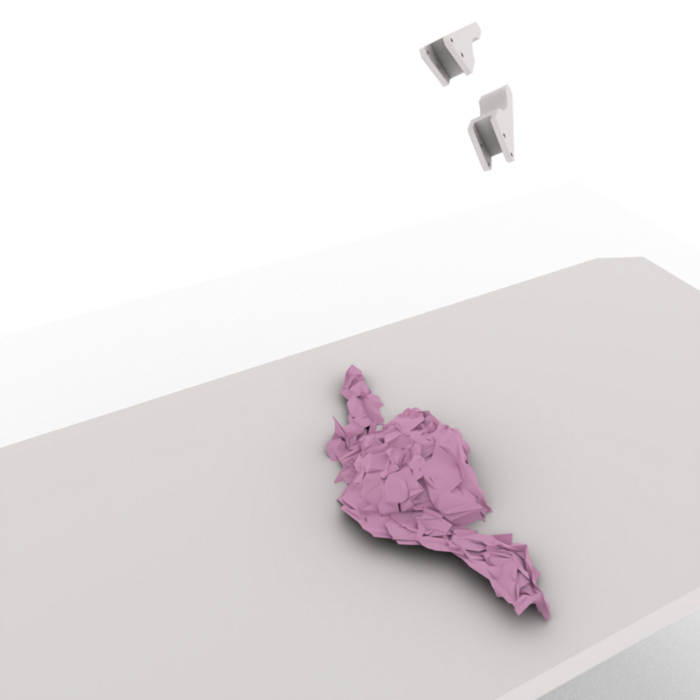} &
        \includegraphics[width=0.18\linewidth]{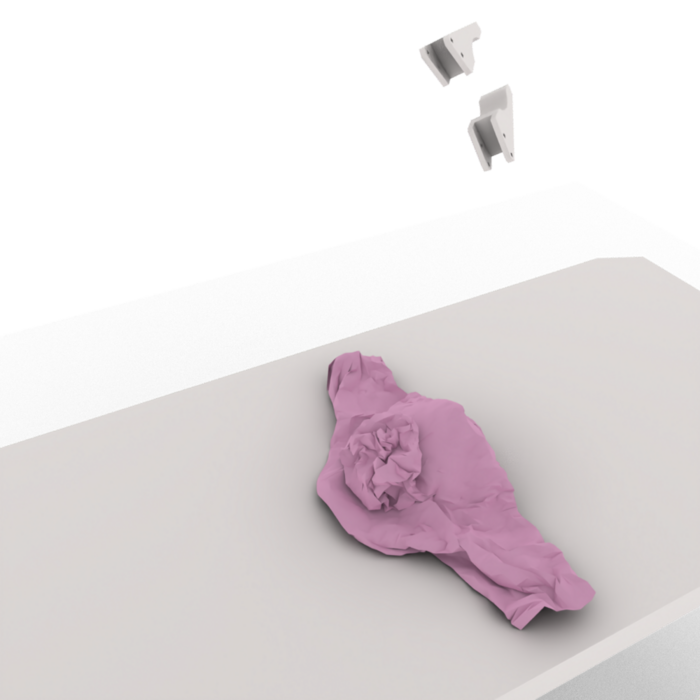} \\
    \end{tabular}
    \caption{Additional qualitative comparisons. Each row shows a different scenario. From left to right: Ground Truth, the SOTA GNN~\cite{DBLP:conf/cvpr/GrigorevBH23, DBLP:conf/siggraph/0002BBHT24}, MAT~\cite{DBLP:journals/cgf/LiWKCS24}, LayersNet~\cite{DBLP:conf/iccv/ShaoL023}, and our method.}
    \label{fig:more_results}
\end{figure}

We present additional qualitative comparisons between our method and three baselines (the SOTA GNN~\cite{DBLP:conf/cvpr/GrigorevBH23, DBLP:conf/siggraph/0002BBHT24}, MAT~\cite{DBLP:journals/cgf/LiWKCS24}, and LayersNet~\cite{DBLP:conf/iccv/ShaoL023}) across diverse scenarios. As shown in Fig.~\ref{fig:more_results}, our method consistently produces more accurate cloth geometry with fewer artifacts.

\end{document}